\documentclass[12pt]{iopart}
\usepackage{amssymb}
\usepackage{feynmp}
\usepackage{epic,eepic}

\usepackage{ulem}
\usepackage{lineno}

\usepackage{wrapfig,epsfig}
\expandafter\let\csname equation*\endcsname\relax
\expandafter\let\csname endequation*\endcsname\relax
\usepackage{amsmath}
\usepackage{color}

\definecolor{mycolor}{rgb}{0.6,0.0,0.4}

\def\gtap{\ \raise.3ex\hbox{$>$\kern-.75em\lower1ex\hbox{$\sim$}}\ }
\def\ltap{\ \raise.3ex\hbox{$<$\kern-.75em\lower1ex\hbox{$\sim$}}\ }

\def\slashchar#1{\setbox0=\hbox{$#1$}
   \dimen0=\wd0 \setbox1=\hbox{/} \dimen1=\wd1
   \ifdim\dimen0>\dimen1 \rlap{\hbox to \dimen0{\hfil/\hfil}} #1
   \else  \rlap{\hbox to \dimen1{\hfil$#1$\hfil}} / \fi}
\def\P{\slashchar{P}}

\usepackage{bm,longtable,latexsym}
\usepackage{graphicx}

\begin{document}
\begin{flushright}
 KEK-TH-1905, J-PARC-TH-0052 
\end{flushright}
%
\title{Towards a Unified Model of Neutrino-Nucleus Reactions for
Neutrino Oscillation Experiments}

\author{
S.X. Nakamura$^1$, H. Kamano$^{2,3}$,
Y. Hayato$^4$, M. Hirai$^5$, \\
W. Horiuchi$^6$, S. Kumano$^{2,3}$, T. Murata$^1$, 
K. Saito$^{7,3}$,\\ M. Sakuda$^8$, T. Sato$^{1,3}$, Y. Suzuki$^{9,10}$
}
\address{
$^1$ Department of Physics, Osaka University, Toyonaka, Osaka 560-0043, Japan\\
$^2$ KEK Theory Center, Institute of Particle and Nuclear Studies, KEK\\
\ \   1-1, Ooho, Tsukuba, Ibaraki, 305-0801, Japan\\
$^3$ J-PARC Branch, KEK Theory Center,
     Institute of Particle and Nuclear Studies,\\ \ \ KEK
     and Theory Group, Particle and Nuclear Physics Division, J-PARC Center,\\
\ \  203-1, Shirakata, Tokai, Ibaraki, 319-1106, Japan\\
$^4$ Kamioka Observatory, Institute for Cosmic Ray Research, 
     University of Tokyo, Kamioka, Japan\\
$^5$ Nippon Institute of Technology, Saitama 345-8501, Japan\\
$^6$ Department of Physics, Hokkaido University, Sapporo 060-0810, Japan\\
$^7$ Department of Physics, Tokyo University of Science, Noda 278-8510, Japan\\
$^8$ Department of Physics, Okayama University, Okayama 700-8530, Japan\\
$^{9}$ Department of Physics, Niigata University, Niigata 950-2181, Japan\\
$^{10}$ RIKEN Nishina Center, Wako 351-0198, Japan
}
 \ead{nakamura@kern.phys.sci.osaka-u.ac.jp}
\vspace{10pt}
\begin{indented}
\item September 30, 2016
\end{indented}

\begin{abstract}
A precise description of neutrino-nucleus reactions
will play a key role in addressing fundamental questions
such as the leptonic CP violation and the neutrino mass hierarchy
through analyzing data from next-generation neutrino oscillation
experiments.
The neutrino energy relevant to the neutrino-nucleus reactions spans
a broad range and, accordingly,
the dominant reaction mechanism varies across the energy region
from quasi-elastic scattering through nucleon resonance excitations
to deep inelastic scattering. This corresponds to transitions of
the effective degree of freedom for theoretical description from nucleons
through meson-baryon to quarks.
The main purpose of this review is to report our recent efforts towards a
unified description
of the neutrino-nucleus reactions over the wide energy range;
recent overall progress in the field is also sketched.
Starting with an overview of the current status of neutrino-nucleus
scattering experiments,
we formulate the cross section to be commonly used for the reactions
over all the energy regions.
A description of the neutrino-nucleon reactions follows and, in particular,
a dynamical coupled-channels model for meson productions in and beyond
the $\Delta$(1232) region is discussed in detail.
We then discuss the neutrino-nucleus reactions,
putting emphasis on our theoretical approaches.
We start the discussion with electroweak processes in few-nucleon systems
studied with the correlated Gaussian method.
Then we describe quasi-elastic scattering with nuclear spectral functions,
and
meson productions with a $\Delta$-hole model.
Nuclear modifications of the parton distribution functions determined
through a global analysis are also discussed.
Finally, we discuss issues to be addressed for future developments.
\end{abstract}

\pacs{13.15.+g, 12.15.Ji, 14.60.Pq, 25.30.Pt}


\vspace{2pc}
\noindent{\it Keywords}: neutrino-nucleus interaction, neutrino oscillation

\submitto{Reports on Progress in Physics}

 
%

\section{Introduction}

Extensive researches on reactors, accelerators, solar and atmospheric neutrinos
have revealed fundamental properties of the neutrino~\cite{PDG14,Gonzalez2016,Capozzi2014,Forero2014}. 
Current objectives of  neutrino experiments are
to precisely determine the neutrino mixing angles and CP violating
 phase, and to solve the neutrino mass hierarchy problem.
Those neutrino properties will be studied with
the long-baseline neutrino oscillation
experiments such as HK~\cite{HK1,HK2} and DUNE~\cite{DUNE} near future.
To extract the neutrino properties from the neutrino oscillation experiments,
one of the major sources of systematic errors is uncertainties in
neutrino-nucleus reaction cross sections.
Actually, these uncertainties are already
one of dominant sources of the systematic errors in the recent
neutrino oscillation experiments like T2K. For example, total
systematic errors of the number of $\nu_e$ and $\bar{\nu_e}$
appearance events are ${\cal O}$(6\%), and about half of the errors is coming from the
uncertainties in the neutrino-nucleus reaction cross
sections~(Table~XX of Ref.~\cite{t2k_nue}; Sec.~V\!~F\!~3 of Ref.~\cite{t2k_bnue}).
Therefore, reducing these uncertainties is one of the most important
tasks for the currently running and also for the future high precision
experiments.
Thus a quantitative understanding of the neutrino-nucleus reactions at
the level of a few percent accuracy is required to achieve the
above-mentioned objectives of the neutrino oscillation
experiments~\cite{HK1,HK2,DUNE,Gallagher2011,FZ_review,Luis2014,Garvey2015,BHMD_review,mosel_review,KM_review}.

\begin{figure}
\begin{center}
\includegraphics[width=0.4\textwidth]{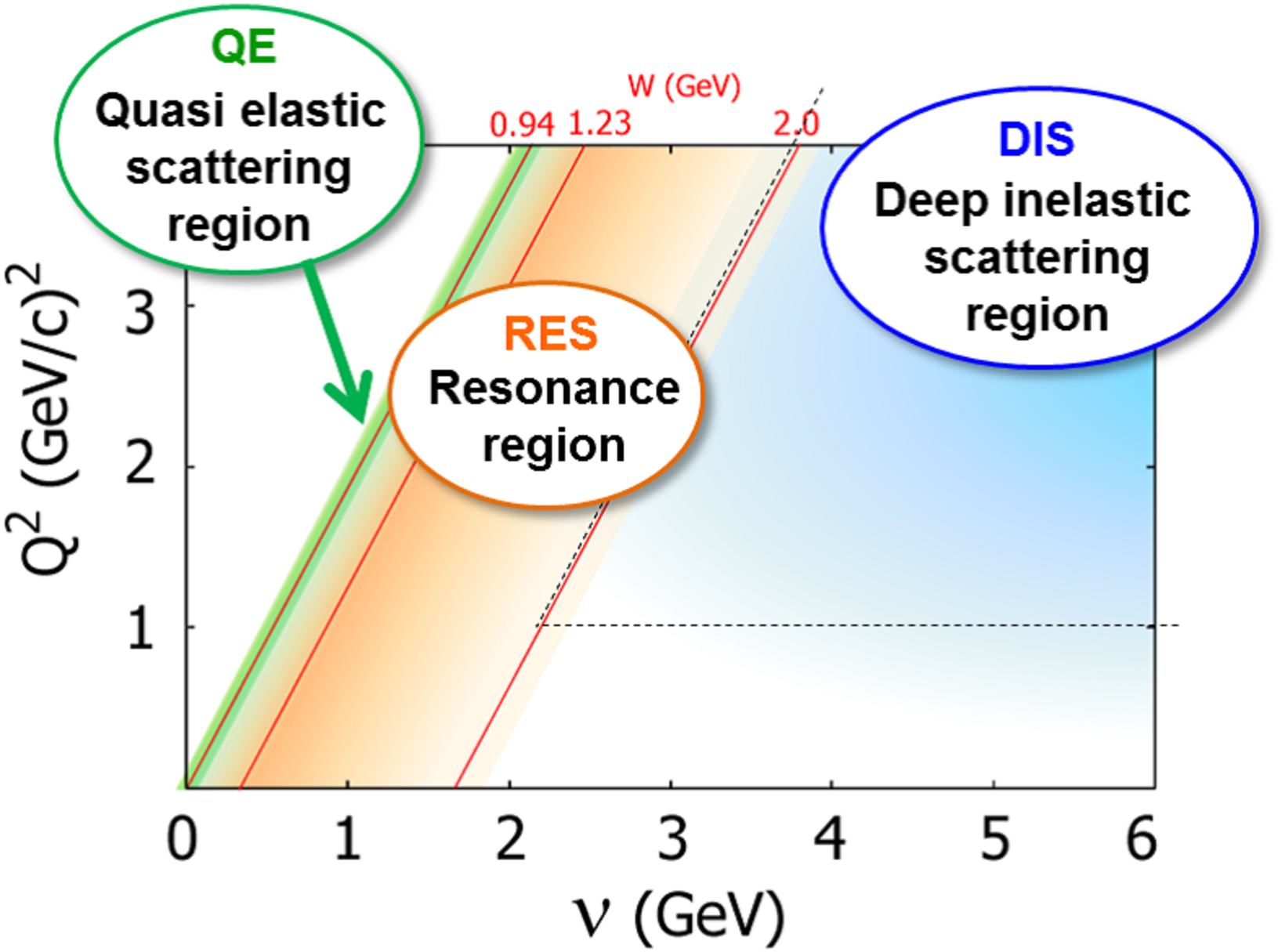}
\caption{
Kinematical regions of the neutrino-nucleus interaction relevant to 
the next-generation neutrino-oscillation experiments.
The energy transfer to a nucleus and the squared four-momentum transfer are denoted 
by $\nu$ and $Q^2$, respectively. 
}
\end{center}
\label{fig:kinem}
\end{figure}

The neutrino energy relevant to the oscillation experiments
spans from several hundred MeV to tens of GeV, and thus
the neutrino-nucleus reactions over a wide kinematical region need to be understood.
From the low to high energy side, the neutrino-nucleus
reaction is characterized by the quasi-elastic (QE), resonance (RES),
and deep inelastic scattering (DIS) regions (Fig.~\ref{fig:kinem}). 
The neutrino-nucleus reactions in each of the regions have quite different
characteristics and, accordingly, effective degrees of freedom for
theoretical descriptions are quite different. 
In the QE region, an incident neutrino interacts with one of nucleons
inside a nucleus quasi-elastically, and thus nucleons are the effective
degrees of freedom.
Meanwhile, in the RES region, the internal structure of a scattered nucleon is excited to a
resonant state that subsequently decays into a meson-baryon final state;
here a meson-baryon dynamics plays a central role.
Finally, in the DIS region, a high-energy neutrino even directly sees
the subcomponent of the nucleon: the quarks and gluons, or collectively
the partons.
Perturbative QCD and non-perturbative parton
distributions in the nucleon (bound in a nucleus)
are basic ingredients for a theoretical
description. 

Although we roughly divided the kinematical region into three based on
the reaction mechanisms, the reality is more complicated. 
For example, it is well-known that the QE and RES regions overlap to
form the so-called `dip' region (the dip between the QE and $\Delta(1232)$
peaks of the nuclear response), and there, 
processes that involve more than single nucleon (often called
two-particle two-hole (2p2h) or $n$-particle $n$-hole ($n$p$n$h) 
processes) give an important contribution.
An example of the 2p2h processes is a process where
a pion produced through a $\Delta(1232)$-excitation
is absorbed by a surrounding nucleon, leading to a two-nucleon emission.
Therefore, in order to understand the dip region,
the elementary single nucleon amplitudes for the QE and RES regions should
 be consistently implemented in a nuclear many-body theory.
We also point out that the kinematical region has not necessarily been 
correctly divided previously.
Namely, multi-pion emission rates beyond the $\Delta(1232)$ but still
within the (higher-)resonance region have been
often estimated with a parton (DIS) model that is extrapolated to the lower $W$ region 
($W$: total hadron energy) where the model is in principle not valid.
It is important to correctly define the RES and DIS regions
considering $W$ and $Q^2$, and construct a model suitable for each of the regions.

Obviously, it is essential to combine different areas of expertise to
construct `a unified model' for the neutrino-nucleus reactions covering
all of the kinematical regions discussed above. 
Here, `a unified model' does {\it not} mean a single theoretical framework 
that works over all the kinematical region in question. 
Rather it is constructed by consistently
combining baseline models for each of the kinematical
regions characterized by the reaction mechanisms,
so that transitions between the different kinematical regions can also
be well described.
In this sense, there already exist several unified models such as
neutrino interaction generators (the NEUT~\cite{neut}, the GENIE~\cite{genie}, and
the NuWro~\cite{nuwro}) that are often used in analyzing data from neutrino
experiments.
The GiBUU~\cite{gibuu} that particularly features a semi-classical hadron
transport can also be regarded as a unified model. 
However, it would be still hoped to develop 
a new unified model that consists of theoretically and
also phenomenologically more well-founded models.
Thus, to tackle this issue,
experimentalists and theorists recently
got together to form a collaboration at the J-PARC Branch of KEK Theory Center~\cite{KEKJ}. 
The ultimate goal of this collaboration is to develop a unified model
that comprehensively describes the neutrino-nucleus reaction over the QE, RES, and DIS regions.
A general outline of our strategy to achieve this goal is the following:

\begin{enumerate}
\item We first develop baseline models describing the QE, RES, and DIS
regions individually,
by applying appropriate physics mechanisms and theoretical treatments,
as mentioned above,
to each kinematical region.

\item We then connect the hadronic model describing the QE and RES regions
to the perturbative QCD model of DIS by matching the cross sections
and/or structure functions
computed from the two at certain points or region in the ($Q^2, \nu$) plane,
where the transition of the basic degrees of freedom (hadrons versus
quarks and gluons)
of the reactions is expected to occur.
\end{enumerate}

The purpose and also the unique feature of this article is 
to report the current status of developing, based on our own approaches, 
the baseline models for each of the kinematical regions
and discuss a future perspective towards a unified 
neutrino-nucleus reaction model that consists of those baseline models.
On the other hand, 
this article is not intended to comprehensively review all the
developments in the field of the neutrino-nucleus reactions on equal
footing, although we also cite and sketch other approaches and their
recent developments. 
There exist several review papers on the neutrino-nucleus
scattering physics~\cite{Gallagher2011,FZ_review,Luis2014,Garvey2015,BHMD_review,mosel_review,KM_review}
as briefly introduced in the following, and we refer readers to those papers to find more
overall developments in the field.
Extensive compilation and explanation of existing data for
neutrino-nucleon and neutrino-nucleus reactions over
the low-energy, QE, RES, and DIS regions are given in Ref.~\cite{FZ_review}.
References~\cite{Gallagher2011,Garvey2015} particularly focused on the
QE processes, summarizing recent theoretical and experimental results,
and issues to be resolved. 
Theoretical approaches to nuclear many-body problems 
particularly relevant to the QE processes 
along with the influence of the theoretical treatment on the
determination of the neutrino oscillation parameters
are discussed in Ref.~\cite{BHMD_review}.
Reference~\cite{mosel_review} discusses neutrino interaction generators
and particularly the transport approach GiBUU~\cite{gibuu}, and their
role on the energy reconstruction of incident neutrinos and thus the
determination of the neutrino oscillation parameters.
References~\cite{Luis2014,KM_review}
focus on the neutrino interactions on nucleon and nucleus
up to a few GeV, putting emphasis on
recent developments in the QE-like processes involving 
multi-nucleon mechanisms;
incoherent and coherent meson and photon
productions are also discussed.

We spend the rest of this section to describe the organization of this
article, and also specify key questions to be
addressed in each section.
\begin{itemize}
 \item[Sec. 2:]
We review the current experiments on neutrino-nucleus reactions and the
understanding of the data in terms of neutrino reaction generators.
Then we summarize open questions and future prospects. 
These form the introduction of this article.
 \item[Sec. 3:]
We present a cross section formula for neutrino-nucleon(nucleus)
	      reactions for all kinematical regions, and discuss
neutrino-nucleon reaction models that are the key building blocks of
neutrino-nucleus reaction models. 
Then we give a dedicated discussion on 
our recent development of a dynamical coupled-channels model for the
	      whole RES region.
There has been a strong demand to develop a model that works
	      well in a region between the $\Delta(1232)$
and the boundary with the DIS where a reliable model has
been missing.
Such a model should be able to describe the resonant character of the
reactions and important two-pion productions. 
Our development meets this demand.
We have achieved, for the first time, to develop a neutrino-nucleon reaction model that fully satisfies
the coupled-channels unitarity. The model is constructed from the analysis of pion-, photon-
and electron-induced reaction data including multi-meson productions. Comparisons with
currently available models are also given.
\end{itemize}
We then move on to the neutrino-nucleus 
reactions. Though a main feature of the neutrino-nucleus reactions in the QE, RES and DIS regions
can be understood qualitatively from the corresponding elementary processes,
an accurate description is very difficult
because of the involved nuclear many-body problem. The experience
and knowledge on both the reactions and structures of nuclei accumulated in the nuclear
physics must be integrated to understand the whole processes of neutrino-nucleus reactions.
\begin{itemize}
 \item[Sec. 4:]
We discuss the low-energy neutrino reactions in
few-nucleon systems. 
This system is particularly attractive as one can
describe the nuclear many-body problem accurately 
with {\it ab initio} calculations. 
A development of the {\it ab initio} calculation up to the QE region is highly
	      hoped because it can test, through a comparison with data,
	      meson exchange currents and nuclear correlations in the
	      energy region relevant to the oscillation experiments. 
	      The {\it ab initio} calculations require a sophisticated
	      technique and experience.
	      Here, we discuss an {\it ab initio} approach formulated by
	      a combination of correlated Gaussian and the complex scaling method.
	      Then we report an application of this approach to 
	      neutrino-$^4$He reactions which has a direct relevance to the neutrino
	      heating in supernova explosions.
 \item[Sec. 5:]
We discuss QE processes. 
It is well-known that the QE process dominates the cross sections 
for neutrino reactions on 
nuclei of $A\gtap$ 10 ($A$: mass number) at the
	      neutrino energies between 0.1 GeV and 1 GeV. 
A challenge here is to accurately take account of nuclear correlations
	      in the initial and final states. We describe the QE process
with the nuclear spectral function and the final state interactions,
and show that this approach can
provide much better description of
data than the conventional Fermi-gas model does.
 \item[Sec. 6:]
We discuss neutrino-nucleus reactions in the RES region.
Here, a question is how we microscopically describe hadron dynamics in a
	      nuclear system. We discuss the nuclear effects, such as
	      the rescattering and absorption of pions and the $\Delta$
	      propagation in nuclei.
The rescattering of a produced pion with a spectator nucleon,  and the
	      final state interaction between nucleons
are examined for the neutrino-deuteron ($\nu$-$d$) reaction. 
The $\nu$-$d$ reactions play key role to determine the axial vector coupling 
of the $N\Delta$ transition which is an input for describing the neutrino-nucleus reactions.
The pion production reactions in nuclei in the $\Delta$ resonance region
have been studied extensively in terms of the $\Delta$-hole approach. 
As an application of this approach, we
discuss neutrino-induced coherent pion-production reactions.
 \item[Sec. 7:]
We discuss neutrino-nucleus reactions in the DIS region.
The nuclear medium effects in the DIS region are an interesting and
	      important question.
An answer to this question is needed for describing
nuclei in terms of quark and gluon degrees of freedom. 
Also, some previous analyses claimed that the nuclear effects can be
	      different between charged lepton and neutrino DIS.
We discuss the current status of
the nuclear parton distribution functions from a global analysis of the world data in
connection with the neutrino-nucleus
reactions. In the region of small $Q^2$ and large $\nu$
(see Fig.~\ref{fig:kinem} for the definition of $Q^2$ and $\nu$),
it is, however, difficult to treat the neutrino-nucleon interaction in terms of perturbative QCD,
and thus we will need a help from some other approach such as those based on the Regge
phenomenology to describe it. 
We will briefly summarize such studies, and introduce
recent parametrizations for the neutrino reactions.
Furthermore, we will also discuss how
the Regge region with the small $Q^2$ and large $\nu$ could be connected to the regions of DIS
and RES.
 \item[Sec. 8:]
We summarize future prospect on how we take an approach towards
a unified understanding of the neutrino-nucleus reactions over the wide $W$ and $Q^2$
regions.
\end{itemize}

\section{Experimental status}

 From early 1970's, neutrino-nucleon/nucleus scatterings were
intensively studied with bubble chambers. The bubble chamber
detector provides clear images of the neutrino interactions.
The charged particles produced in the detectors are identified
efficiently and momentum thresholds of the particles are quite
low. Also, these detectors are magnetized and charge and momentum 
of a particle could be measured by the trajectory. Type of 
a particle could be identified with the thickness of the 
trajectory, which corresponds to the energy deposit per unit 
length. Therefore, it is possible to reconstruct nucleon 
resonance mass with observed charged pion and proton. 
On the other hand, detection efficiency of gamma
was not so high because of the limited size of the detector
and there are some difficulties in differentiating low momentum 
pions from muons by thickness of the track because the masses 
of these two particles are quite similar. Also,
all the images were scanned manually and thus, statistics 
is limited. Understandings of the incident neutrino fluxes 
were not satisfactory compared to the standard today.
Still, the data sets from the bubble chamber are
valuable because recent experiments use different 
detectors and thus, characteristics of the detector is
completely different.
These bubble chamber experiments have measured not only
total cross sections but also differential 
cross sections, for example, $d\sigma/dQ^2$, $d\sigma/dW$ 
and so on. Furthermore, some of the
bubble chamber experiments have used the Deuterium target
(ANL~\cite{Barish:1977qk,Radecky:1981fn}, 
BNL~\cite{Baker:1981su,Kitagaki:1986ct},
BEBC~\cite{Allasia:1990uy}
and FNAL~\cite{Kitagaki:1983px}) and they provide neutrino interaction 
with quasi-free neutron, which could not be achieved by 
the other later experiments. The other experiments
used heavier gases, like Neon, Propane and Freon.
These experiments used wide variety of neutrino beam,
ranging from a few hundreds of MeV to several tens of GeV.
Therefore, various interactions like quasi-elastic, single pion 
production and deep inelastic scattering of both charged and
neutral currents are studied.
 There are several other neutrino experiments, which have used
high energy neutrino beam to study weak interaction, nuclear 
structure ( structure function  $xF_3$ measurements ) or short
baseline neutrino oscillations. Among them, 
CHORUS experiment~\cite{KayisTopaksu:2007pe}
used the emulsion detector with a calorimeter and a muon spectrometer. 
The emulsion detector provides precise particle track information
even around the vertex. The NOMAD experiments~\cite{Wu:2007ab}
used the low 
averaged drift chambers as the active target. These experiments 
provided not only the differential cross sections but also
charged hadron multiplicities. These multiplicity information 
are also useful to understand the neutrino interactions at 
higher $W$ region. 
CDHS~\cite{Berge:1991hr},
CCFR~\cite{Yang:2000ju}
and NuTeV~\cite{Fleming:2000bg,Bodek:2000pj}
used similar detectors 
but optimized for the beamline of each experiment to measure
the total cross sections and differential cross sections
to extract the structure function, $xF_3$.
Results from these
experiments are basically well explained by a simple
model of neutrino-nucleon or neutrino-nucleus reactions
within the statistics and the systematic errors.

\begin{figure}
\begin{center}
\includegraphics[width=0.4\textwidth]{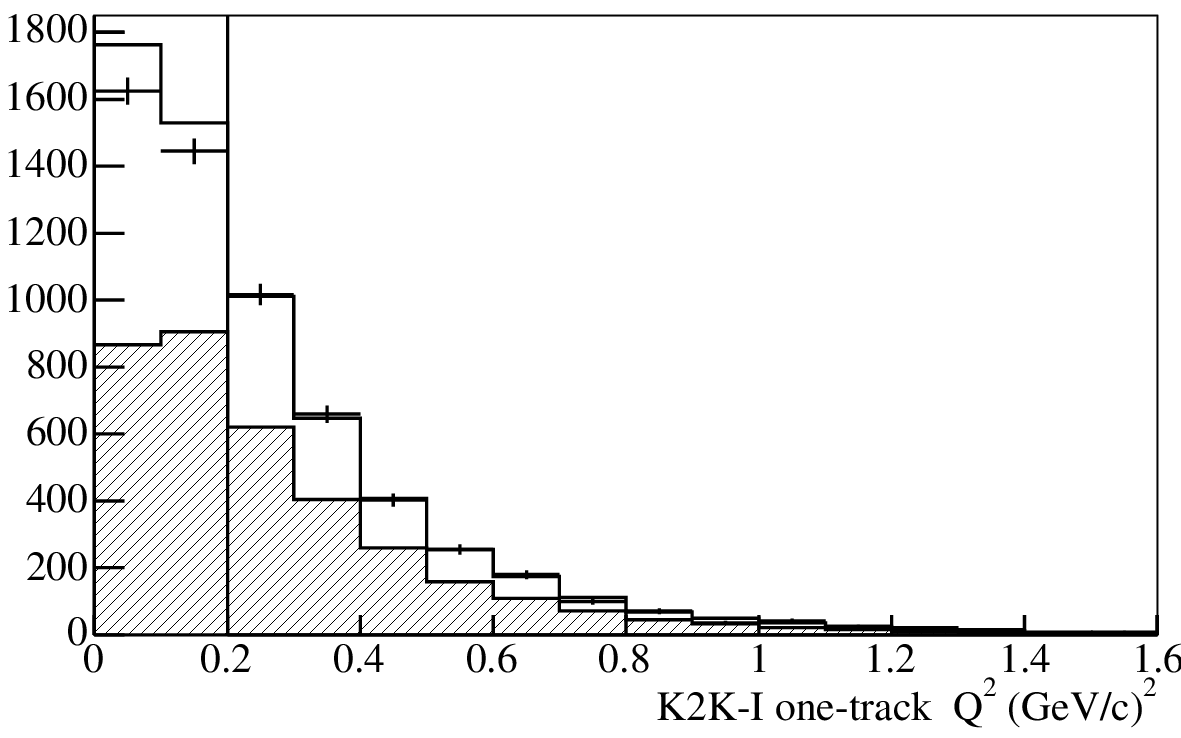}
\includegraphics[width=0.4\textwidth]{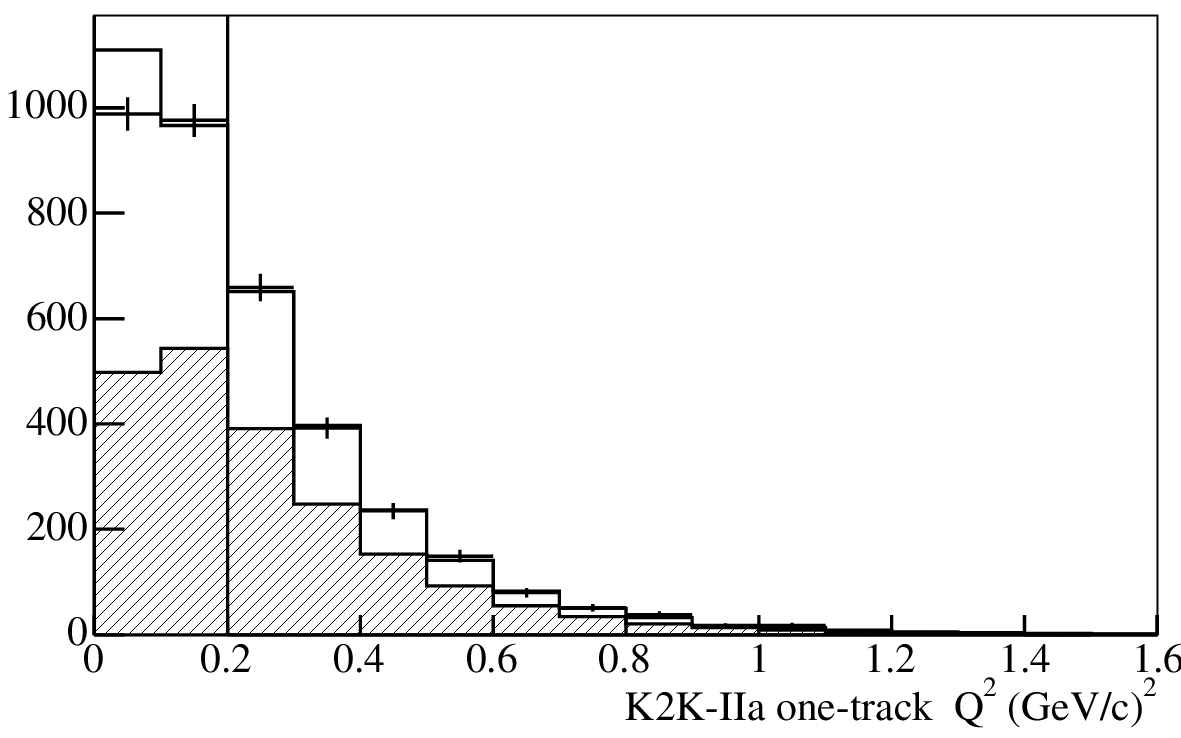}
\end{center}
\caption{
The data and the best fit $Q^2_{rec}$ distributions for K2K-1 1 track data 
(left) and K2K-IIa data (right) from SciFi detector. The shaded region shows 
the QE fraction of each sample, estimated from the MC. The lowest two data 
points in each plot are not included in the fit, due to the large uncertainty 
from various nuclear effects.
The best fit value is $M_A =$ 1.20 GeV/$c^2$ for CCQE.
(c) APS~\cite{Gran:2006jn}
}
\label{fig:kine}
\end{figure}

In 1999, the K2K experiment, the first long-baseline 
neutrino oscillation experiment to confirm the atmospheric neutrino
oscillation, started data taking and collected neutrino 
interaction data with the near detectors. 
They found that the forward going muons are much fewer
than expected. This observation was found not only in 
the 1kt Water Cherenkov detector but also in the scintillating 
fiber tracker detector (SciFi)~\cite{Gran:2006jn}
and the full active scintillator bar detector 
(SciBar)~\cite{Hasegawa:2005td}. 
The forward deficit 
was well explained by increasing the axial coupling
parameter ($M_A$) for charged current QE (CCQE)
and CC resonance production,
and also by applying the correction to the parton distribution
function suggested by Bodek and Yang~\cite{Bodek:2002vp}. 
The K2K experiment
did not publish the absolute cross section but they have
extracted $M_A$ by fitting the shape of $d\sigma/dQ^2$.
The extracted $M_A$ value was $\sim$ 20 \% larger than
the nominal value, $\sim 1.0$ GeV/$c^2$ (Fig.~\ref{fig:kine})~\cite{Gran:2006jn}.

 From 2008, the MiniBooNE experiment started 
publishing the results of various cross section 
measurements~\cite{AguilarArevalo:2010zc,Aguilar-Arevalo:2013dva}.
This experiment utilizes relatively low energy
neutrino beam 
(average $E_\nu \sim 0.8$ GeV) and they used the oil Cherenkov detector. 
They confirmed that the forward going muons are fewer than 
expected, as observed in K2K. Interestingly, the observed number
of CCQE-like events are a few tens of \% larger than simple 
Relativistic Fermi-Gas model prediction. Even after considering 
the uncertainty 
of the absolute beam flux, the number of CCQE-like events 
are significantly larger than the simple model predictions.
A similar small $Q^2$ deficit was also observed in the 
MINOS experiment~\cite{Adamson:2014pgc}
and the best fit $M_A$ value is almost
the same as the one from the K2K experiment~\cite{Gran:2006jn}.

 These results are interesting in various aspects. The forward
going muon deficit or small $Q^2$ deficit are likely to be 
due to the inappropriate model for the CCQE reaction on a
nucleus and it is necessary to have a more sophisticated model
compared to the simple Fermi-Gas model. However, most of
the CCQE cross sections from sophisticated models are
smaller than those from the simple Fermi-Gas model. On the other hand,
the observed results are larger than the Fermi-Gas model gives,
and this implies that these
analyses may be missing some mechanisms. One of the candidates
of the `missing' components is multi-nucleon interactions.
These have been observed in the electron-nucleus scattering
experiments and thus, it is quite natural to observe them also
in the neutrino-nucleus scattering experiments. The MINER$\nu$A
experiment tried to identify this kind of interactions~\cite{Rodrigues:2015hik}.

 The pion productions via resonance for both nucleon and
nucleus target have been studied with the bubble chamber 
experiments. However, statistics was not sufficient especially
for this interaction mode. Then K2K~\cite{Nakayama:2004dp} and
MiniBooNE~\cite{AguilarArevalo:2009ww}
experiments measured the neutral-current (NC) $\pi^0$ production. The number of events,
momentum and angular distributions agree quite well
with the expectations and past experiments. On the other
hand, CC $\pi^+$ production measured in MiniBooNE~\cite{AguilarArevalo:2010bm}
did not agree with the expectation, and the pion momentum 
distribution is also different from those from 
MINER$\nu$A~\cite{Eberly:2014mra}.
The source of these differences is not clear and
further experimental data are still needed.
These differences are
expected to be related to the pion re-scattering both
in nucleus and in the detector. Therefore, it is crucial
to understand not only the initial momentum and directional
distributions of pions but also pion interactions.
Another interesting topic is the coherent pion production,
which is the interaction of neutrino and nucleus without
breaking up the nucleus. This interaction produces just 
lepton and pion in the final state and no nucleons are 
emitted. Experimentally, this interaction has been studied
by searching for the pion + lepton events without nucleon 
emission. The K2K~\cite{Hasegawa:2005td}
and SciBooNE~\cite{Hiraide:2008eu}
experiments found that the
cross section of CC coherent pion production
in $\sim$ 1~GeV region, is much smaller than the simple PCAC 
based model~\cite{Rein:1982pf}. The MINER$\nu$A
 experiment measured the cross section around a few GeV region,
and found them consistent with the
recent calculations~\cite{Higuera:2014azj}. 
The interesting point is that the
cross section for the NC coherent pion production
was observed to be consistent with the same simple PCAC 
model in K2K~\cite{Nakayama:2004dp} 
and MiniBooNE~\cite{AguilarArevalo:2009ww}.

Neutrino-nucleus deep inelastic scattering has been used to 
determine the structure function $xF_3$ and parton distribution 
functions. In the past experiments, experimental data are 
corrected and analyzed to determine the structure functions
of iso-scalar nucleus or nucleon~\cite{Berge:1991hr,Yang:2000ju}.
Recently, MINER$\nu$A experiment
are collecting a large amount of data from 5 GeV to 50 GeV
and started studying the partonic nuclear effects~\cite{Mousseau:2016snl}. 
They have observed the deficit in the small $x$ region
which is so-called shadowing region. It is important to
understand the nuclear dependences of DIS
in the experiments
where the neutrino beam of this energy range is utilized.

\section{Neutrino-nucleon reactions}

In this section, we first present a general formula that represents
cross section for neutrino-nucleon and neutrino-nucleus reactions for
all kinematical regions within the standard model. 
The cross section formula is written in terms of the structure
functions.
We will briefly sketch how the structure functions are 
modeled and evaluated in different kinematical regions such as the QE,
RES, and DIS regions.
Then we spend a substantial portion of this section to discuss our own
work on the dynamical coupled-channels model for the RES region.

\subsection{Cross section formula}

The charged current (CC) and neutral
current(NC) semi-leptonic reactions on a nucleon or on a nucleus
are described by the effective
interaction from the standard model as
\begin{eqnarray}
   {\cal L}^{\rm CC} & = &
 - \frac{G_F}{\sqrt{2}}\frac{m_W^2}{m_W^2 + Q^2}
\int d^4x [ J^{\rm CC}_\mu(x) l^{\rm CC\, \mu}(x) + {\rm h.c.} ],\\
   {\cal L}^{\rm NC} & = &
- \frac{G_F}{\sqrt{2}}\frac{m_Z^2}{m_Z^2 + Q^2}
\int d^4x  J^{\rm NC}_\mu(x) l^{\rm NC\, \mu}(x),
\label{eq:weak_lag}
\end{eqnarray}
where $G_F$ is the Fermi coupling constant, and
$l^\mu$ and $J^\mu$ are lepton current and quark current, respectively;
$m_W$ and $m_Z$ are the weak boson masses.
The quark currents are given as follows,
\begin{eqnarray}
J_\mu^{\rm CC}(x) & = & \bar{u}(x)\gamma_\mu(1-\gamma_5)d'(x)
                  +  \bar{c}(x)\gamma_\mu(1-\gamma_5)s'(x)\ , \\
J_\mu^{\rm NC}(x) & = & 
\sum_{q=u,c} \bar{q}\gamma_\mu
\left(\frac{1}{2}(1- \gamma_5) -\frac{4}{3}\sin^2\theta_W \right)
q(x)
 \nonumber \\
 & + & 
\sum_{q=d,s} \bar{q}\gamma_\mu
\left(-\frac{1}{2}(1-\gamma_5) +\frac{2}{3}\sin^2\theta_W\right) 
q(x) \ ,
\, 
\end{eqnarray}
where 
we kept terms relevant to our following discussions.
The weak eigenstates, $d'$ and $s'$,  are written in terms of 
mass eigenstates and Cabibbo-Kobayashi-Maskawa (CKM) matrix;
$\theta_W$ is the Weinberg angle.
For an analysis of neutrino-nucleus reaction
in the QE and RES region, it is convenient to write
the quark currents  with the vector ($V_\mu$) and axial
($A_\mu$) currents as,
\begin{eqnarray}
\label{eq:jcc}
J_\mu^{\rm CC}(x) & =& V_{ud} (V_\mu^+(x) - A_\mu^+(x))  
\ , \\
\label{eq:jnc}
J_\mu^{\rm NC}(x)
& =&  (1 - 2\sin^2\theta_W) V_\mu^3(x) - 2 \sin^2\theta_W V^{\rm s}_\mu(x) - A_\mu^3(x) 
\\
& = &  V_\mu^3(x) - 2 \sin^2\theta_W J^{\rm EM}_\mu(x) - A_\mu^3(x) 
\ , 
\end{eqnarray}
where the superscript $+(-)$ indicates the isospin raising (lowering)
current, '3' is the third component of the isovector, and 's' is the
isoscalar current; '${\rm EM}$' indicates the electromagnetic current.
The lepton currents are given as
\begin{eqnarray}
 l_\mu^{\rm CC}(x) & = & \sum_{l=e,\mu} \bar{l}(x)\gamma_\mu (1 - \gamma_5) \nu_l(x), \\
 l_\mu^{\rm NC}(x) & = & \sum_{l=e,\mu,\tau} \bar{\nu}_l(x)\gamma_\mu (1 - \gamma_5) \nu_l(x).
\end{eqnarray}

\begin{figure}[t]
\vspace*{5mm}
\begin{center}
\begin{fmffile}{./neutrino-nucleon-dis}
\begin{fmfchar*}(110,110)
\fmfstraight
  \fmfleft{l1,l2}
  \fmfright{r1,r2}
    \fmf{fermion,tension=1.0}{l1,i1,l2} 
    \fmf{photon,tension=1.0}{i1,i2}  
    \fmf{phantom,tension=1.0}{r1,i2,r2} 
    \fmf{fermion,tension=0.0}{r1,i2} 
\linethickness{0.3mm}
 \drawline(83.0,55.0)(109.0,127.8)
 \drawline(83.0,55.0)(111.0,121.8)
 \drawline(83.0,55.0)(113.0,115.8)
 \drawline(83.0,55.0)(115.0,109.8)
 \drawline(83.0,55.0)(117.0,103.8)
 \drawline(83.0,55.0)(119.0,97.8)
 \filltype{shade}
  \put(-7,-7){$l$}
  \put(-7,113){$l'$}
  \put(108,-9){$i$}
  \put(116,112){$f$}
  \put(32,41){$W^+$ or $Z$}
  \put(52,62){$q$}
  \put(-1,25){$p_l$}
  \put(-1,80){$p_{l'}$}
  \put(103,26){$P$}
  \put(75,80){$p_{f}$}
\end{fmfchar*}
\end{fmffile}
\end{center}
\caption{\label{fig:neutrino-hadron}
The neutrino-nucleon (nucleus) reaction. 
The participating particles are the initial ($l$) and final ($l'$) leptons,
and the target ($i$) and final $(f)$ hadrons, and the weak boson 
($W^+$ or $Z$)
is exchanged between the lepton and the hadron.
Beside the line for each of the particles,
its four-momentum is indicated.
}
 \end{figure}
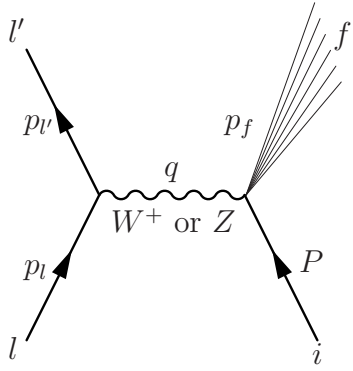
%
We consider a neutrino-nucleon (neutrino-nucleus)
reaction that can be diagrammatically
represented in Fig.~\ref{fig:neutrino-hadron}.
With a matrix element of the above weak interaction
and also with kinematical variables defined in Fig.~\ref{fig:neutrino-hadron}, 
we can write down the cross section in the laboratory frame as 
\begin{eqnarray}
\frac{d\sigma^\alpha}{d\Omega_{l'} dE_{l'}} & = &
\frac{G_F^2 C_\alpha^2}{8\pi^2}
\frac{|\bm{p}_{l^\prime}|}{ |\bm{p}_l|}
 L^{\mu\nu}W^{\alpha}_{\mu\nu}, \label{cc-crs}
\end{eqnarray}
where $\alpha=$ CC or NC;
$Q^2=-q^2$ and 
$C_\alpha = 1/(1 + Q^2/m_W^2)$ $[C_\alpha = 1/(1 + Q^2/m_Z^2)]$ for $\alpha=$ CC [NC];
$L^{\mu\nu}$ and $W^{\mu\nu}$ are
the lepton and hadron tensors, respectively.
The lepton tensor is written as
\begin{eqnarray}
L^{\mu\nu} 
 & = & 2[p_l^\mu p_{l'}^\nu + p_l^\nu p_{l'}^\mu 
    - g^{\mu\nu}((p_l\cdot p_{l'}) - m_l m_{l'})
    \pm i \epsilon^{\mu\nu\alpha\beta}p_{l,\alpha}p_{l',\beta} ],  \label{lepton-cc}
\end{eqnarray}
where $+(-)$ in the last term is for neutrino (anti-neutrino) reactions.
The hadron tensor is defined by 
\begin{eqnarray}
W_{\mu\nu}^\alpha & = & \bar{\sum_i} \sum_f (2\pi)^3 V \frac{E_T}{M_T}
 \delta^{(4)}(P + q - p_f) \langle f|J_\mu^\alpha(0)|i\rangle 
\langle f|J_\nu^\alpha(0)|i\rangle ^*\ , \label{hadron}
\end{eqnarray}
where $V$ is the quantization volume that disappears in final
results; $E_T$ and $M_T$ are the energy and the mass of target hadron;
$\bar{\sum}_i$ is the average of the spin states of the target hadron;
$\langle f|J_\mu^\alpha(0)|i\rangle$ is 
a matrix element of the quark currents between hadronic states, 
$|i\rangle$ and $|f\rangle$.
The hadron tensor 
includes all information of the hadron response to the current $J_\mu^\alpha$.
For an inclusive reaction, the hadron tensor can be expressed using
two available vectors, the momentum of the target (with mass $M_T$)
$P$ and the momentum transfer $q$, as
\begin{eqnarray}
W^{\alpha,\mu\nu} & = & - g^{\mu\nu}W_1^\alpha + \frac{W_2^\alpha}{M_T^2}P^\mu P^\nu
  +i \frac{W_3^\alpha}{2M_T^2} \epsilon^{\mu\nu\rho\sigma}P_\rho q_\sigma
 \nonumber \\
& & + \frac{W_4^\alpha}{M_T^2}q^\mu q^\nu + \frac{W_5^\alpha}{M_T^2}
(P^\mu q^\nu + q^\mu P^\nu) + \frac{W_6^\alpha}{M_T^2}(P^\mu q^\nu - q^\mu P^\nu),
\label{hadron-tensor}
\end{eqnarray}
where we have introduced six structure functions
$W_i(\nu,Q^2)$ where $\nu\equiv p_l^0-p_{l'}^0$.
Then the neutrino-hadron inclusive reaction cross section 
for the laboratory frame is given with the structure functions as
\begin{eqnarray}
\label{eq:cross}
\frac{d\sigma^{\alpha}}{d\Omega_{l'} dE_{l'}} &  &=
\frac{G_F^2 C_\alpha^2 |\bm{p}_{l^\prime}|E_{l'}}{2\pi^2}  
\left[2W_1^{\alpha} \sin^2\frac{\chi}{2} + W_2^{\alpha} \cos^2 \frac{\chi}{2}
\nonumber \right.\\
&& \pm \frac{W_3^{\alpha}}{M_T}\left( (E_l + E_{l'})\sin^2\frac{\chi}{2}-
 \frac{m_{l'}^2}{2E_{l'}} \right)
\left.+ \frac{m_{l'}^2}{M_T^2}W_4^{\alpha} \sin^2 \frac{\chi}{2} 
 - \frac{m_{l'}^2}{M_T E_{l'}}W_5^{\alpha}
\right] \ ,
\nonumber \\
\end{eqnarray}
where $\cos\chi = |\bm{p}_{l'}|/E_{l'} \cos\theta$
with $\theta$ being the lepton scattering angle, and
$\pm$ are for neutrino and anti-neutrino reactions, respectively.
The contributions of $W_4^\alpha$ and $W_5^\alpha$ are proportional to the lepton mass
and can be neglected in the high energy reactions. 
$W_6^\alpha$ term does not contribute to the cross section.
Now the problem is to model and evaluate the structure functions.
Depending on $\nu$ and $Q^2$, by which a reaction can be categorized into
either of QE, RES, or DIS region, 
the structure functions need to be modelled
with different effective degrees of freedom, as we will discuss in the
next subsection for neutrino-induced reactions on a
single nucleon. 

\subsubsection{Multipole expansion of structure functions}
\label{sec:mult-exp}

We introduce standard multipole expansions of weak hadronic current~\cite{Walecka75,NSGK}.
The  Coulomb $T^{JM}_C$, electric $T^{JM}_E$, longitudinal $T^{JM}_L$
and  magnetic $T^{JM}_M$ multipole operators
of the weak hadronic current $J^\mu$ are defined as
\begin{eqnarray}
T_C^{JM}({ J}) & = & \int d\bm{x} j_J(qx)Y_{JM}(\hat{\bm{x}}){ J}_0(\bm{x}), \\
T_E^{JM}({ J}) & = & 
  \frac{1}{q} \int d\bm{x} 
  \bm{\nabla}\times [j_J(q x)\bm{Y}_{JJM}
(\hat{\bm{x}})]    \cdot \bm{{ J}}(\bm{x}), \\
T_M^{JM}({ J}) & = & 
   \int d\bm{x} 
  j_J(q x)\bm{Y}_{JJM}(\hat{\bm{x}})
   \cdot \bm{{ J}}(\bm{x}),\\
  T_L^{JM}({ J}) & = & 
  \frac{i}{q} \int d\bm{x} 
  \bm{\nabla} [j_J(q x)Y_{JM}(\hat{\bm{x}})]
   \cdot \bm{{ J}}(\bm{x}), \label{eq_op-longi}
\end{eqnarray}
where $\bm{Y}_{JLM}(\hat{\bm{x}})$ are vector spherical harmonics.

The structure functions, $W_1^\alpha, W_2^\alpha, W_3^\alpha$ 
($\alpha =$ CC, NC), 
are expressed using the reduced matrix element
between an initial state of angular momentum and parity $J^\pi=J_i^{\pi_i}$ 
and a final state $J_f^{\pi_f}$ as
\begin{eqnarray}
 2 W_1^\alpha& =& \sum_f \frac{4\pi}{2J_i+1} \delta(E_l + E_T - E_{l'} - E_f) R^\alpha_T,\\
   W_2^\alpha& =& \sum_f \frac{4\pi}{2J_i+1} \delta(E_l + E_T - E_{l'} - E_f)( R^\alpha_L + \frac{Q^2}{2|\bm{q}|^2}R^\alpha_T),\\
 \frac{W_3^\alpha}{M_T} & = &
  - \sum_f \frac{4\pi}{2J_i+1} \delta(E_l + E_T - E_{l'} - E_f)\frac{R^\alpha_{T'}}{|\bm{q}|},
\end{eqnarray}
where $R^\alpha_T$, $R^\alpha_L$, and $R^\alpha_{T'}$ are respectively given as
\begin{eqnarray}
R^\alpha_L & = & \sum_J|<T^J_C(J^{\alpha,A}) + \frac{\omega}{|\bm{q}|}T^J_L(J^{\alpha,A})>|^2
   + |\frac{Q^2}{|\bm{q}|^2}<T^J_C(J^{\alpha,V})>|^2 ,
 \label{eq:rtrl}
\\
R_T^\alpha & = & \sum_J\sum_{\beta=M,E}[|<T^J_{\beta}(J^{\alpha,V})>|^2 + |<T^J_{\beta}(J^{\alpha,A})>|^2],\\
R_{T^{\prime}}^\alpha & = & \sum_J2{\rm Re}[<T^J_M(J^{\alpha,V})><T^J_E(J^{\alpha,A})>^*
\nonumber\\
&& + <T^J_M(J^{\alpha,A})><T^J_E(J^{\alpha,V})>^* ]\ .
\end{eqnarray}
$T^J(J^{\alpha,V})$ [$T^J(J^{\alpha,A})$] denotes
multipole operators for the vector [axial] part of the hadron current
$J^\alpha$ in Eqs.~(\ref{eq:jcc})-(\ref{eq:jnc}).
The reduced matrix element of the multipole operator 
$<T^J> = 
\left<J_f\left\|T^J\right\|J_i\right>
$ is defined as
\begin{eqnarray}
  \left<J_f,M_f\left|T^{JM}\right|J_i,M_i\right>
  & = & \frac{(J_i,M_i,J,M|J_f,M_f)}{\sqrt{2J_f+1}}
   \left<J_f\left\|T^J\right\|J_i\right>.
\end{eqnarray}
Here the matrix element of the
longitudinal multipole operator of the vector current is 
eliminated in Eq.~(\ref{eq:rtrl}) with the help of
the current conservation relation of the vector current, $q \cdot V = 0$.

\subsection{Neutrino-induced reactions on nucleon in the QE, RES, and DIS regions}
\label{sec:nu-nucleon}

For the QE scattering, 
the response of the nucleon to the weak current is represented
by nucleon form factors. The matrix elements of the vector and axial currents
evaluated with nucleon states are generally parametrized in terms of the
form factors as follows:
\begin{eqnarray}
\langle N(p')|V_\mu^\pm(0) |N(p)\rangle & = &
   \bar{u}_N(\bm{p}')\left[ f_V(Q^2) \gamma_\mu + i {f_M(Q^2)\over 2 m_N}
		 \sigma_{\mu\nu}q^\nu \right] \tau^\pm u_N(\bm{p}) \ , 
\label{eq:vec}\\
\langle N(p')|A_\mu^\pm(0) |N(p)\rangle & = &
   \bar{u}_N(\bm{p}')\left[ f_A(Q^2) \gamma_\mu\gamma_5 + f_P(Q^2) \gamma_5
		 q_\mu \right]\tau^\pm u_N(\bm{p}) \ ,
\label{eq:axial}
\end{eqnarray}
where we have omitted the second class currents. 
For a recent investigation of possible effects from the second class
current on neutrino-nucleus scatterings, see Ref.~\cite{SCC}.
The nucleon spinor with the momentum $\bm{p}$ is denoted by
$u_N(\bm{p})$ and the isospin spinor, on
which the isospin raising (lowering) operator 
$\tau^\pm\equiv (\tau^1\pm i\tau^2)/2$ acts, is also
implicitly included.
The quantities, $f_V(Q^2)$, $f_M(Q^2)$, $f_A(Q^2)$, and $f_P(Q^2)$ are
the form factors, and $m_N$ denotes the nucleon mass.
The matrix elements of the third component of the isovector currents
are obtained by simply replacing $\tau^\pm$ with $\tau^3/2$.
Similarly, the isoscalar current is also parametrized with different form factors as
\begin{eqnarray}
\langle N(p')|V_\mu^s(0) |N(p)\rangle & = &
   \bar{u}_N(\bm{p}')\left[ f^s_V(Q^2) \gamma_\mu + i {f^s_M(Q^2)\over 2 m_N}
		 \sigma_{\mu\nu}q^\nu \right] {1\over 2}u_N(\bm{p}) \ .
\label{eq:vec_is}
\end{eqnarray}
The form factors for the vector current
are determined by analyzing electron-nucleon scattering data. 
Regarding the axial current,
the axial form factor $f_A(Q^2)$ is conventionally parametrized in a
dipole form as 
\begin{eqnarray}
f_A(Q^2) = g_A \left( {1\over 1+Q^2/M_A^2}\right)^2 \ ,
\label{eq:fa}
\end{eqnarray}
with $g_A$=1.27 determined by the neutron life time~\cite{PDG14}. 
The axial mass $M_A$ has been determined either by neutrino-deuteron
QE scattering data or by the pion electroproduction data near threshold, 
and its value has been estimated to be $M_A=1.026\pm 0.021$~GeV~\cite{bernard}.
The induced pseudoscalar form factor $f_P$ is often related to $f_A$ by
the PCAC relation and the pion-pole dominance. 
In addition to the above-described currents,
the strange component of the nucleon contributes to the NC neutrino nucleus/nucleon reactions.
In particular, the strange axial vector current contribution has been 
investigated~\cite{Agulilar-NC2010,Agulilar-NC2015,Ankowski2012,Hobbs2016}.
The iso-scalar axial current is parametrized as
\begin{eqnarray}
\langle N(p')|A_\mu^s(0) |N(p)\rangle & = &
   \bar{u}_N(\bm{p}') {1\over 2} f^{\rm s}_A(Q^2) \gamma_\mu\gamma_5  u_N(\bm{p}) \ ,
\label{eq:axi_is}
\end{eqnarray}
with
\begin{eqnarray}
f^{\rm s}_A(Q^2) & = & \frac{\Delta s}{( 1 + Q^2/M_A^2)^2} \ .
\end{eqnarray}
The experimental value of $\Delta s$ is 
$\sim -0.1$, while lattice QCD and hadron model calculations
suggest a smaller magnitude~\cite{Hobbs2016}.
With the matrix elements of Eqs.~(\ref{eq:vec})-(\ref{eq:vec_is}), we
can construct the hadron tensor of Eq.~(\ref{hadron}), and also the
structure functions $W_i^\alpha$ in the cross section formula,
Eq.~(\ref{eq:cross}).

In the RES region, the weak current can excite a nucleon to its
resonant states ($N^*$), which is followed by a deexcitation through meson
emissions. 
The main process of this kind in the neutrino-nucleon scattering is 
a single-pion production for which the $\Delta(1232)$ resonance
gives a dominant contribution.
As the nucleon gets excited to a higher resonance beyond $\Delta(1232)$,
the double-pion production becomes comparable or even more important than
the single pion production. 
Also, $\eta N$, $K\Lambda$, and $K\Sigma$ are produced with
probabilities suppressed by an order of magnitude.
In these meson-production processes, 
the different meson-baryon channels are strongly coupled with each other
in the final state interaction.

Theoretical descriptions of these meson production processes can be
categorized into two approaches. 
One is to relate the divergence of the axial current amplitude with the
pion-nucleon reaction amplitude via the PCAC relation at $Q^2\sim 0$.
Because, at $Q^2=0$, 
only $W^\alpha_2$ among the structure functions 
gives nonzero contribution
and is solely determined by the divergence of the axial current
amplitude,
the cross section for the neutrino-induced reaction at $Q^2=0$
can be written with that of the pion-nucleon reaction.
This approach has been taken in Ref.~\cite{paschos}.
However, the validity of this approach is limited to very small $Q^2$
region, and the extrapolation of the cross sections from $Q^2=0$ to
finite $Q^2$ is difficult to control.
Another approach
is to model the processes microscopically with hadronic
degrees of freedom.
A pioneering work has been done by Adler~\cite{adler68} who analyzed
the pion production mechanisms with
a model based on the dispersion theory for a unified description of weak and 
electromagnetic pion production reactions.
Then several
models~\cite{LPP,valencia1,valencia2,indiana1,indiana2,giessen,sul,msl},
which we will briefly review later, 
have been developed so far, and 
some of them are focused
on the $\Delta(1232)$ region because of its important relevance to the
oscillation experiments.
Recently, three of the present authors 
developed a dynamical coupled-channels (DCC) model that
includes all relevant resonance contributions of $W\le 2$~GeV, and 
takes account of $\pi N, \pi\pi N, \eta N, K\Lambda, K\Sigma$ 
coupled-channels in the hadronic rescattering~\cite{nks}. 
We will discuss the DCC model in detail in the following subsection.
Key quantities for the hadronic models are 
form factors analogous to those in
Eqs.~(\ref{eq:vec})-(\ref{eq:vec_is}) but
associated with $N$-$N^*$ transitions.
For example, the $N$-$\Delta(1232)$ transition matrix element is often
parametrized as:
\begin{eqnarray}
 \label{eq:delta}
\langle \Delta (p_\Delta\!=\!p\!+\!q)\, | J^{\rm CC}_\mu(0)| N(p) \rangle 
&=&\bar u_\Delta^\alpha(\bm{p}_\Delta) \Gamma_{\alpha\mu}\left(p,q \right)
T^\pm u_N(\bm{p}\,)
V_{ud} \ ,
\end{eqnarray}
where $u_\Delta^\alpha(\bm{p}_\Delta)$ and $T^\pm$
are the $\Delta$ vector spinor and the isospin transition operator,
respectively, and
\begin{eqnarray}
\Gamma_{\alpha\mu} (p,q) &=&
\left [ \frac{C_3^V}{m_N}\left(g_{\alpha\mu} \slashchar{q}-
q_\alpha\gamma_\mu\right) + \frac{C_4^V}{m_N^2} \left(g_{\alpha\mu}
q\cdot p_\Delta- q_\alpha p_{\Delta\,\mu}\right) \right. \nonumber\\
&& \left. + \frac{C_5^V}{m_N^2} \left(g_{\alpha\mu}
q\cdot p- q_\alpha p_\mu\right) + C_6^V g_{\mu\alpha}
\right ]\gamma_5 
+ \left [ \frac{C_3^A}{m_N}\left(g_{\alpha\mu} \slashchar{q}-
q_\alpha\gamma_\mu\right) 
\right.
\nonumber\\
&& \left.
+ \frac{C^A_4}{m_N^2} \left(g_{\alpha\mu}
q\cdot p_\Delta- q_\alpha p_{\Delta\,\mu}\right) 
+ C_5^A g_{\alpha\mu} + \frac{C_6^A}{m_N^2} q_\mu q_\alpha
\right ] \ ,   \label{eq:del_ffs}
 \end{eqnarray}
where $C_i^V$ and $C_i^A$ ($i=3,4,5,6$) 
that depend on $Q^2$
are vector and axial form
factors, respectively. 
With well-controlled form factors, we can apply the model to
the neutrino-induced meson productions of the whole $Q^2$ region.
The vector form factors can be reasonably determined by analyzing a large amount of
data for single-pion photo- and electro-production off the nucleon. 
The axial form factors are difficult to determine because of the
shortage of experimental information.
Thus the axial form factors, those associated with $N$-$\Delta(1232)$ transition
in particular, have been estimated with 
quark models~\cite{nimi,hemmert95}, chiral perturbation
theory~\cite{geng08,procura08}, and lattice QCD~\cite{dina11}.
However, experimental inputs are still very valuable. 
For the moment, 
only the axial $N$-$\Delta(1232)$ transition form factors can be constrained by 
analyzing the deuterium bubble chamber data~\cite{Kitagaki:1986ct,anl}.
In analyzing the data, however,
a complication could arise due to a significant effect from the $NN$
final state interaction as pointed out in Ref.~\cite{wsl} and will be
discussed in Sec.~\ref{sec:deuteron};
the previous analyses neglected this effect.
For the other axial $N$-$N^*$ form factors, the PCAC relation to the 
$\pi NN^*$ couplings is conventionally invoked at $Q^2=0$, and a certain
$Q^2$-dependence is assumed.


The DIS region is usually specified
by the kinematical conditions, $W^2 \ge 4$ GeV$^2$ and $Q^2 \ge 1$ GeV$^2$,
as shown in Fig.\,\ref{fig:kinem}.
However, different boundaries may be taken depending on researchers.
For example, there are some people to take lower $W^2$ 
({\it e.g.} $W^2 \ge 3.5$ GeV$^2$), and higher $Q^2$ values 
({\it e.g.} $Q^2 \ge 4$ GeV$^2$) could be taken to avoid higher-twist effects.
In the DIS, the Bjorken scaling variable $x$ is used instead of the energy 
transfer $\nu$, and it is defined by 
$x=Q^2 / (2 p \cdot q) = Q^2 / (2 m_N \nu)$.
Furthermore, the structure functions $F_1$, $F_2$, and $F_3$ are usually
used instead of $W_1$, $W_2$, and $W_3$ defined in the hadron tensor
of Eq.\,(\ref{hadron-tensor}), and they are given by
\begin{eqnarray}
\! \! \! \! \! \! \! \! \! \! \! \! \! \! \! \! 
\! \! \! \! \! \! \! \! \! \! \! \! \! \! \! \! 
\! \! \! 
 F_1^{\,\alpha}(x,Q^2) = m_N W_1^{\,\alpha}(\nu,Q^2), \ 
 F_2^{\,\alpha}(x,Q^2) =  \nu W_2^{\,\alpha}(\nu,Q^2), \ 
 F_3^{\,\alpha}(x,Q^2) =  \nu W_3^{\,\alpha}(\nu,Q^2).
\label{eq:f123}
\end{eqnarray}

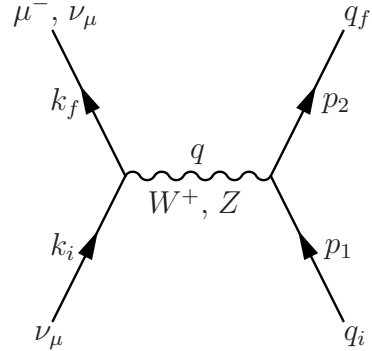
\begin{wrapfigure}[11]{r}{0.40\textwidth}
   \vspace{+0.10cm}
\begin{center}
\begin{fmffile}{./neutrino-quark}
\begin{fmfchar*}(110,110)
\fmfstraight
  \fmfleft{l1,l2}
  \fmfright{r1,r2}
    \fmf{fermion}{l1,i1,l2} 
    \fmf{photon,tension=1.0}{i1,i2}  
    \fmf{fermion}{r1,i2,r2} 
  \put(-7,-7){$\nu_\mu$}
  \put(-15,114){$\mu^-,\hspace{0.05cm}\nu_\mu$}
  \put(110,-7){$q_i$}
  \put(110,114){$q_f$}
  \put(36,41){$W^+,\hspace{0.05cm}Z$}
  \put(52,62){$q$}
  \put(-1,25){$k_i$}
  \put(-1,80){$k_f$}
  \put(103,26){$p_1$}
  \put(102,81){$p_2$}
\end{fmfchar*}
\end{fmffile}
\\
\vspace{+0.3cm}
\hspace{-2.40cm}
   \begin{minipage}{0.53\textwidth}
\caption{\label{fig:neutrino-quark}
Neutrino-quark scattering}
   \end{minipage}
\vspace{-0.2cm}
\end{center}
\end{wrapfigure}
\normalsize

If $Q^2$ is large, the neutrino-nucleon DIS cross section is described
by the simple addition of the $W$ or $Z$ interaction cross sections with 
individual partons ($i$): $d \sigma (\nu N) = \sum_i  d \sigma (\nu \, i)$
as shown in Fig.~\ref{fig:neutrino-quark}.
It is called impulse or incoherent assumption, which is valid
in the DIS region by considering that partons do not interact
with each other, namely frozen, when the $W$ or $Z$ interact with a quark. 
By this parton model, the structure functions are expressed 
in the leading order (LO) of $\alpha_s$ 
and also in the leading twist as~\cite{nu-n-sfs,qcd-collider-book}
\begin{align}
2xF_1^{\,\alpha}(x,Q^2)_{\rm LO} & = F_2^{\,\alpha}(x,Q^2)_{\rm LO}, \\
  F_2^{\,\text{CC}(\nu p)}(x,Q^2)_{\rm LO} & = 2 x \left [ \, d(x,Q^2) + s(x,Q^2) 
                           + \bar u (x,Q^2) + \bar c (x,Q^2) \, \right ], 
\label{eqn:f2-nu} \\
x F_3^{\,\text{CC}(\nu p)}(x,Q^2)_{\rm LO} & = 2 x \left [ \, d(x,Q^2) + s(x,Q^2) 
                           - \bar u (x,Q^2) - \bar c (x,Q^2) \, \right ], 
\label{eqn:f3-nu} \\
F_2^{\,\text{NC}(\nu p)}(x,Q^2)_{\rm LO} & 
  =  x \left [ \, \left\{ (g_V^u)^2 + (g_A^u)^2 \right\} \right.
      \left\{ u^+ (x,Q^2) + c^+ (x,Q^2) \right\}      
\nonumber \\
   & \ \ \ \, \, \, \left.  + \left\{ (g_V^d)^2 + (g_A^d)^2 \right\}
      \left\{ d^+ (x,Q^2) + s^+ (x,Q^2) \right\} \, \right ], 
\label{eqn:f2-nu-nc}
\\
x F_3^{\,\text{NC}(\nu p)}(x,Q^2)_{\rm LO} & 
   = 2 x \left [ \, g_V^u \, g_A^u  \left\{ u^- (x,Q^2) + c^- (x,Q^2) \right\}  \right.   
\nonumber \\
   & \ \ \ \ \ \  \left.  + \, g_V^d \, g_A^d
      \left\{ d^- (x,Q^2) + s^- (x,Q^2) \right\}   \, \right ] .
\label{eqn:f3-nu-nc}
\end{align}
Here, 
the parton distribution functions (PDFs) are denoted by $q(x,Q^2)$ and
$\bar q(x,Q^2)$ ($q=u,d,s,c$), and 
$q^\pm$ are given by 
$q^\pm (x,Q^2) \equiv q (x,Q^2) \pm \bar q (x,Q^2)$, so that
the $q^-$ distributions are valence-quark distributions by definition.
The strange and charm valence-quark distributions are considered to
be small $|s^-|,\, |c^-| \ll |u^-|,\, |d^-|$, so that they are usually neglected.
There is some indication on $s_v (x) \equiv s^- (x) \ne 0$ from
opposite-sign dimuon production in neutrino reactions; however,
its measurements are not accurate enough to determine the distribution.
Here, the couplings of neutral-current interactions are given by
$g_V^{\,q} =T_q^{\,3} - 2 e_q \sin^2 \theta_W $ and $g_A^{\,q} =T_q^{\,3}$
by the third component of the isospin $T_q^3$ and the quark charge $e_q$.
The bottom quark contributions are neglected in the expressions,
and they can be included by replacing $s (x,Q^2)$ by 
$s (x,Q^2) + b (x,Q^2)$.
The structure functions in the antineutrino reaction,
$F_2^{\,\text{CC}(\bar\nu p)}(x,Q^2)_{\rm LO}$ and 
$F_3^{\,\text{CC}(\bar\nu p)}(x,Q^2)_{\rm LO}$,
can be obtained by the changes, $d \to u$, $s \to c$, 
$\bar u \to \bar d$, and $\bar c \to \bar s$ 
in Eqs.~(\ref{eqn:f2-nu})-(\ref{eqn:f3-nu-nc}).

By including higher-order $\alpha_s$ effects,
we have the expressions 
\begin{align}
\bar F_n^{\,\alpha}(x,Q^2) & 
=  C_n^q (x,Q^2) \otimes \bar F_n^{\,\alpha}(x,Q^2)_{\rm LO}
   + C_n^g (x,Q^2) \otimes xg(x,Q^2), \nonumber \\
\bar F_1 & = x F_1, \ \ \bar F_2 = F_2, \ \ \bar F_3 = x F_3 ,
\label{eqn:f2-higher-order}
\end{align}
in terms of the coefficient functions $C_n^q (x,Q^2)$ and $C_n^g (x,Q^2)$,
and the symbol $\otimes$ indicates the convolution integral
$f (x) \otimes g (x) = \int _x^1 (dy / y) f (x/y) g(y)$.
Explicit expressions of the coefficient functions are, for example, found 
in Ref.\,\cite{qcd-collider-book}. 
Neutrino scattering measurements have been done often 
at a relatively low-energy scale of $Q^2 \sim 1$ GeV$^2$, where 
higher-twist effects could be conspicuous. Considering such effects in the form 
of longitudinal-transverse structure function ratio 
$R \equiv F_L / (2xF_1) = \left [ (1+4m_N^2 x^2/Q^2) F_2 - 2xF_1 \right]/(2xF_1)$,
we express $F_1$ in terms of $R$ and $F_2$ as
\begin{equation}
2 x F_1(x,Q^2) = \frac{1 + 4m_N^2 x^2/Q^2}{1 + R(x,Q^2)} F_2(x,Q^2) .
\label{f1-r-f2}
\end{equation}
In handling the small $Q^2$ ($\sim 1$ GeV$^2$) data, 
Eq.~(\ref{f1-r-f2}) is usually used with the structure function 
$F_2$, which is calculated in terms of the PDFs, by
Eq.~(\ref{eqn:f2-higher-order}) together with
Eqs.~(\ref{eqn:f2-nu}) and (\ref{eqn:f2-nu-nc}).
The function $R(x,Q^2)$ is known in charged-lepton DIS~\cite{r199x},
and the same function is often used also in neutrino DIS.
Using these structure functions with appropriate 
PDFs, we can calculate
the neutrino-nucleon or nucleus cross sections.
In the neutrino-nucleon case, the nucleonic 
PDFs~\cite{nucleon-pdfs}
should be used
in calculating the structure functions, whereas
the neutrino-nucleus cross sections can be calculated simply
by replacing the nucleonic PDFs with the nuclear parton distribution
functions (NPDFs).
The NPDFs are modified from the corresponding nucleonic PDFs,
and the modifications are discussed in Sec.\,\ref{nu-A-dis}.


\subsection{Dynamical coupled-channels model for neutrino-induced meson productions}

In this subsection, we mainly discuss our own work on a
dynamical coupled-channels (DCC) model for neutrino-induced meson
productions off the nucleon.
First, we briefly review previous microscopic models for the
pion productions.
Next, we present an overall picture of the DCC model without
going into detailed expressions and equations.
For a full 
presentation of the DCC model used for the neutrino
reactions, see Refs.~\cite{nks,knls}.
Then we present some selected results from our DCC model-based analysis of
$\pi N, \gamma^{(*)} N\to \pi N, \pi\pi N, \eta N, K\Lambda, K\Sigma$
reactions data.
Through the analysis, all model parameters that govern hadronic
interactions and vector form factors are determined. 
The $\pi NN^*$ couplings are related to the axial $N$-$N^*$ transition
strength at $Q^2$=0 through the PCAC relation.
Thus most of parameters needed to calculate the neutrino-induced
processes are determined through the analysis.
Because of the scarce neutrino data, it is important to have 
the analysis done before applying the DCC model
to the neutrino reactions.
With the parameters determined in the analysis and an assumed
$Q^2$ dependence of the axial form factors, we predict cross sections
for the neutrino-induced meson productions that are compared with
available experimental data.

\subsubsection{Microscopic models for neutrino-nucleon reactions in the RES region}


The previous models for the resonance region
can be classified into three categories depending on
dynamical contents included in the models.
Models of the first category consist of a sum of
the Breit-Wigner amplitudes that represent resonant contributions.
A recent model of this category is found in Ref.~\cite{LPP} where 
$\Delta(1232)3/2^+$, $N(1535)1/2^-$, $N(1440)1/2^+$ and $N(1520)3/2^-$ 
resonances are considered.
Models of the second category consider tree-level non-resonant
mechanisms along with resonant ones of the Breit-Wigner type, 
and are developed in
Refs.~\cite{valencia1,valencia2,indiana1,indiana2}.
The authors of these references considered tree-level
non-resonant mechanisms derived from a chiral Lagrangian in addition to 
$\Delta(1232)$ of the Breit-Wigner type.
A more extended model in the second category was developed in
Ref.~\cite{giessen} where all 4-star resonances with masses below 1.8
GeV and rather phenomenological non-resonant contributions were
considered. 
The so-called Rein-Sehgal model~\cite{RS,RS2}, which has been often used
in analyzing data from neutrino experiments, also belongs to the second
category, and includes higher resonances whose 
axial-vector couplings had been estimated with a quark model.
In the third category, a model further takes account of the hadronic rescattering,
thereby maintaining the unitarity of amplitudes.
Such a model in the $\Delta(1232)$ region was developed in Refs.~\cite{sul,msl}.
The DCC model discussed below can be regarded as an extension of the 
model of Refs.~\cite{sul,msl};
the Fock space of the $\pi N$ channel
is extended to include
more hadronic two-body and $\pi\pi N$ channels
and higher resonances beyond $\Delta(1232)$.


\subsubsection{Overview of Dynamical Coupled-Channels model}

The starting point of the DCC model is a set of phenomenological
Lagrangians giving interactions among mesons, baryons and external
currents. Couplings of the octet pseudoscalar mesons are consistent with
those from a chiral Lagrangian at the low-energy limit. 
We derive a set of meson-baryon interaction potentials, acting on a
given Fock space, from the Lagrangians using
a unitary transformation method~\cite{UT-method,sl}. 
The potentials obtained in this way
are energy independent, and thus unitary amplitudes can be calculated in
a straightforward manner. 
For our particular model, we choose 
the Fock space that consists of 
meson-baryon states ($\pi N, \eta N, K\Lambda,K\Sigma$ and $\pi\pi N$
states) and 'bare' excited states ($N^*,\Delta,\rho,\sigma$).
The bare $N^*$ state represents a quark core component of a
nucleon resonance of a given spin-parity, and is dressed by the meson cloud to form the
resonance.
Now our Hamiltonian reads
\begin{eqnarray}
H & = & H_0 + v + \Gamma  \ ,
\label{eq:effectiveH}
\end{eqnarray}
where $H_0$ is the free Hamiltonian of mesons and baryons, and $v$ is
non-resonant interaction potentials between two-body meson-baryon states
and also between $\pi\pi$ states.
The non-resonant interactions are from
$s$-, $t$-, and $u$-channel hadron-exchange and contact mechanisms.
$\Gamma$ describes a
transition between the bare excited states
and two-body states such as $\Delta \leftrightarrow \pi N$ and 
$\rho \to \pi\pi$.
With this Hamiltonian, we solve the coupled-channels Lippmann-Schwinger
equation that reads as
\begin{eqnarray}
T_{\beta\alpha} (\bm{p}',\bm{p};W) &=& 
V_{\beta\alpha}(\bm{p}',\bm{p}) 
\nonumber\\
&&+ \sum_\gamma \int d\bar{\bm{p}}\;
V_{\beta\gamma}(\bm{p}',\bar{\bm{p}})\; G_\gamma(\bar{\bm{p}},W)\; 
T_{\gamma\alpha}(\bar{\bm{p}},\bm{p};W)\ ,
\label{eq:LS-eq}
\end{eqnarray}
where each of the indices $\alpha$, $\beta$, and $\gamma$ specifies one of the channels
included in the Fock space.
The scattering amplitude ($T$-matrix element) is denoted by $T_{\beta\alpha}$
and the Green's function for a channel $\gamma$ by $G_\gamma$.
The interaction potential $V_{\beta\alpha}$ is 
either $v_{\beta\alpha}$ or $\Gamma_{\beta\alpha}$
in Eq.~(\ref{eq:effectiveH}).
As mentioned above, thanks to the energy-independent potential
$V_{\beta\alpha}$, it is easily proved that the scattering amplitude
$T_{\beta\alpha}$ satisfies the multichannel unitarity.
The quantity
$W$ is the total energy of the hadronic system while
$\bm{p}$ and $\bm{p}'$ are the incoming and
outgoing momenta; for $\pi\pi N$ channels, it is understood that 
$\bm{p}$ implicitly denotes two independent momenta. 
Observables such as cross sections for meson-baryon scattering are
calculated with the scattering amplitudes in a straightforward manner. 

\begin{figure}[t]
\begin{center}
\includegraphics[width=78mm]{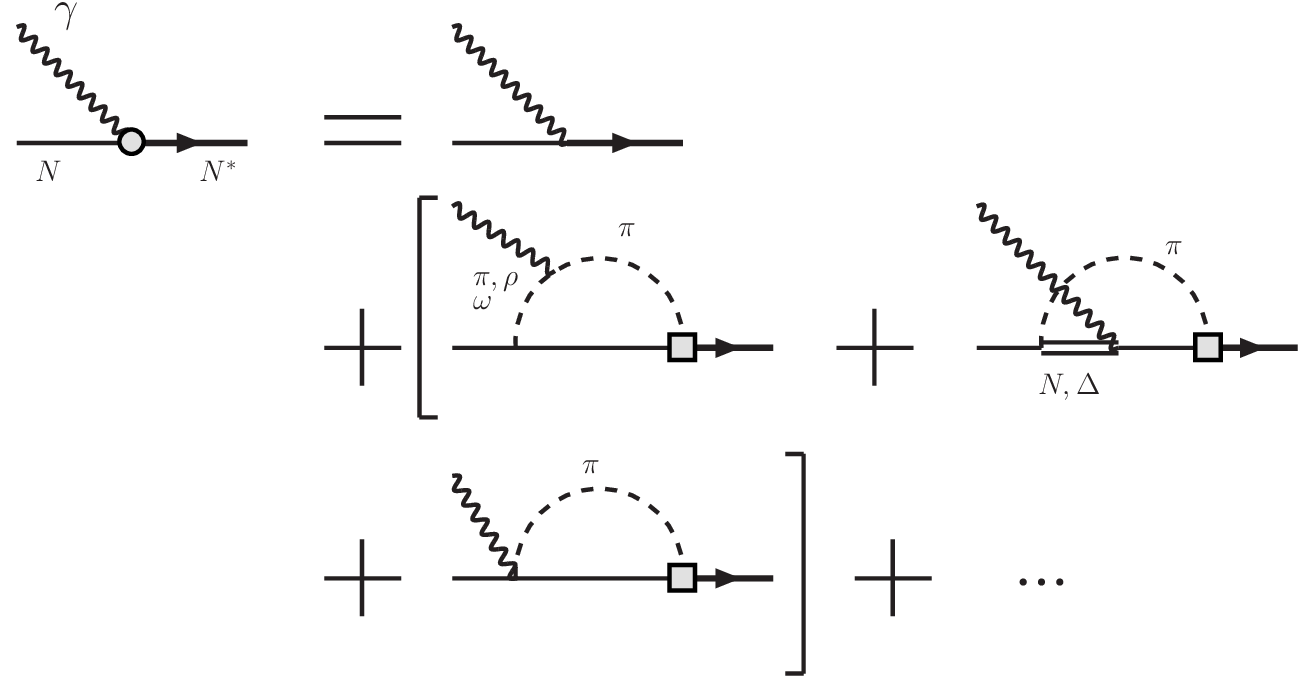}
\end{center}
\caption{
Dressed $\gamma N\to N^*$ vertex diagrams.
Figure taken from Ref.~\cite{nks}. 
Copyright (2015) APS.
}
\label{fig:dcc-diagram}
\end{figure}

Now let us move on to electroweak processes on a single nucleon.
We again use the unitary transformation method to
derive electroweak interaction potentials from the Lagrangians
that have couplings of external currents to hadrons.
Then we describe the electroweak processes with these
perturbative potentials followed by hadronic rescattering;
the rescattering is described by
the scattering amplitudes from Eq.~(\ref{eq:LS-eq}).
Thus, the electroweak amplitudes are given by 
\begin{eqnarray}
A_{\alpha\lambda} (\bm{p}',\bm{q};W,Q^2) &=& 
j_{\alpha\lambda}(\bm{p}',\bm{q},Q^2)
\nonumber\\
 &+& \sum_\gamma \int d\bar{\bm{p}}\;
T_{\alpha\gamma}(\bm{p}',\bar{\bm{p}};W) G_\gamma(\bar{\bm{p}},W)\; 
j_{\gamma\lambda}(\bar{\bm{p}},\bm{q},Q^2)
\ ,
\label{eq:ew-amp}
\end{eqnarray}
where the index $\lambda$ specifies either of $\gamma^{(*)}N$, $W^\pm N$, or
$ZN$ channels with a certain polarization, $\bm{q}$ is the momentum
brought into the hadronic system from the current. 
The electroweak interaction potentials are denoted by 
$j_{\alpha\lambda}$.
The electroweak amplitude denoted by 
$A_{\alpha\lambda}$
corresponds to 
$\langle f|J^\mu_\alpha(0)|i\rangle$ in Eq.~(\ref{hadron}),
and thus we can easily see the connection between 
$A_{\alpha\lambda}$ and 
the cross section formula of Eq.~(\ref{eq:cross})
 for the neutrino-induced meson productions.
Some diagrams with which $A_{\alpha\lambda}$ is build up
are shown in Fig.~\ref{fig:dcc-diagram}.

\subsubsection{DCC analysis of data for pion, photon, and electron-induced
   meson productions off the nucleon}

We have performed a combined analysis of 
$\pi N, \gamma p\to \pi N, \eta N, K\Lambda, K\Sigma$ reaction data
with the DCC model up to $W\le$ 2.1~GeV 
(up to $W\leq 2.3$~GeV for $\pi N\to \pi N$)~\cite{knls}.
With suitably adjusted model parameters, 
the DCC model is able to give reasonable fits to
$\sim$23,000 data points. 
For an extensive presentation of the DCC-based description of the data, 
see Ref.~\cite{knls}.
The model parameters associated with the hadronic interactions and the
vector $N$-$N^*$ transition strengths at $Q^2=0$ have been fixed through
the combined analysis. 
The DCC model obtained above was also applied to $\pi N\to \pi\pi N$ reactions,
and the model predictions were found to give a reasonable description of
the data~\cite{kamano-pipin}.
Before applying the DCC model to the neutrino-induced reactions, 
we need to determine the $Q^2$ dependence of the vector form
factors associated with $N$-$N^*$ transitions, and also need to separate
the vector form factors into isovector and isoscalar parts. 
The $Q^2$ dependence can be determined by analyzing electron-induced
reaction data. 
For the isospin separation, we need to analyze data for 
photon and electron-induced reactions on the neutron.
We have done these analyses to determine the vector form factors for
$W\le$ 2~GeV and $Q^2\le$ 3~GeV$^2$~\cite{nks}.
This covers the whole kinematical region appearing in
the neutrino reactions for $E_\nu\le$ 2~GeV.

\begin{figure}[t]
\begin{center}
\includegraphics[width=78mm]{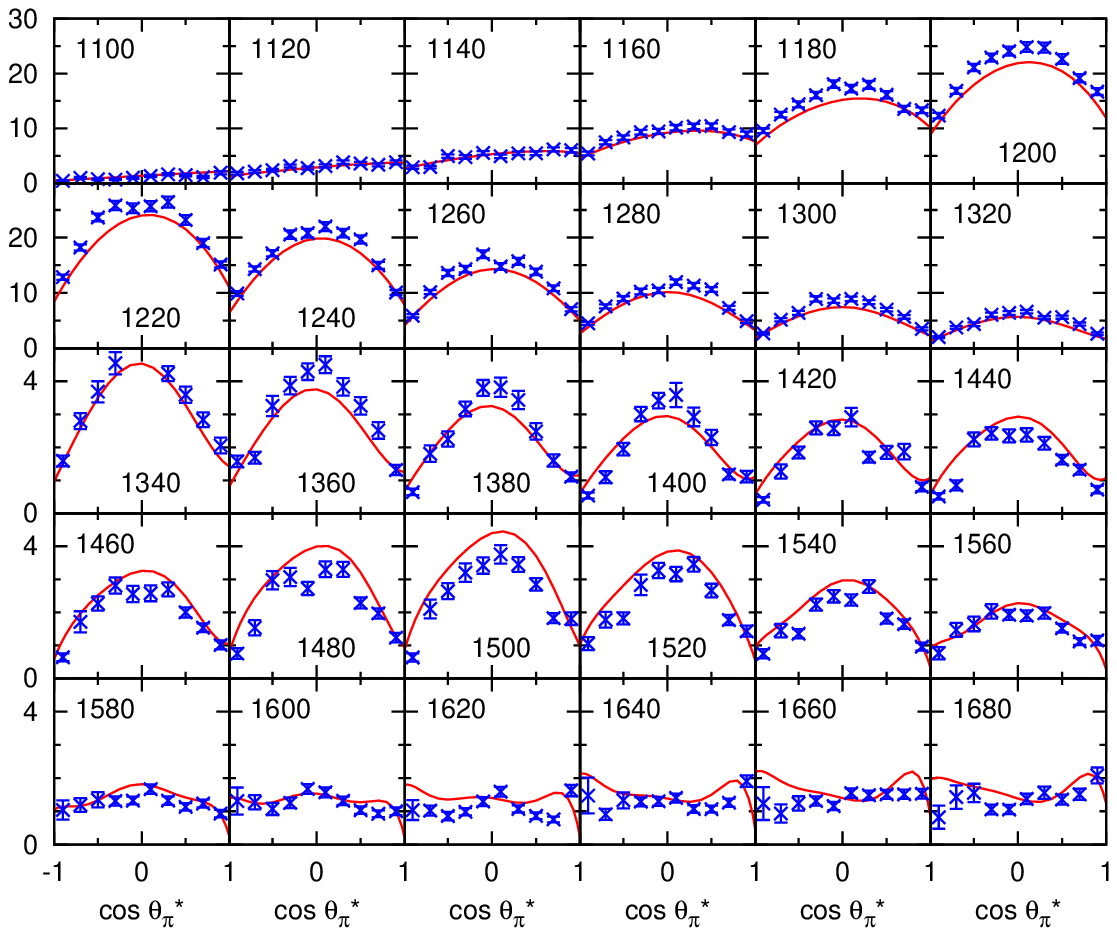}
\end{center}
\caption{(Color online)
The virtual photon cross section 
$d\sigma_T/d\Omega^*_\pi+\epsilon\, d\sigma_L/d\Omega^*_\pi$ ($\mu$b/sr)
at $Q^2$=0.40~GeV$^2$ for
$p(e,e'\pi^0)p$ from the DCC model.
The number in each panel indicates $W$ (MeV).
The data are from Ref.~\cite{eepi-joo-prl}.
Figure taken from Ref.~\cite{nks}. 
Copyright (2015) APS.
}
\label{fig:eepi-0.40}
\end{figure}
Here we present some selected results from the DCC analysis. 
In Fig.~\ref{fig:eepi-0.40}, we present
the virtual photon cross sections 
at $Q^2$=0.40~GeV$^2$ for
$p(e,e'\pi^0)p$ from the DCC model in comparison with the data. 
The agreement with the data is reasonable.
\begin{figure}[t]
\begin{center}
\includegraphics[width=0.34\textwidth]{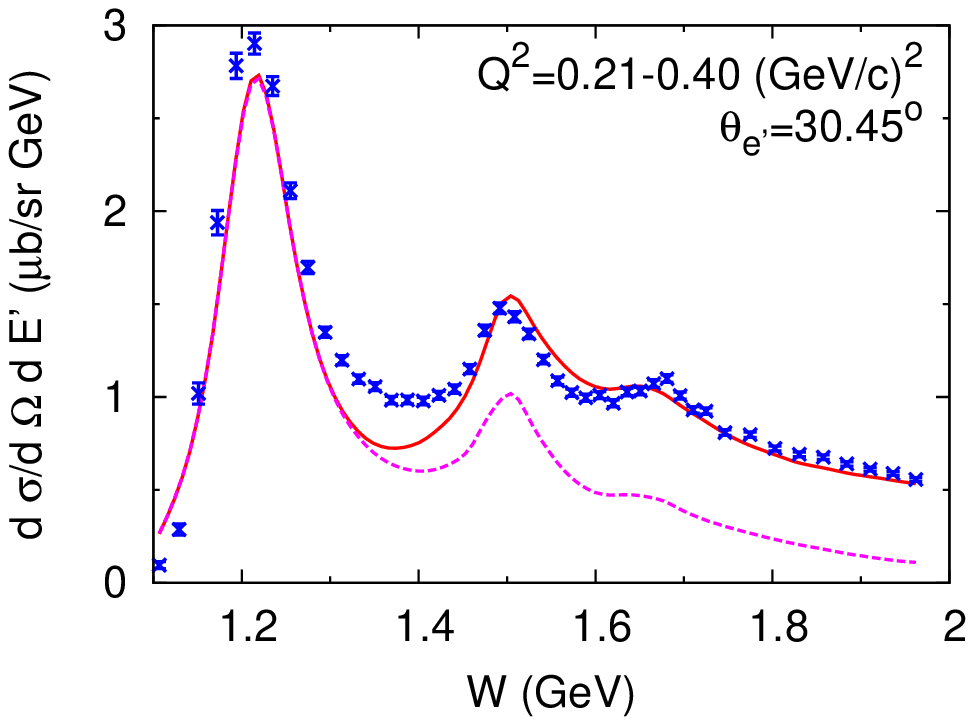}
\hspace{-5mm}
\includegraphics[width=0.34\textwidth]{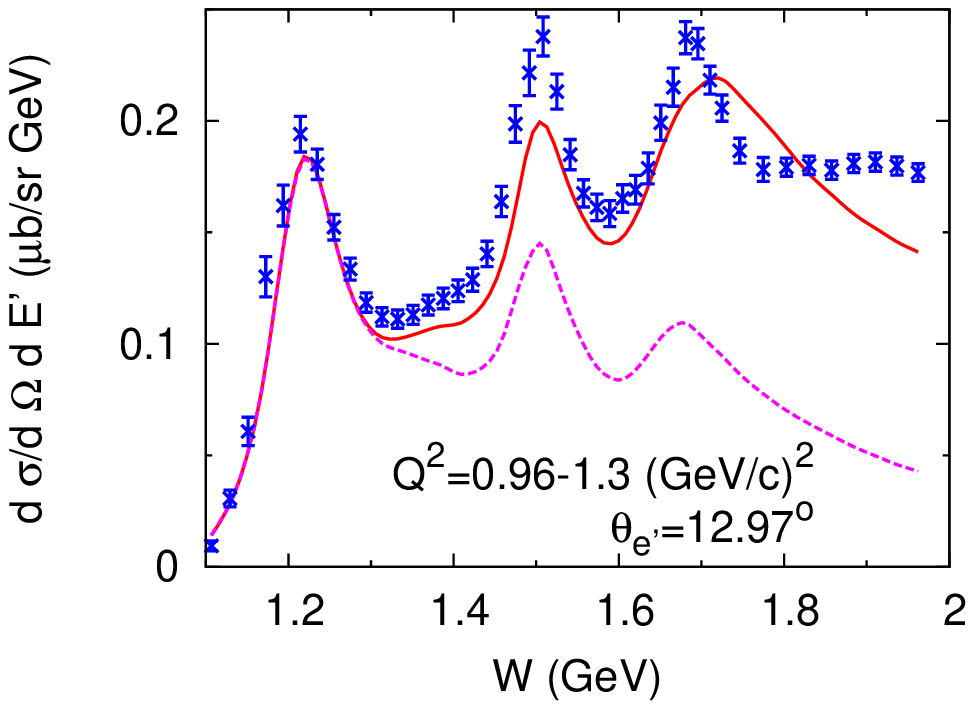}
\end{center}
\caption{(Color online)
Comparison of DCC-based calculation with data for inclusive electron-proton
 scattering at $E_e$=5.498~GeV.
The red solid curves are for inclusive cross sections while the magenta
dashed-curves 
are for contributions from the $\pi N$ final states.
The range of $Q^2$ and the electron scattering angle
 ($\theta_{e'}$) are indicated in each panel.
The data are from Ref.~\cite{E00-002}.
Figures taken from Ref.~\cite{nks}. 
Copyright (2015) APS.
}
\label{fig:eepi-incl}
\end{figure}
Next we present in Fig.~\ref{fig:eepi-incl} 
differential cross sections for
the inclusive electron scattering from the DCC model, 
and compare them with the data. 
In the same figure, we also present 
the single pion electroproduction cross sections from the DCC
model.
The range of $Q^2$ is indicated in each of the panels, 
 and $Q^2$ monotonically decreases as $W$ increases.
Overall, we see a reasonable agreement between the DCC model with the
data.
Also, contributions from the multi-pion production processes are
increasing above the $\Delta(1232)$ resonance region.
We however find 
a discrepancy between the model and data in $W=1.3\sim 1.45$~GeV
at $Q^2\sim$ 0.3~GeV$^2$.
Because the DCC model reasonably describes 
the single pion electroproduction data as seen in 
Fig.~\ref{fig:eepi-0.40}, the discrepancy seems to be from 
a problem of the model in describing double-pion electroproduction in
this kinematics, which might call for a combined analysis including
double-pion production data.
Our purpose is to develop 
a neutrino reaction model in the RES
region that has a comparable quality to neutrino scattering data that
are available in the near future. 
For this purpose, 
we believe that the quality of the fits to the 
electron-induced reactions data at the level seen in the figures
should be enough.

\subsubsection{DCC model for
neutrino-induced meson productions off the nucleon}
\label{sec:dcc}

We now apply the DCC model to the neutrino-induced meson productions. 
Before doing so, we need to fix the remaining unknown piece, the axial
current. 
The nonresonant axial current can be derived from a chiral Lagrangian on
which our $\pi N$ interaction potentials are based. 
By construction, 
the nonresonant axial current and 
the $\pi N$ potentials are related by the PCAC relation at $Q^2=0$.
For the resonant part, namely, the axial $N$-$N^*$ transition strengths
at $Q^2=0$, we relate them to the corresponding $\pi NN^*$ couplings via
the PCAC relation. 
The advantage of our approach over the existing models is that we have 
the $\pi NN^*$ couplings from our DCC model, and thus we can uniquely fix not
only the axial coupling strengths but also their phases. 
In this way, we can make the
interference between the resonant and nonresonant axial amplitudes under
control within the DCC model. 
The $Q^2$ dependences of the axial couplings are difficult to determine
because of the lack of experimental information. 
Here we assume that all of the axial couplings have the same dipole
$Q^2$ dependence of $1/ (1+Q^2/M_A^2)^2$ with $M_A=1.026$~GeV.
With this setup, we make predictions for the neutrino-induced meson
productions, results of which are presented below. 

\begin{figure*}[t]
\includegraphics[height=0.24\textwidth]{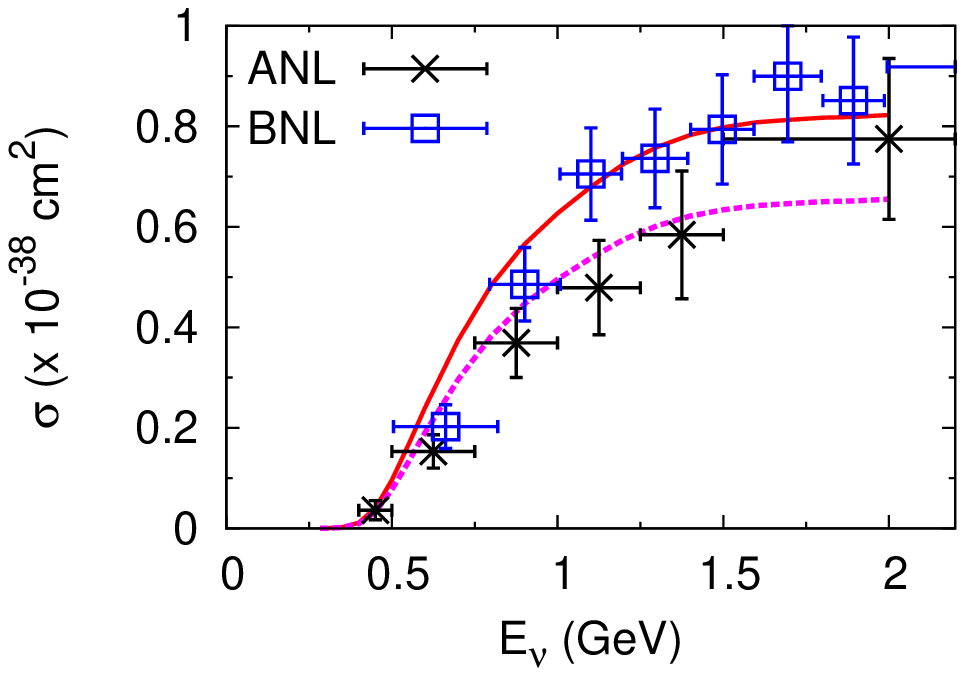}
\hspace{-5mm}
\includegraphics[height=0.24\textwidth]{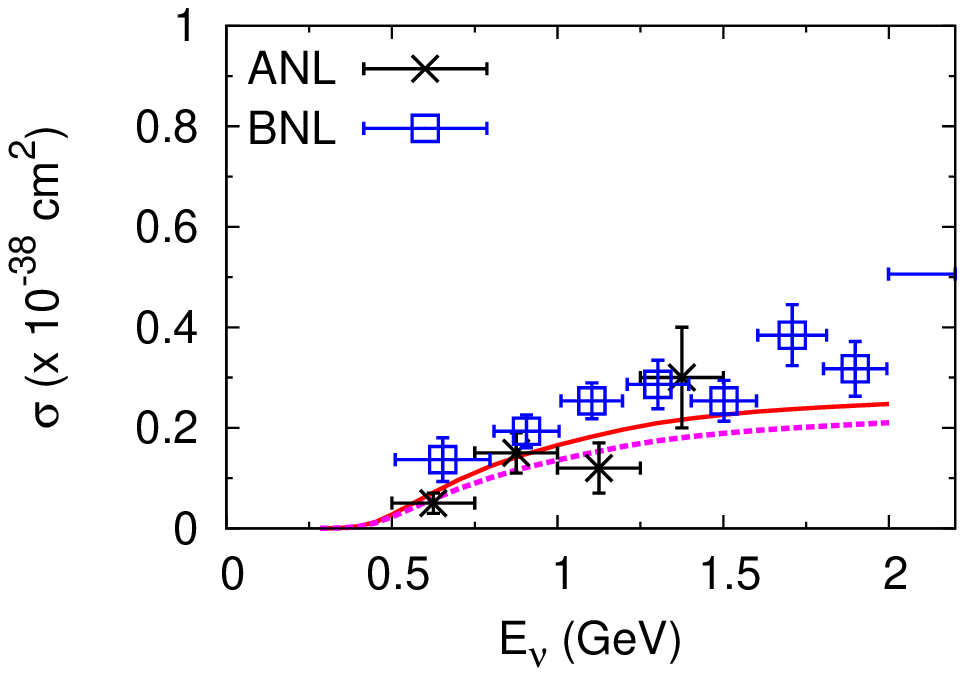}
\hspace{-5mm}
\includegraphics[height=0.24\textwidth]{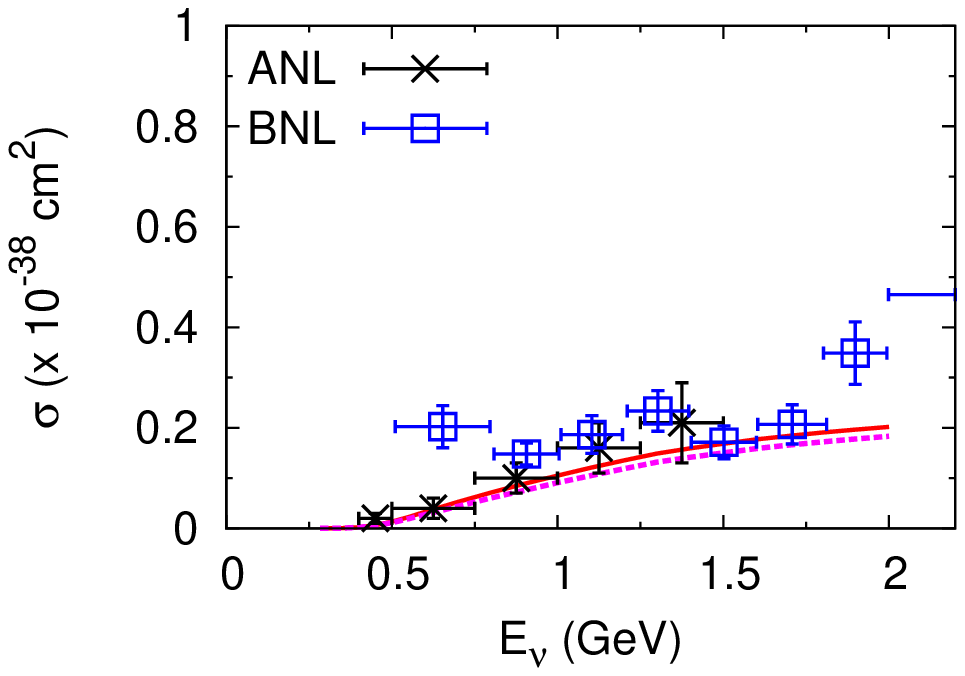}
\caption{(Color online)
The DCC-based calculation (red solid curves) for
$\nu_\mu\, p\to \mu^- \pi^+ p$ (left),
$\nu_\mu n\to \mu^- \pi^0 p$ (middle)
and $\nu_\mu n\to \mu^- \pi^+ n$ (right)
in comparison with data.
The result obtained with $0.8\times g_{AN\Delta(1232)}^{\rm PCAC}$
is also shown (magenta dashed curve).
ANL (BNL) data are from Ref.~\cite{anl} (\cite{Kitagaki:1986ct}).
Figures taken from Ref.~\cite{nks}. 
Copyright (2015) APS.
}
\label{fig:neutrino-tot-data}
\end{figure*}
We present in Fig.~\ref{fig:neutrino-tot-data} the total cross sections
for the single pion productions in comparison with the ANL~\cite{anl}
and BNL~\cite{Kitagaki:1986ct} data. 
Our result obtained with the PCAC-based axial $N$-$\Delta(1232)$ transition
strength ($g_{AN\Delta(1232)}^{\rm PCAC}$) is consistent with the BNL data for 
$\nu_\mu\, p\to \mu^- \pi^+ p$ (Fig.~\ref{fig:neutrino-tot-data} (left)),
and somewhat overestimates the ANL data. 
For the neutron target processes shown in
Fig.~\ref{fig:neutrino-tot-data} (middle, right), 
our result is consistent with both of the ANL and BNL data. 
In a recent reanalysis of the ANL and BNL data~\cite{reanalysis},
it was found that the discrepancy between the two datasets can be
resolved, and the resulting cross sections are reasonably consistent
with the previous ANL data.
Therefore, $g_{AN\Delta(1232)}^{\rm PCAC}$ may be too large, and we are
tempted to adjust it to fit the ANL data. 
Thus we present also in Fig.~\ref{fig:neutrino-tot-data} the total cross
sections obtained with $g_{AN\Delta(1232)}^{\rm PCAC}$ multiplied by 0.8.
Now our cross sections for $\nu_\mu\, p\to \mu^- \pi^+ p$ are consistent
with the ANL data, and those for $\nu_\mu\, n\to \mu^- \pi N$ are not
largely changed because mechanisms other than the $\Delta(1232)$
excitation are also important for these processes on the neutron
target. 
In our present calculations, we do not consider nuclear effects that
must exist in the deuterium target processes. 
Because Ref.~\cite{wsl} showed a large nuclear effect, it will be
important to analyze the deuterium bubble chamber data~\cite{Radecky:1981fn,Kitagaki:1986ct,anl}
with the nuclear effects taken into account. 

\begin{figure*}[t]
\includegraphics[height=0.24\textwidth]{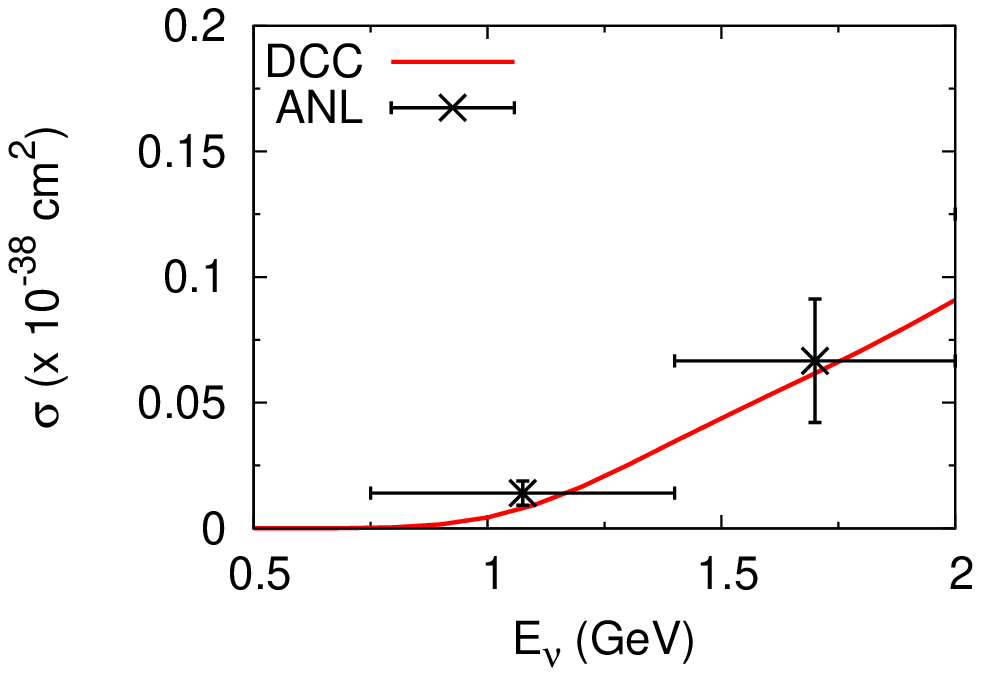}
\hspace{-5mm}
\includegraphics[height=0.24\textwidth]{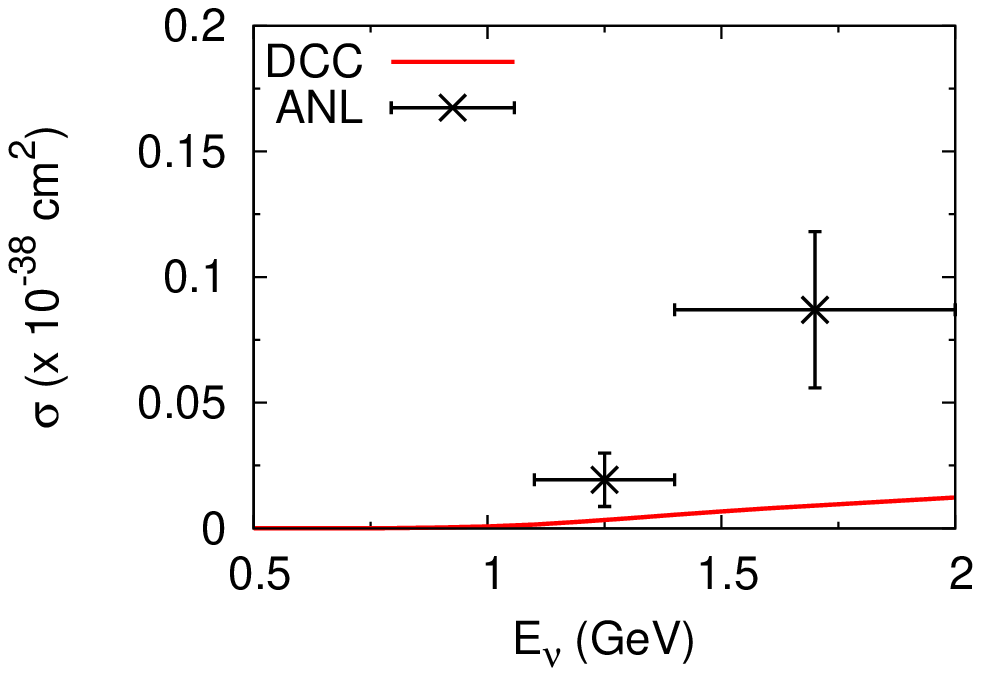}
\hspace{-5mm}
\includegraphics[height=0.24\textwidth]{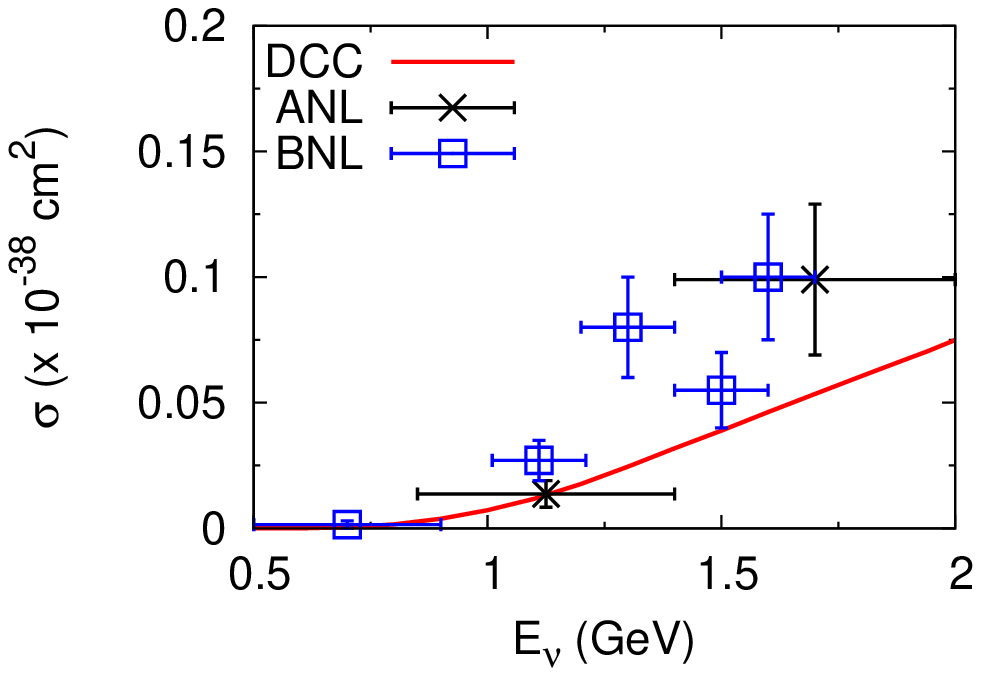}
\caption{(Color online)
The DCC-based calculation for 
$\nu_\mu\, p\to \mu^- \pi^+\pi^0 p$ (left),
$\nu_\mu\, p\to \mu^- \pi^+\pi^+ n$ (middle)
and $\nu_\mu n\to \mu^- \pi^+\pi^- p$ (right)
in comparison with data.
ANL (BNL) data are from Ref.~\cite{anl2} (\cite{Kitagaki:1986ct}).
Figures taken from Ref.~\cite{nks}. 
Copyright (2015) APS.
}
\label{fig:neutrino-pipin-data}
\end{figure*}
We next discuss double pion productions for which our
predictions are presented in Fig.~\ref{fig:neutrino-pipin-data}
in comparison with data~\cite{Kitagaki:1986ct,anl2}.
Our calculation for these processes has been done 
with contributions from all relevant resonances  below $W=2$~GeV taken into account 
for the first time; 
other previous models~\cite{biswas,adjei,spain-pipin}
consist of dynamical contents that were valid only near the threshold.
In comparison with the data, we obtained a good agreement for 
$\nu_\mu\, p\to \mu^- \pi^+\pi^0 p$
and $\nu_\mu n\to \mu^- \pi^+\pi^- p$.
However, the cross sections for 
$\nu_\mu\, p\to \mu^- \pi^+\pi^+ n$ are rather underestimated.
Because the statistics of the data is rather limited, we do not attempt
to fit our model to the data. 
\begin{figure*}[t]
\includegraphics[height=0.24\textwidth]{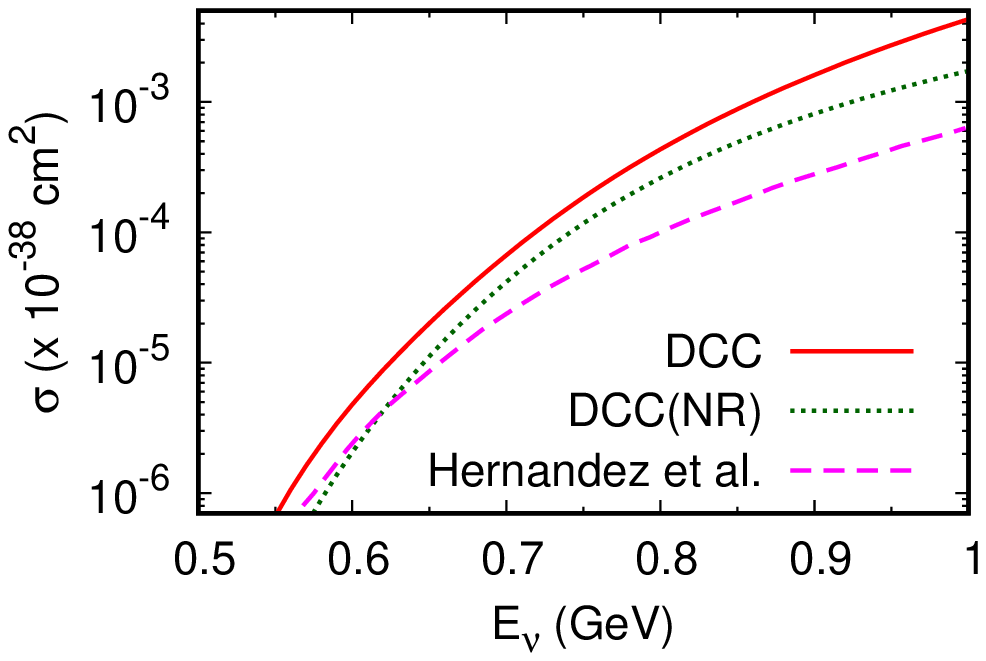}
\hspace{-5mm}
\includegraphics[height=0.24\textwidth]{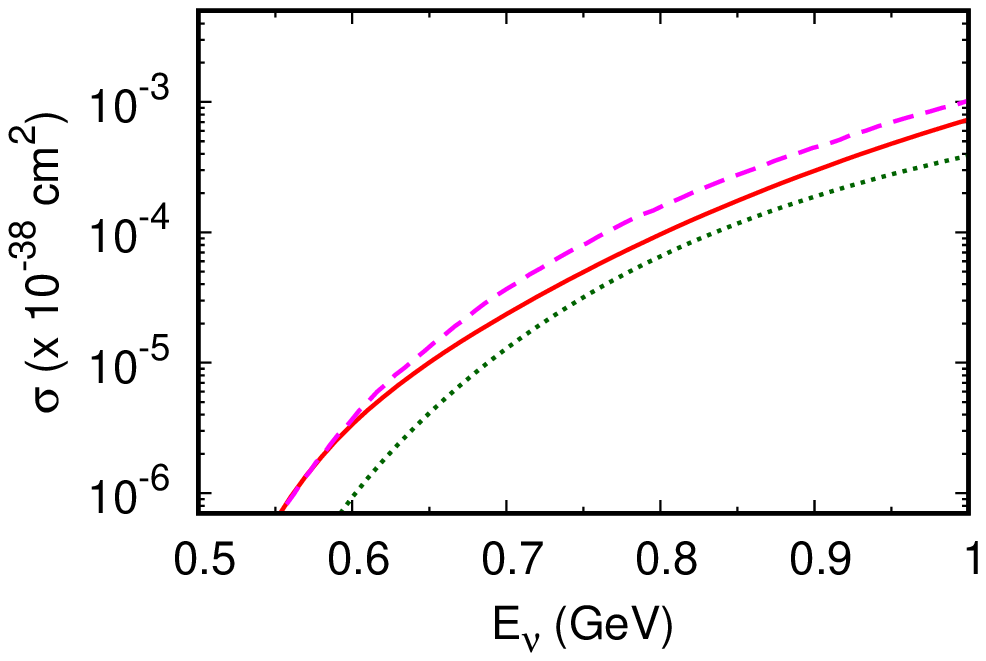}
\hspace{-5mm}
\includegraphics[height=0.24\textwidth]{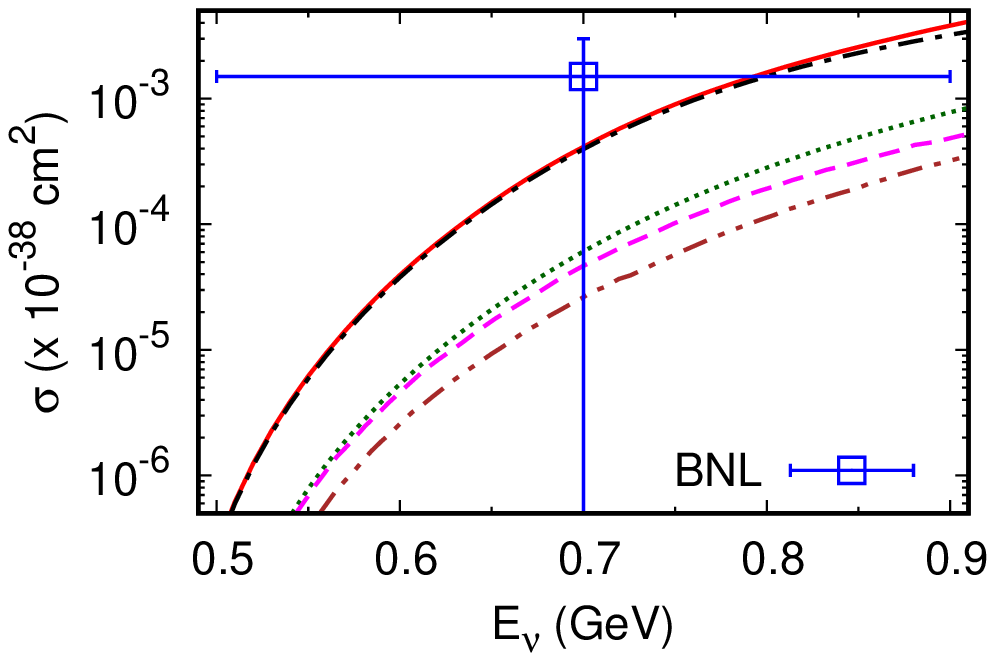}
\caption{(Color online)
Comparison of cross sections from the DCC model with those from 
the Full model (FF4) of Hern\'andez et al.~\cite{spain-pipin} for 
$\nu_\mu\, p\to \mu^- \pi^+\pi^0 p$ (left),
$\nu_\mu\, p\to \mu^- \pi^+\pi^+ n$ (middle)
and $\nu_\mu n\to \mu^- \pi^+\pi^- p$ (right) near the threshold.
Contributions from the non-resonant mechanisms are indicated by 'NR'.
In the right panel, the black dash-dotted curve is from the NR
and the $P_{11}$ resonant contributions of the DCC model,
while the brown dash-two-dotted curve is from the NR of 
Ref.~\cite{spain-pipin}.
In the left and middle panels, the results of Ref.~\cite{spain-pipin}
includes only the NR contributions because $N(1440)1/2^+$, which is the
only resonance considered in Ref.~\cite{spain-pipin}, does not
contributes to these channels.
BNL data are from Ref.~\cite{Kitagaki:1986ct}.
}
\label{fig:neutrino-pipin-data-thres}
\end{figure*}
We also compare
the result from the DCC model with those from
Ref.~\cite{spain-pipin} in Fig.~\ref{fig:neutrino-pipin-data-thres}.
The model of Ref.~\cite{spain-pipin} consists of non-resonant
mechanisms derived from a chiral Lagrangian, and a resonant mechanism
associated with an excitation of the Roper resonance ($N(1440)1/2^+$).
Because this dynamical content is expected to be valid only near the
threshold, we limit the comparison to $E_\nu\ltap 1$~GeV.
For a detailed comparison, non-resonant contributions 
from the two models are also shown.
In $\nu_\mu\, p\to \mu^- \pi^+\pi^0 p$ (Fig.~\ref{fig:neutrino-pipin-data-thres}\,(left)) and
$\nu_\mu\, p\to \mu^- \pi^+\pi^+ n$ (Fig.~\ref{fig:neutrino-pipin-data-thres}\,(middle)) processes,
where $I=1/2$ ($I$: isospin) resonances (and thus the Roper) are not excited,
only the non-resonant mechanisms contribute in the model of
Ref.~\cite{spain-pipin}.
The non-resonant mechanisms of the two models give rather different
contributions to each of the channels, but it is difficult to identify the
origin of the difference from this comparison only.
The resonant contributions are also quite different between the two
models.
While the Roper resonance in the model of
Ref.~\cite{spain-pipin} enhances the $\nu_\mu n\to \mu^- \pi^+\pi^- p$
cross sections by $\sim$\,20\,-\,30\,\% (Fig.~\ref{fig:neutrino-pipin-data-thres}\,(right)),
the contribution seems much smaller than that
from the DCC's resonant $P_{11}$ partial wave amplitude where the Roper exists.
The $I=3/2$ resonances, not considered in Ref.~\cite{spain-pipin} but in
the DCC model, also give significant contributions as seen in 
Fig.~\ref{fig:neutrino-pipin-data-thres}(left,middle) even near the threshold.

\begin{figure}[t]
\begin{center}
\includegraphics[height=0.33\textwidth]{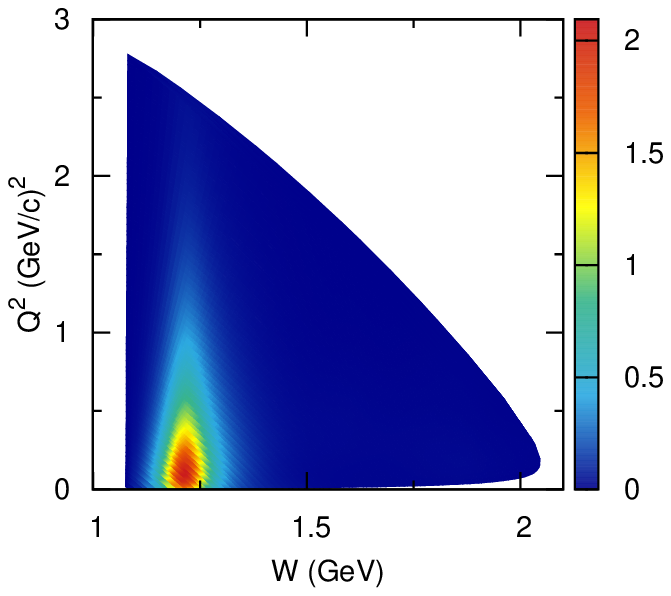}
\hspace{10mm}
\includegraphics[height=0.33\textwidth]{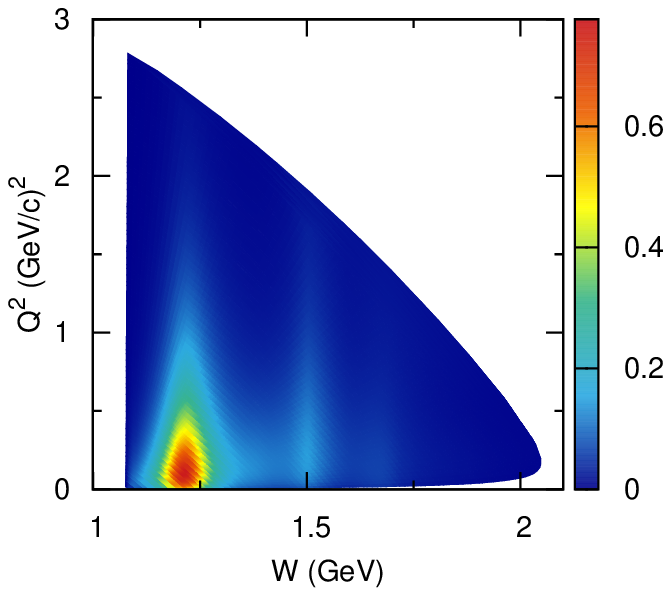}
\end{center}
\caption{(Color online)
$d\sigma/dWdQ^2$ for
$\nu_\mu\, p\to \mu^- \pi^+ p$ (left)
and $\nu_\mu n\to \mu^- \pi N$ (right)
at $E_\nu=2$~GeV
in contour plots.
Figures taken from Ref.~\cite{nks}. 
Copyright (2015) APS.
}
\label{fig:ds_dw_dq2_pin}
\end{figure}
\begin{figure}[t]
\begin{center}
\includegraphics[height=0.33\textwidth]{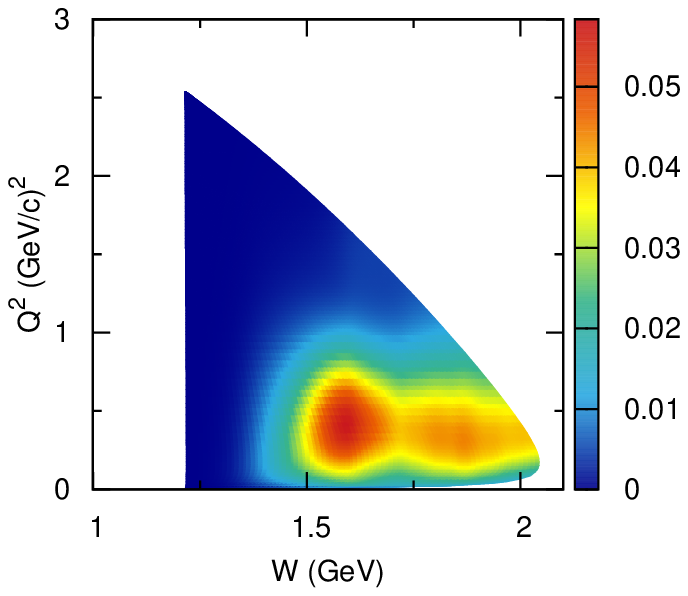}
\hspace{10mm}
\includegraphics[height=0.33\textwidth]{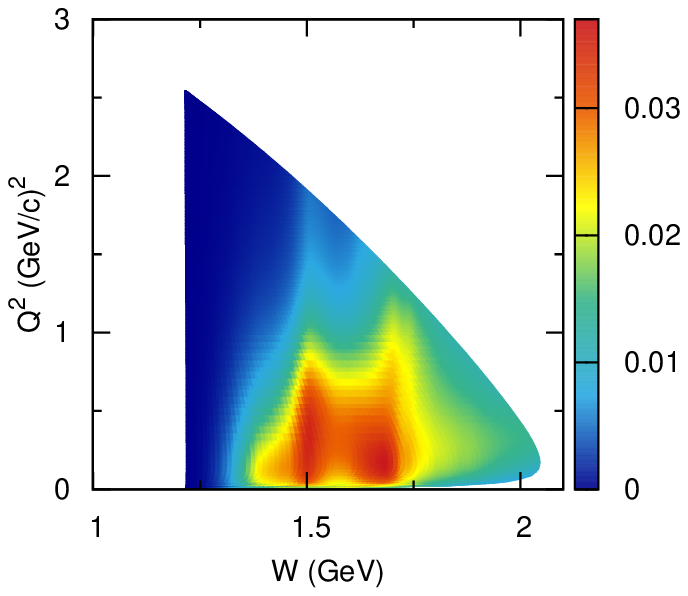}
\end{center}
\caption{(Color online)
$d\sigma/dWdQ^2$ for
$\nu_\mu\, p\to \mu^- \pi^+ \pi^0 p$ (left)
and $\nu_\mu n\to \mu^- \pi^+\pi^- p$ (right)
at $E_\nu=2$~GeV
in contour plots.
Figures taken from Ref.~\cite{nks}. 
Copyright (2015) APS.
}
\label{fig:ds_dw_dq2_pipin}
\end{figure}
Now let us examine the double differential cross sections, 
$d\sigma/dWdQ^2$, shown in Fig.~\ref{fig:ds_dw_dq2_pin}
for the single pion productions and in Fig.~\ref{fig:ds_dw_dq2_pipin}
for the double pion productions at $E_\nu$=2~GeV.
The figures clearly show the resonant behavior.
For the single pion productions, the $\Delta(1232)$ excitation creates
the prominent peak, with a long tail toward the higher $Q^2$ region.
For the neutron target process (Fig.~\ref{fig:ds_dw_dq2_pin} (right)),
the second resonances at $W\sim$ 1.5~GeV also create the noticeable peak.
For the double pion productions, the situation is completely different. 
We now do not have the $\Delta(1232)$ peak because it is below the 
threshold for the double pion productions, and the main contributors are
the $N^*$s in the so-called second and third resonance regions as
clearly seen in Fig.~\ref{fig:ds_dw_dq2_pipin}.

\begin{figure}[t]
\begin{center}
\includegraphics[height=0.33\textwidth]{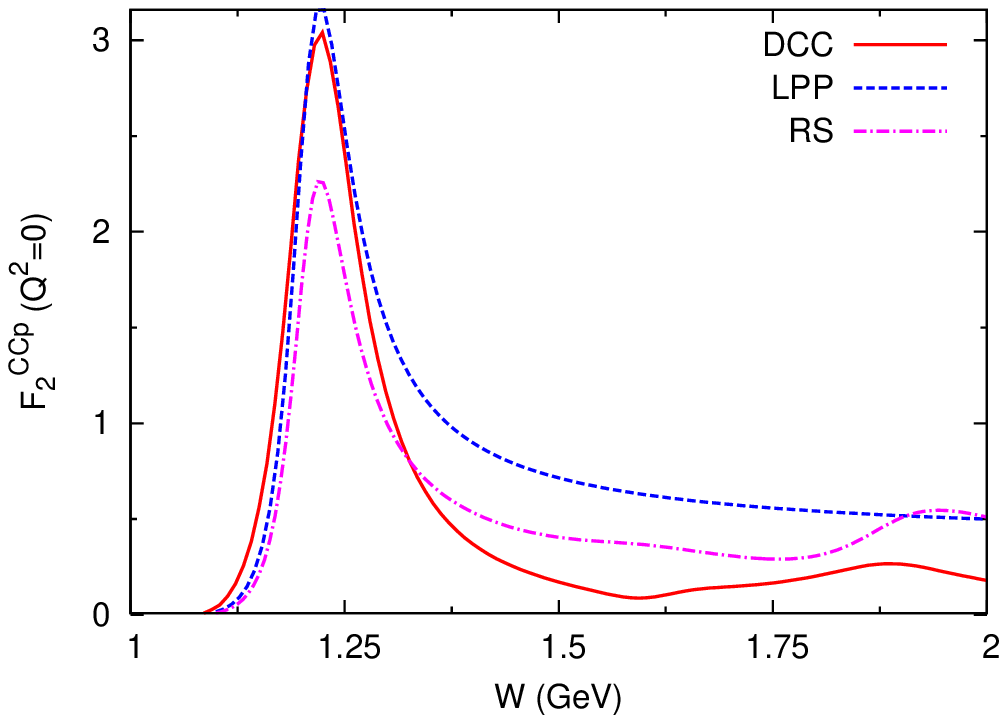}
\includegraphics[height=0.33\textwidth]{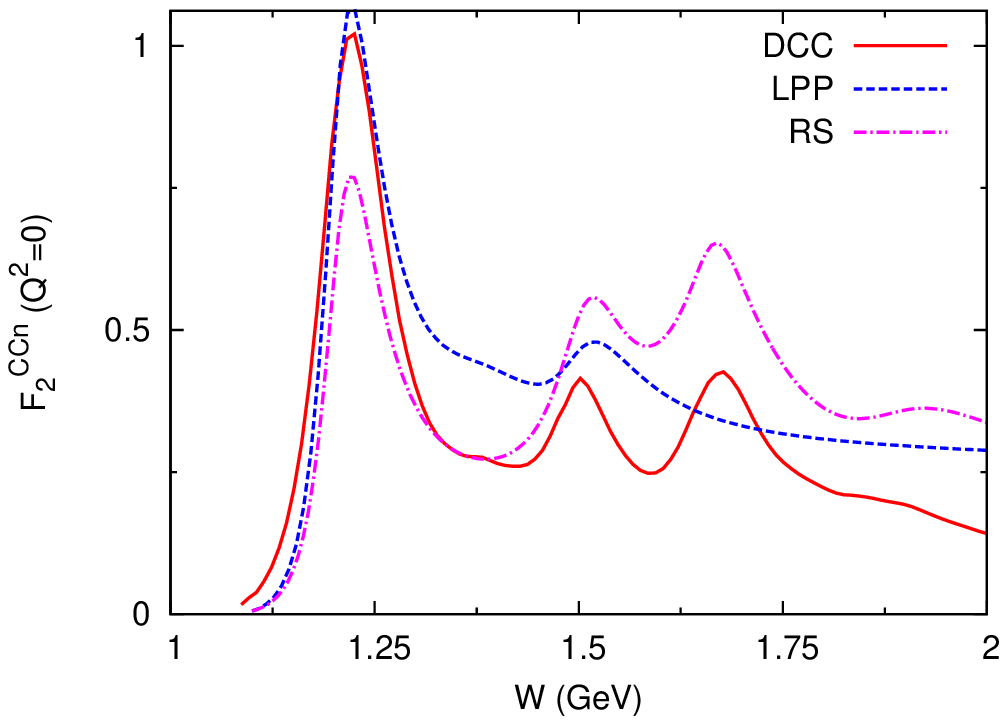}
\end{center}
\caption{(Color online)
$F^{\rm CC}_2$ at $Q^2=0$ 
for the single pion production.
The DCC model is compared with
Lalakulich et al. (LPP) model~\cite{LPP}
and Rein-Sehgal (RS) model~\cite{RS,RS2}.
The left (right) panel is for the CC
$\nu_\mu\, p$ ($\nu_\mu\, n$) reaction.
Figures taken from Ref.~\cite{nks}. 
Copyright (2015) APS.
}
\label{fig:LP}
\end{figure}
Finally, we compare our predictions from the DCC model with those from
other models developed by Lalakulich et al. (LPP)~\cite{LPP}
and Rein-Sehgal (RS)~\cite{RS,RS2}.
The LPP model consists of four Breit-Wigner amplitudes
for $\Delta(1232)3/2^+$, $N(1535)1/2^-$, $N(1440)1/2^+$ and $N(1520)3/2^-$ resonances
with no background.
The RS model has 18 Breit-Wigner amplitudes plus a non-interfering
non-resonant background of $I=1/2$.
We show in Fig.~\ref{fig:LP}
$F^{\rm CC}_2$ (see Eq.~(\ref{eq:f123}) for definition)
at $Q^2=0$ that includes contributions from 
the single pion production only. 
Near the $\Delta(1232)$ peak, we find a good agreement between the LPP
and the DCC
models while the RS model makes a significant underestimation. 
In the higher energy region, both the LPP and RS models rather
overestimate the result from the DCC model.
According to the PCAC relation,
$F^{\rm CC}_2$ at $Q^2=0$ is related to the $\pi N$ cross sections, and
thus is given almost model-independently.
Within our DCC model, 
the axial current satisfies the PCAC relation to
the precise $\pi N$ model by construction, and therefore 
$F^{\rm CC}_2$ from the DCC model agree well with those from the 
$\pi N$ cross sections.
On the other hand, the other models did not fully implement this
consistency required by the PCAC relation, and as a consequence, 
we found the difference in $F^{\rm CC}_2(Q^2=0)$
between the DCC model and the LPP and the RS models.

\clearpage
\section{Neutrino reactions in few-nucleon systems}

A precise description of neutrino-nucleus reactions especially in the QE
region is crucial for analyzing data from the long-baseline neutrino
oscillation experiments.
A practical theoretical description of the one-nucleon knock-out reaction in the 
impulse approximation will be discussed in the next section. However,
due to a wide energy band of the neutrino flux, reactions somewhat off the
QE peak
region become relevant for precisely determining the neutrino properties.
In this region, nuclear correlations such as two-particle two-hole effects
including meson-exchange currents play an important  role.
Since precise neutrino-nucleus reaction data
comparable to electron scattering are not available,
further efforts to reduce systematic
uncertainties originating from theoretical treatments of
nuclear electroweak break-up reactions are highly called for.
An {\it ab initio} approach would be a promising option in this regard.

In the {\it ab initio} approach,
nuclear many-body problems are solved
in principle 'exactly', once nuclear interactions
such as realistic nucleon-nucleon potential and nuclear currents 
(impulse and meson-exchange currents) are set.
Approaches of this kind have been extensively applied
to low-energy ($E_\nu \ltap 100$~MeV) break-up reactions in few-nucleon systems.
This is partly because the knowledge of low-energy electroweak processes,
neutrino-nucleus reactions in particular,
are of great importance to understand astrophysical phenomena
such as the neutrino heating in core-collapse supernovae of massive
stars~\cite{bethe85,Balasi15,haxton1988,nakamuraetal, nasuetal},
nucleosynthesis via neutrino reactions~\cite{Woosley90,suzuki2006},
and yields of light isotopes from which 
the neutrino properties can be extracted~\cite{Yoshida2006}.
In the next two paragraphs, we briefly summarize previous
{\it ab initio} calculations for electroweak processes in few-nucleon systems.
An extension of these {\it ab initio} approaches to higher energy, e.g., the QE peak region,
is highly desirable, because it would
provide information on the role of nuclear correlations and
nuclear currents in the energy region relevant to the neutrino
oscillation experiments.
Such a calculation for electron scatterings on $^4$He and $^{12}$C has
just been done recently based on
the Green's function Monte Carlo approach, and indeed, leading
to interesting findings on those nuclear many-body mechanisms~\cite{gfmc}.
Yet, it would be desirable to confirm the results with an independent
{\it ab initio} calculation.

Electroweak reactions in two-nucleon systems at low energies have been
studied with the conventional nuclear physics approach (CNPA)
that consists of high precision nucleon-nucleon potential and one- and two-nucleon 
electroweak currents~\cite{NSGK,Netal}. The approach has been
successful in describing electron scattering~\cite{Tamura92} and photo-reactions~\cite{Sato95}
with the electromagnetic current, 
while the nuclear axial vector current has been tested by the muon
capture rate~\cite{Doi90}.
Meanwhile, the effective field theory approach~\cite{Kubodera2004}, equipped with a 
 systematic expansion scheme, has been applied to the 
 $\nu$-$d$ reaction~\cite{Butler2000} and recently to the $pp$ fusion reaction~\cite{Marucci2013}.
Both approaches agree with each other for the low-energy neutrino-deuteron reactions~\cite{Netal}.
Recently, the neutrino-deuteron reaction has been studied up to $E_\nu \ltap 1$ GeV 
region with the CNPA~\cite{shen2012}.

{\it Ab initio} calculations of electroweak reactions on $A \ge 3$ nuclei
including multi-nucleon break-up channels have been 
carried out with various approaches.
Electromagnetic reactions on three-nucleon systems have been extensively studied
based on the Faddeev calculations~\cite{Golak2005}.
The Lorentz integral transformation method has been applied 
to electron scattering and photo reactions~\cite{Bacca14}, 
and also to neutrino reactions in the supernova environment~\cite{gazit2007,oconner07}.
The Green's function Monte Carlo method was used for electromagnetic 
and NC neutrino reactions~\cite{Lovato2015,carlson2015}.

In what follows, we discuss a promising alternative 
{\it ab initio} approach formulated with a combination of 
the correlated Gaussian (CG) and 
the complex scaling method (CSM), and its application 
to the dipole and spin-dipole responses of $^4$He in electroweak processes.

\subsection{Calculation of nuclear strength functions}

The nuclear excitation processes are described with nuclear strength
(response) functions. The nuclear strength functions for the electroweak 
reactions reflect important information on resonant and continuum structure of the nuclear system.
The nuclear strength function with the excitation energy $E$
for an operator ${\mathcal{O}}$
  characterized by the angular momentum and isospin 
labels, $\lambda$ and $p$, is defined by
\begin{equation}
S(p,\lambda,E)=\mathcal{S}_{f\mu}\left|\right<\Psi_f|
\mathcal{O}_{\lambda\mu}^p
|\Psi_0\left>\right|^2
\delta(E_f-E_0-E),
\end{equation}
where $\Psi_0$ ($\Psi_f$) is the ground (final) state
wave function with the energy $E_0$ ($E_f$),
and $\mathcal{S}_{f\mu}$ denotes the summation over all the final states
as well as the $z$-component of the angular momentum, $\mu$.
The label $p$ distinguishes different types of isospin operators, e.g., 
isoscalar, 1 ($p=$IS), isovector, $\tau_{3}$ ($p=$IV0), 
charge-exchange $\tau^\mp$ ($p=$IV$\mp$),
and electric $\frac{1}{2}(1+\tau_{3})$ types ($p=E\lambda$).
(Here we adopt convention of isospin described in Sec.~\ref{sec:nu-nucleon} throughout
this paper, while convention in Ref.~\cite{Horiuchi13b}
is $\tau^{+}\left|p\right> = \left|n\right>$ and  $\tau_0=-\tau_3$.)
Taking the summation over the final states makes it possible to 
rewrite the strength function as 
\begin{equation}
S(p,\lambda,E)=-\frac{1}{\pi}{\rm Im}\sum_{\mu}\left<\Psi_0\right|
  \mathcal{O}_{\lambda\mu}^{p\dagger}
  \frac{1}{E-H+E_0+i\epsilon}  \mathcal{O}_{\lambda\mu}^p\left|\Psi_0\right>,
\label{resp2.eq}
\end{equation}
where $H$ is the nuclear Hamiltonian and $1/(E-H+E_0+i\epsilon)$ is 
the many-body Green's function. 
A positive infinitesimal $\epsilon$ is put to 
ensure the outgoing wave after the excitation of the initial state.

Though an explicit construction of the final states is avoided in 
Eq.~(\ref{resp2.eq}), an evaluation of Eq.~(\ref{resp2.eq})
is still in general difficult
because of the presence of the Green's function that involves
complicated many-body correlations and boundary conditions. We 
employ the complex scaling method (CSM)~\cite{Moiseyev98}
to avoid these complications.
In the CSM, a particle coordinate (momentum),
$\bm{r}$ ($\bm{k}$), is rotated on the complex plane
by a positive angle $\theta$ as
$\bm{r}{\rm e}^{i\theta}$ ($\bm{k}{\rm e}^{-i\theta}$).
Under this transformation,
the asymptotics of the wave function
damps exponentially at large distances,
which allows us to represent the Green's function in the 
expansion by the eigenstates of the complex-rotated
Hamiltonian, $H(\theta)$,
\begin{equation}
  H(\theta)\Psi_k(\theta)=E_k(\theta)\Psi_k(\theta).
\end{equation}
This class of complex eigenvalue problems is solved
with a set of square-integrable ($\mathcal{L}^2$) basis functions.
Since the resonant and continuum states are treated
in a manner similar to a bound-state problem,
the method has widely been applied to calculating the
strength functions~\cite{CSM}.
The accuracy of the CSM calculation crucially depends on how
completely the ${\cal L}^2$ basis functions are prepared.
In principle, if the model space is complete,
the result would not depend on the scaling angle in some limited range of $\theta$.
Practically, the $\theta$ value is determined 
by examining the stability of $S(E)$ against changing $\theta$.

\subsubsection{Correlated Gaussian method}

As the basis functions we employ  
correlated Gaussians (CG)~\cite{Varga95,SVM,Suzuki08},
which are flexible enough to describe 
different types of structure and correlated motion of particles.
Many examples have confirmed that the CG method can describe,
e.g., short-range repulsion and tensor correlations
in the nuclear force~\cite{Suzuki08,Kamada01,Feldmeier10},
and both cluster and shell-model
configurations~\cite{Horiuchi14}. See also a recent review~\cite{Suzuki16}.
Because of its flexibility,
the basis functions have been applied to describe not only nuclear physics 
but also other quantum physics~\cite{Mitroy13}.

The total wave f{unction with the angular momentum $J$, its $z$-component 
$M_J$, parity $\pi$, and isospin quantum numbers $T,\, M_T$ is expressed 
as a combination of many basis functions. Each basis function is 
given} in $LS$ coupling scheme
\begin{equation}
\Phi_{(LS)JM_J, TM_T}^{\pi}=\mathcal{A}\left[\phi_L^{\pi}\times
\chi_S\right]_{JM_J}\eta_{TM_T},
\label{LScoupling}
\end{equation}
where $\mathcal{A}$ is the antisymmetrizer, and the symbol 
$\left[L \times S\right]_{JM_J}$ 
stands for  the angular momentum coupling. 
The total spin (isospin) function $\chi_S$ ($\eta_{T})$ is constructed
by a successive coupling of the spin (isospin) functions of all the nucleons.

For the spatial part of the basis function, $\phi_L^\pi$, we use the CG.
Let $\bm{x}$=($\bm{x}_i$) denote a
set of the Jacobi coordinates excluding the center-of-mass coordinate. 
We express $\phi_L^{\pi}$ as~\cite{Varga95,SVM}
\begin{equation}
\phi_{LM_L}^{\pi}(A,\bm{x})
=\exp(-\tilde{\bm{x}}A\bm{x})
\theta_{LM_L}^\pi(\bm{x}),
\label{GVR.eq}
\end{equation}
where $\tilde{\bm{x}}A\bm{x}=\sum_{i,j}A_{ij}\bm{x}_i\cdot\bm{x}_j$
with a positive-definite symmetric matrix $A$.
The angular part $\theta_{LM_L}^\pi(\bm{x})$
is expressed by coupling the solid harmonics, 
$\mathcal{Y}_{lm_l}(\tilde{u}\bm{x})=
|\tilde{u}{\bm{x}}|^l Y_{lm_l}(\widehat{\tilde{u}\bm{x}})$, where 
$\tilde{u}\bm{x}=\sum_{i}(u)_i\bm{x}_i$ is a global vector. 
The reader is referred to Refs.~\cite{SVM, Suzuki08, Aoyama12}
for details of single-, double-, and triple-global-vector
representations. 
It should be noted that all coordinates are explicitly correlated through
the $A$ and $u$ ($u$'s).
An advantage of the representation is that it keeps its functional form 
under any linear transformation of the coordinates,
which is a key to describing many-body bound and unbound states
in a unified manner. 

\subsection{{\it Ab initio} calculation for $^{4}$He}

\subsubsection{Hamiltonian and spectrum of $^{4}$He}

The Hamiltonian of an $N$-nucleon system consists of
two- and three-nucleon forces
\begin{equation}
  H=\sum_{i=1}^NT_i-T_{cm}+\sum_{i<j}V_{ij}^{(2)}+\sum_{i<j<k}V_{ijk}^{(3)},
  \end{equation}
where $T_i$ is the single-nucleon kinetic energy and the center-of-mass 
kinetic energy is subtracted to ensure the nuclear intrinsic motion.
We employ Argonne $v$8$^\prime$~\cite{AV8p} (AV8$^\prime$)
potential which contains central, tensor and spin-orbit components.
Since it is vital to reproduce the threshold energies
in the calculation of the strength function,
a central three-body interaction (3NF)~\cite{Hiyama04} is employed
to reproduce the binding energies of the three- and four-nucleon bound states.

The ground state wave function of $^{4}$He is obtained
with a superposition of many CG functions of Eq.~(\ref{LScoupling}).
The set of the variational parameters,
$A$, $u$'s, $L$, spin and isospin configurations,
 are determined by
the stochastic variational method (SVM)~\cite{SVM},
which allows us to get a precise solution of a many-body Schr\"odinger
equation in a relatively small number of bases.
The ground state energy agrees with the one obtained by other methods~\cite{Suzuki08, Kamada01}.

The first excited $0^+$ and the seven negative-parity states are  observed 
below the excitation energy of 26 MeV~\cite{Tilley92}
and all of them are reproduced
very well~\cite{Horiuchi08,Horiuchi13a}.
It should be noted that the level ordering of $^{4}$He 
can be reproduced only when the realistic nuclear interaction is employed. 
If one uses an effective interaction that consists of a central term alone,
the negative parity levels would be almost degenerate,
and no correct level ordering could be obtained~\cite{Horiuchi13a}.
The tensor term  plays a decisive role, for example,
the lowering of $J^\pi T= 0^-0$ state is understood 
by a strong coupling between different angular momentum channels
due to the tensor force~\cite{Suzuki08}.

\subsubsection{Dipole-type excitations of $^{4}$He}

The dipole- and spin-dipole(SD)-type operators
are main pieces to determine neutrino-$^4$He reaction cross sections
at low energies. 
Though the SD operators belong to a class of the first forbidden transition,
they can dominantly contribute to the reaction on $^4$He because
 the Fermi and Gamow-Teller type transitions are strongly suppressed
due to the closed shell nature.

The spectrum of the four-nucleon system is closely related to
the dipole-type electroweak responses. 
The seven negative-parity states of $^4$He can be excited by  six SD
and  one dipole operators~\cite{Horiuchi13b}. 
Basis functions for the final states reached by these operators are constructed
by paying attention to two points: the sum rule 
of the electroweak strength functions 
and the decay channels~\cite{Horiuchi13b,Horiuchi12}. In fact the basis 
functions are constructed in three types: 
(i) a single-particle excitation built on the $^4$He ground-state
wave function multiplied by $\mathcal{O}_{\lambda\mu}^p$.
(ii) a $3N$+$N$ ($^3$H+$p$ and $^3$He+$n$) two-body disintegration.
(iii) a $d$+$p$+$n$ three-body disintegration.
The basis (i) is useful for satisfying the sum rule, 
and the bases (ii) and (iii) take care of the two- and 
three-body decay asymptotics. These cluster configurations 
are better described using appropriate relevant coordinates 
rather than the single-particle coordinate.
The relative motion between the clusters is described
with several Gaussians. For the wave functions of $d$ and $3N$
subsystems, we use a set of the bases obtained 
by the two- and three-body calculations
with the SVM algorithm~\cite{Varga95,SVM}, which
greatly reduces the total dimension of the matrix elements.
The expression is again given in the CG with the global vectors
and the matrix elements can be evaluated without any change
of the formulas.

\subsubsection{Photoabsorption of $^4$He}

First, we discuss photoabsorption reactions
of $^4$He to see the reliability of the method.
There has been a controversy in the low-lying photoabsorption cross section,
that is, the experimental data are in serious disagreement~\cite{Shima05,Nakayama07}.
In the energy region around 26 MeV, the photoabsorption reaction 
takes place mainly through the electric-dipole ($E1$) transition.
The cross section $\sigma_\gamma (E_{\gamma})$ 
can be calculated by the formula~\cite{RingSchuck}
\begin{equation}
  \sigma_\gamma (E_{\gamma})=\frac{4\pi^2}{\hbar c}E_{\gamma}\frac{1}{3}S(E_{\gamma}),
  \label{photo-abs.eq}
\end{equation}
where $S(E)$ is the strength function for the $E1$ transition
with the $E1$ operator $\sum_i (\bm{\xi}_i)_\mu\frac{1}{2}(1+\tau_{3i})$
where $\bm{\xi}_i = \bm{r}_i - \bm{x}_N$, and
$\bm{x}_N$ denotes the center-of-mass coordinate
of the $N$-nucleon system.

\begin{figure}[ht]
  \begin{center}
  \includegraphics[width=8.5cm]{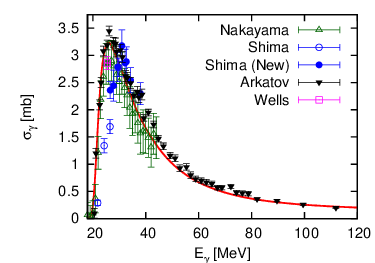}
\end{center}
  \caption{Photoabsorption cross sections of $^{4}$He obtained with the
    AV8$^\prime$+3NF interaction.  The figure is drawn based on the data of
    Ref.~\cite{Horiuchi12}. The experimental data are taken from Refs.~\cite{Arkatov79, Wells92, Shima05, Nakayama07, Shima.new}.}
  \label{E1.fig}
\end{figure}

Figure~\ref{E1.fig} compares the theoretical and experimental photoabsorption cross
sections $\sigma_{\gamma}(E_{\gamma})$.
The calculation predicts a sharp rise of the cross section 
from the threshold, which is observed by several 
measurements~\cite{Nakayama07, Arkatov79} but not in the 
data of Ref.~\cite{Shima05}.
Our result satisfies almost 100\% of
the non-energy-weighted sum rule (NEWSR),
and this is also consistent with the cross sections obtained by
the Lorentz Integral Transform calculations~\cite{Gazit06,Quaglioni07},
especially in the cross section near the threshold.
The low-lying photoabsorption cross sections are 
mostly understood by
the excitation of the $3N+N$ relative motion.
In fact, the $3N+N$ contribution dominates
in the low-lying $E1$ strength~\cite{Horiuchi12}.
Thus, it is hard to understand the low-lying behavior of Ref.~\cite{Shima05},
though all the data are consistent above 30 MeV.

\subsubsection{Spin-dipole excitations}

\begin{figure}[ht]
\begin{center}
  \includegraphics[width=8.5cm]{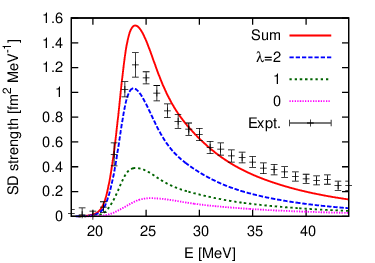}
  \caption{Spin-dipole strength functions
    of $^{4}$He to $^{4}$H charge-exchange process with
    the AV8$^\prime$+3NF interaction.
    The figure is drawn based on the data of
    Ref.~\cite{Horiuchi13b}.
    The `experimental data'
    are taken from Ref.~\cite{Nakayama07} for reference. See text for a detail.
  }
  \label{SD.fig}
\end{center}
\end{figure}

We have confirmed the 
reliability and potential predictive power of our approach.
It is interesting to apply it to the SD response of $^{4}$He because the 
relevant operators are closely related to those
of the neutrino-nucleus reaction.
Figure~\ref{SD.fig} exhibits the SD strength
functions of IV$-$
type, $\mathcal{O}_{\lambda\mu}^{{\rm IV}-}=\sum_i 
\left[\bm{\xi}_i\times\bm{\sigma}_i\right]_{\lambda\mu}\tau^-_i$, which excites
the ground state of $^4$He to the excited states of $^4$H.
The NEWSR for the SD operators are fully satisfied,
and it is a very interesting observable that
can reveal the role of the tensor force in the ground state~\cite{Horiuchi13b}.
The peak positions well correspond to the observed excitation 
energies of the three negative-parity states of $^4$H~\cite{Tilley92}.
The ratio of the strengths for $J^\pi=0^-$ ,$1^-$, and $2^-$
is roughly 1:3:5 following their multipolarity $2J+1$
but the ratio is actually modified to approximately 1:2:4 due to the
tensor force~\cite{Horiuchi13b}.
We can also estimate the decay width of the resonance by taking
the difference of two excitation energies at which
the strength becomes half of the maximum strength at the peak.
The agreement between theory and experiment 
is very satisfactory.
The strength functions can also be compared with the spin-flip cross sections
of the $^4$He($^7$Li,$^7$Be$\gamma$) measurement~\cite{Nakayama07}.
Since the absolute value of the SD component
was not determined experimentally,
the experimental distribution is normalized to the sum of the
theoretical strength for $\lambda=0,1,2$ integrated from 18 to 44 MeV
where the experimental data are available.
The comparison between the theory and experiment is qualitative, but
the experimentally observed peak apparently agrees
with the calculated one and it is dominated by
the $J^\pi T=2^-1$ state of $^4$H.

\subsubsection{Neutrino-$^4$He reactions}

The typical temperature of the core-collapse supernova is around 10 MeV
and the energy of neutrino is rather low $E_\nu < 100$ MeV. 
In this energy region below the  pion production threshold,
the neutrino (anti-neutrino)-$^4$He CC and NC reactions lead 
to the following continuum states:
\begin{eqnarray}
\nu_e + {\rm ^4He} &\rightarrow& e^- + 3p + n, \ e^- + 2p + d,\  e^- + p + {\rm ^3He} \\
\bar{\nu}_e + {\rm ^4He} &\rightarrow& e^+ + p + 3n, \ e^+ + 2n + d,\ e^+ + n + {\rm ^3H}\\
\nu + {\rm ^4He} &\rightarrow& \nu + 2p + 2n, \ \nu + p + n + d, \  \nu + n + {\rm ^3He}, \nonumber \\
   & &  \nu + p + {\rm ^3H}, \ \nu + 2d
\end{eqnarray}
The inclusive neutrino-nucleus cross section including
 all continuum final states can be studied  using the
strength functions from the {\it ab initio} approach.

The transition matrix elements for the low-energy neutrino-$^4$He reaction
would be dominated by the first-forbidden transitions leading to the negative 
parity states. The allowed Gamow-Teller
and Fermi transitions are expected to be very weak because of the
doubly-closed shell
structure of $^4$He. In this work, we consider following one nucleon 
vector and axial vector currents,
\begin{eqnarray}
V_{IA,0}^{\alpha}(\bm{X}) & = & 
\sum_i f_V \tau_i^{\alpha V}\delta(\bm{X}-\bm{\xi}_i), \label{cc-time-v}\\
\bm{A}_{IA}^{\alpha}(\bm{X}) & = & 
\sum_i f_A \bm{\sigma}_i \tau_i^{\alpha A}\delta(\bm{X}-\bm{\xi}_i).
\end{eqnarray}
Here the isospin operators are
$\tau_i^{\alpha V}=\tau_i^{\alpha A}= \tau_i^{\pm}$ for CC
$\nu$ and $\bar{\nu}$ reactions, and 
$\tau_i^{\alpha V}=(1 - 2 \sin^2\theta_W)\tau^3/2 - 2\sin^2\theta_W$, $\tau_i^{\alpha A}=\tau_i^3/2$
for NC reactions. Here $\theta_W$ is Weinberg angle.
The one-nucleon operators for the first-forbidden transition in the long wavelength 
approximation are given as
\begin{eqnarray}
\mathcal{O}_{JM} =\sum_i f_A [\bm{\xi}_i \otimes \bm{\sigma}_i]^{J}_M \tau^{\alpha A}_i
\end{eqnarray}
for the axial vector current and
\begin{eqnarray}
\mathcal{O}_{1M} =\sum_i f_V (\bm{\xi}_i)_M \tau^{\alpha V}_i
\end{eqnarray}
for the vector current.  It is noticed that 
only $T=1$ states are excited because the isoscalar dipole operator is reduced
to the center of mass coordinate.

In the following we focus on the cross section formula of CC reactions.
The inclusive cross sections for $\nu/\bar{\nu}$-$^4$He CC reactions in the low-energy region
are given with the following strength functions $S_x(p,J,\omega)$:
\begin{align}
S_{sd}(p,J,\omega) & =  \mathcal{S}_{f}
|\left<\Psi_f\right\|\sum_j[\bm{\xi}_j\otimes \bm{\sigma}_j]_{(J)}\tau^{\pm}_j\left\|\Psi_{{\rm ^4He}}\right>|^2 
\delta(E_f - E_0 - \omega), \\
S_{d}(p,1,\omega) & =  \mathcal{S}_{f} 
| \left<\Psi_f\right\|\sum_j\bm{\xi}_j\tau^{\pm}_j\left\|\Psi_{{\rm ^4He}}\right>|^2 
\delta(E_f - E_0 - \omega), \\
S_{sd-d}(p,1,\omega) & =  \mathcal{S}_{f}
\left<\Psi_f\right\|\sum_j[\bm{\xi}_j\otimes \bm{\sigma}_j]_{(1)}\tau^{\pm}_j
\left\|\Psi_{{\rm ^4He}}\right>
\left<\Psi_f\right\|\sum_{j'}\bm{\xi}_{j'}\tau^{\pm}_{j'}
\left\|\Psi_{{\rm ^4He}}\right>^* \nonumber \\
 &  \times \delta(E_f - E_0 - \omega),
\end{align}
where $\omega = E_\nu - E_e$.
 The last term $S_{sd-d}(p,1,\omega)$ is the interference
term of the vector and axial vector currents.
Here $p=$IV$\pm$ for CC neutrino and anti-neutrino reactions.
Using the  standard multipole expansion formula of neutrino reactions
in Sec.~\ref{sec:mult-exp},
the structure functions $W_i$ are given in terms of the strength functions as
\begin{eqnarray}
  W_2 & = & V_{ud}^2 \left(W_L + \frac{Q^2}{2|\bm{q}|^2} W_T \right), \\
2 W_1 & = & V_{ud}^2 \, W_T,
\end{eqnarray}
where
\begin{eqnarray} 
 W_T 
     & = & f_A^2 |\bm{q}|^2 \left( \frac{1}{3}S_{sd}(p,1,\omega) + \frac{1}{5}S_{sd}(p,2,\omega)\right)
      + f_V^2 \frac{2\omega^2}{3}S_{d}(p,1,\omega), \\
 W_L 
     & = & f_A^2 \omega^2 \left( \frac{1}{3}S_{sd}(p,0,\omega) + \frac{2}{15}S_{sd}(p,2,\omega)\right)
      + f_V^2 \frac{Q^4}{3|\bm{q}|^2}S_{d}(p,1,\omega),\\
\frac{W_3}{M_T} 
              & = & \frac{2\sqrt{2}}{3}f_Af_V \omega S_{sd-d}^1(p,1,\omega).
\end{eqnarray}
Similar expressions can be obtained for NC reactions.

\begin{figure}[htb]
\includegraphics[width=12cm]{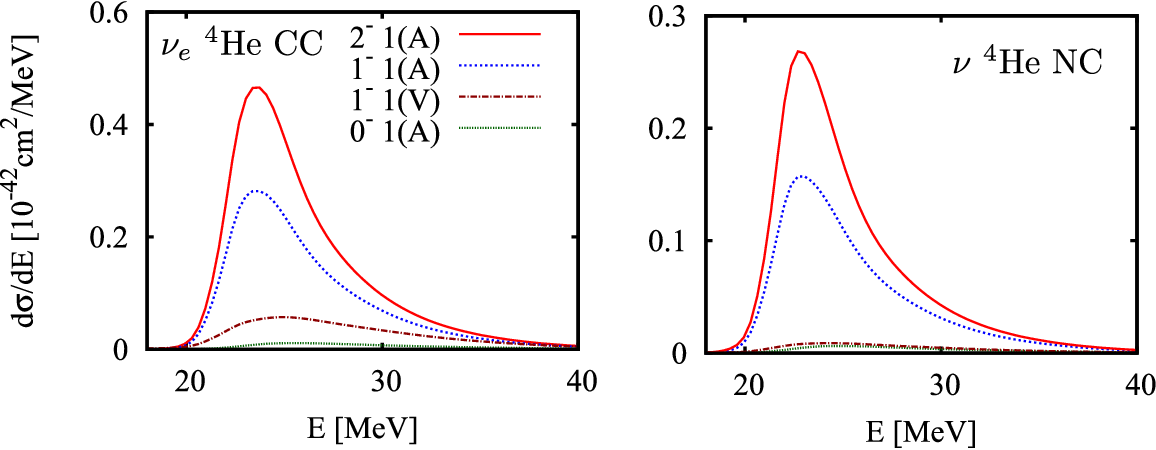}
\caption{The differential cross sections $d\sigma/dE$ of  neutrino-$^4$He CC(left) and NC(right)
reactions at $E_\nu=50$MeV.
Contributions of axial vector spin-dipole operators
  as a function of nuclear excitation energy $E$ are shown in solid
 (red) ($J^\pi T= 2^-1(A)$),
 long-dashed (blue) ($1^-1(A)$) and short-dashed (green) curves ($0^-1(A)$).
The contribution of dipole operator of vector current is shown in
 dash-dotted (brown) ($1^-1(V)$) curve.}
\label{fig:dsde4he}
\end{figure}

\begin{figure}[htb]
\includegraphics[width=15cm]{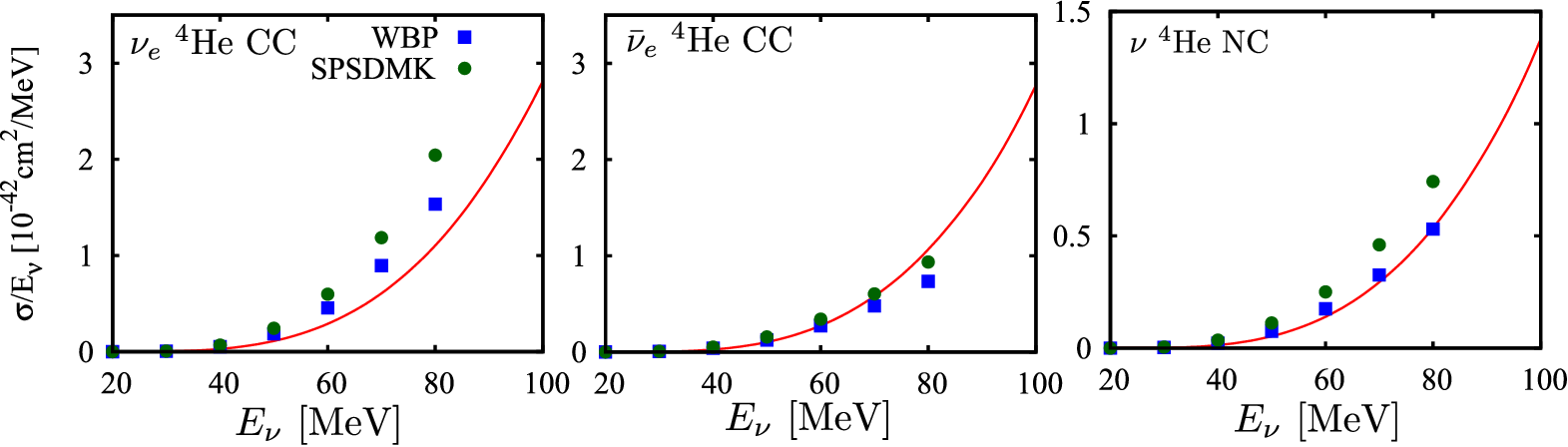}
\caption{The total cross section of CC reactions
 ($\nu_e+{\rm ^4He}\rightarrow e^-+X$ (left panel) and 
$\bar{\nu}_e+{\rm ^4He}\rightarrow e^++X$ (middle panel))
and NC reaction ($\nu+{\rm ^4He}\rightarrow \nu + X$ (right panel)).
 The solid curves show the result of this work and are compared with the
shell model calculation~\cite{Yoshida2008} with the
WBP(green circle) and SPSDMK(blue squares) shell model interactions.}
\label{fig:crossection}
\end{figure}

The cross sections $d\sigma/dE$ of  neutrino-$^4$He CC (left) and NC (right)
reactions as a function of excitation energy ($E$) at $E_\nu=50$ MeV are shown 
in Fig.~\ref{fig:dsde4he}.
Here the cross sections are obtained by using   the strength functions
calculated in Ref.~\cite{Horiuchi13b}. 
As shown in Fig.~\ref{fig:dsde4he}, the contributions of spin-dipole operators 
of $J^\pi T=$ $2^-1$ and $1^-1$ 
give main strength of the neutrino reactions.
The contribution of $0^-1$ states is small for both CC and NC reactions, while
the dipole operator gives non-negligible contribution for the CC reaction.
The energy dependence of the total cross sections for CC~$\nu_e-^4$He,  CC~$\bar{\nu}_e-^4$He 
and NC~$\nu_e-^4$He reactions are shown in Fig.~\ref{fig:crossection}. 
The cross section of the {\it ab initio} calculation is shown in solid curve.
For comparison, results of shell-model calculation~\cite{Yoshida2008}
with the WBP~\cite{WBP}(green circle) and SPSDMK~\cite{SPSDMK}(blue square)
shell model interactions are also shown in Fig.~\ref{fig:crossection}.
The {\it ab initio} calculation of NC reaction 
agrees with the shell model calculation of SPSDMK,
while for CC reaction, the $S_{sd-d}$ term,
which has not been included in our current calculation, 
may give a sizable contribution because of non-negligible 
contribution of the dipole operator;
the $S_{sd-d}$ term is 
the $V$-$A$ interference term in which
the matrix elements 
of the vector current and the axial vector current interfere.

Before predicting the temperature average cross section for the 
simulation of supernova explosion, 
we have to include $V$-$A$ interference term for $1^-1$ final states,
meson-exchange current for the axial vector current and
possible contribution of the recoil order term to the 
time component of the axial vector current
$A_0 \sim \bm{p}\cdot \bm{\sigma}/m_N$, which has been studied
for the nuclear muon capture reaction~\cite{Koshigiri79}.
Taking into account those effects, the {\it ab initio}
study of the neutrino reaction is of great interest to help 
to clarify the role of light nuclei for the heating mechanism of 
the core-collapse supernova.

\clearpage
\section{Quasi-Elastic Interactions}

As stressed in the Introduction, the next-generation neutrino
oscillation experiments that aim at extracting the 
neutrino-mass hierarchy and leptonic CP violation would require a quantitative understanding of the 
neutrino-nucleus interactions at the level of a few \% accuracy or
better. In the neutrino energy region from 0.2 to 2.0 GeV, 
the charged-current quasi-elastic (CCQE) process gives the largest contributions
than the other reaction mechanisms induced by neutrinos.
In neutrino oscillation experiments utilizing neutrinos in this energy
region, the neutrino energy is often reconstructed using a formula based
on the QE kinematics with the initial nucleon at rest as follows:
\begin{equation}\label{eq:recE}
E_\nu^\textrm{Rec}= \frac{2E_{l'}\tilde m_N-(m_{l'}^2+\tilde m_N^2 - m_N^2)}
{2 ( \tilde m_N - E_{l'} + p_{l'}\cos\theta_{l'} )},
\end{equation}
where $\tilde m_N=m_N -\epsilon$ with $\epsilon$ being the separation energy;
$E_{l'}=\sqrt{p_{l'}^2+m_{l'}^2}$.
With this formula, the neutrino energy is reconstructed with 
the final lepton kinematics ($p_{l'}$ and $\theta_{l'}$) measured in the
experiments.
However, the relation of Eq.~(\ref{eq:recE}) could be considerably modified by the
Fermi motion of the initial nucleon and the final state interaction (FSI)
of the outgoing nucleon.
Therefore, a reliable modeling of the CCQE process is of particular
importance to extract  precise neutrino flux from the data.

In this section, we describe
the inclusive lepton-nucleus reactions
in the 'impulse approximation (IA) scheme'~\cite{QE1,QE2,QE3}.
The IA scheme is basically the plane wave impulse approximation, where
a target nucleus can be seen as a collection of individual nucleons
by an electroweak probe with the large spatial momentum $\bm q$ that has 
a spatial resolution of  $\sim1/|\bm{q}|$ sufficiently finer than a typical 
inter nucleon distance in nuclei,
and the struck nucleon and the $(A-1)$ spectator nucleons can be
treated as independent systems.
In the IA scheme, nuclear response is described in terms of 
the nuclear spectral function (SF)~\cite{QE1,QE2,QE3}. 
The following two subsections are devoted to discuss 
the cross section of the inclusive reaction in the IA scheme,
and a presentation of the nuclear SF.
The interaction between the struck nucleon and the
remaining $(A-1)$ nucleons, i.e., FSI, 
is taken into account in the convolution formula explained in the
next subsection.
Then final subsection follows to present numerical results where 
the IA scheme with FSI is confronted with precise 
electron scattering data.

\subsection{Impulse approximation and cross section formula}

Within the IA scheme,
cross sections for the QE lepton-nucleus scattering process can be given
by an incoherent sum of
the cross sections for the individual nucleons as
\begin{eqnarray}
\frac{d\sigma^{\rm IA}_{l A}}{dE_{l'} d\Omega_{l'}} &=& 
\int d^3 p\, dE\, P_{\rm h}(\bm p,E)
\frac{m_N}{E_p}\left[Z\frac{d\sigma _{l p}}{dE_{l'} d\Omega_{l'}} 
+ (A-Z)\frac{d\sigma _{l n}}{dE_{l'} d\Omega_{l'}}\right]
\nonumber \\
&&\times P_{\rm p} (\bm p + \bm q, q^0 -E - t_{A-1}) \ , 
\label{eq:dsig_qe}
\end{eqnarray}
where 
$t_{A-1}$ is the recoil energy of the residual nucleus
and $d\sigma_{l N}/dE_{l'} d\Omega_{l'}$ ($N=p,n$)
is the elementary cross section 
stripped off the energy-conserving $\delta$-function
(see Ref.~\cite{QE1,QE3} for detail);
the elementary cross section is given by the single nucleon matrix
elements presented in Eqs.~(\ref{eq:vec})-(\ref{eq:vec_is}).
The hole SF is denoted by 
$P_{\rm h}(\bm p,E)$, and is discussed in detail in the next subsection.
The particle SF denoted by $P_{\rm p} (\bm{p}^{\prime}, T')$
describes the propagation of the struck nucleon that carries 
the momentum $\bm{p}^{\prime}$ and the kinetic energy $T'$.
We assumed that the spectral functions are the same for protons and
neutrons in Eq.~(\ref{eq:dsig_qe}).

For the particle SF, we use the following two approximations.
The simplest option is to account for 
Pauli blocking  using 
the Heaviside step function,  as in the Fermi gas model, as
\begin{eqnarray}
\label{eq:oldPB}
P^\theta_\textrm{p}(\bm{p}^{\prime},\mathcal{T'})=\delta(E_{\bm{p}^{\prime}}-m_N-\mathcal{T'})\big[1-\theta(\overline p_F-|\bm{p}^{\prime}|)\big],
\end{eqnarray}
where $\overline p_F=211$ MeV has been determined from the 
local density approximation (LDA) average,
\begin{eqnarray}
\overline p_F=\int d^3r \rho(r) p_F(r),
\label{eq:LDA_ave}
\end{eqnarray}
with $p_F(r)=(3\pi^2A\rho(r)/2)^{1/3}$. 
In another option, the particle SF is based on 
the LDA treatment of Ref.~\cite{ref:shape}, 
and is calculated from 
the momentum distribution ($n^\textrm{NM}_\rho(\bm{p}^{\prime})$)
of isospin-symmetric nuclear matter at uniform density $\rho$ as
\begin{eqnarray}
\label{eq:LDAPB}
P^\textrm{LDA}_\textrm{p}(\bm{p}^{\prime},\mathcal{T'})&=&\delta(E_{\bm{p}^{\prime}}-m_N-\mathcal{T'})\nonumber\\
&&\times\left[1-\int d^3r\rho(r)C_\rho n^\textrm{NM}_\rho(\bm{p}^{\prime})\right],
\end{eqnarray}
where $C_\rho=4\pi p^3_F(r)/3$;
$C_\rho n^\textrm{NM}_\rho(\bm{p}^{\prime})$ corresponds to
$\theta\big(p_F(r)-|\bm{p}^{\prime}|\big)$ in the local Fermi gas
model~\cite{Moniz}.

\subsection{Nuclear spectral function}

\begin{figure}
\includegraphics[width=0.50\textwidth]{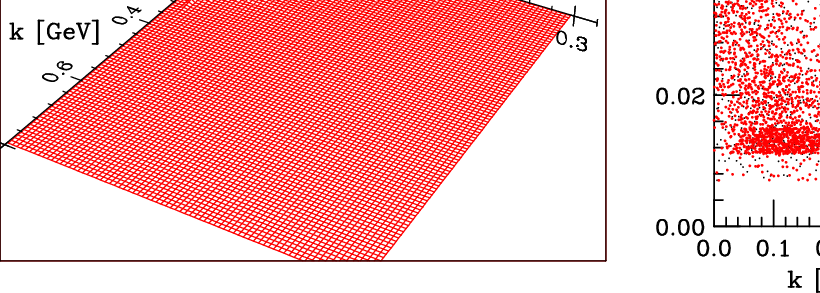}
\includegraphics[width=0.50\textwidth]{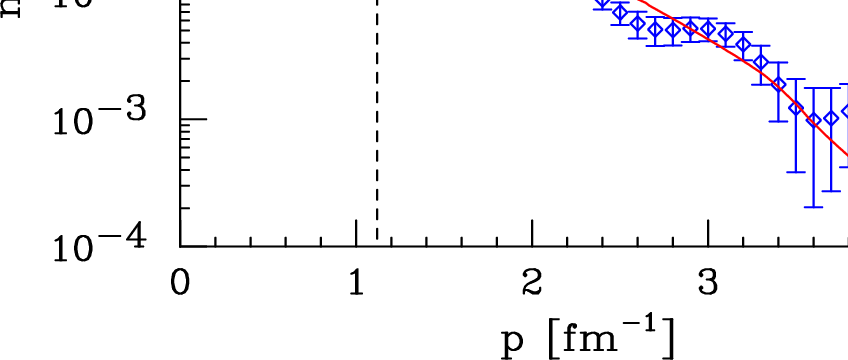}
\caption{\label{fig:sp_16O}(color online).
(Left) 3D plot of the spectral function for the $^{16}$O ground state
based on the LDA approximation.
(Middle) 
Scatter plot of the function shown in the left panel.
(Right)
The momentum distribution of nucleons
in the $^{16}$O ground state.
Solid line: LDA approximation.
Dashed line: FG model with Fermi momentum $p_F$ = 221~MeV. 
Diamonds: Monte Carlo calculation based on 
the wave function of Ref.~\cite{pieper}.
Figures taken from Ref.~\cite{QE1}. 
Copyright (2005) APS.}
\end{figure}

The nuclear (hole) SF $P(\bm p,E)$ (subscript 'h' is omitted for
simplicity in what follows) is the probability of
removing a nucleon of the momentum $\bm p$
from a ground-state nucleus with an excitation energy $E$ as defined by
\begin{eqnarray}
P(\bm p,E) &=& \sum_{\cal R} |\langle 0|{\cal R},-\bm p; N, \bm p \rangle |^2
\delta (E - m_N +E_0 -E_{\cal R}) \ ,
\end{eqnarray}
where $|0\rangle$ and $E_0$ are respectively the state vector
for the ground state of the target nucleus
and its energy eigenvalue, while 
$|{\cal R},-\bm p \rangle$ and $E_{\cal R}$ are an $(A-1)$-body state
vector with the CM momentum $-\bm p$ and its energy eigenvalue, respectively;
$|N,\bm p \rangle$ is a single nucleon state vector with the momentum
$\bm p$.
In the following calculation, we use 
the spectral function calculated with the LDA~\cite{LDA}.
The LDA consists of: 
(i) the mean field contribution
that is based on the experimental
information obtained from $(e,e'p)$ measurements;
(ii) the $NN$ correlation of a uniform nuclear matter.
The LDA-based spectral function for $^{16}$O is shown in
Fig.~\ref{fig:sp_16O} (left and middle panels).
The nuclear matter results of Ref.~\cite{LDA} and the Saclay
$(e,e'p)$ data~\cite{saclay} are encoded in the spectral function.
The mean-field contribution
amounts to $\sim$~80~\% while 
the remaining $\sim$~20~\% are from the correlation.
The large $p$ 
($p\gg p_F$; 
the momentum $p$ is denoted by $k$ in
Fig.~\ref{fig:sp_16O} (left, middle))
and large $E$ ($E\gg e_F$) components of the spectral function are highly
correlated.
It is clear that a relativistic
Fermi Gas model (RFG) assuming a uniform momentum distribution up to the
Fermi momentum ($p_F$) in a uniform potential gives
an inadequate description of the spectral function.
It would be also informative to present 
the nucleon momentum distribution defined by 
\begin{eqnarray}
\label{eq:np}
n(\bm p) &=&
\int dE\, P(\bm p,E)
\\ 
\label{eq:np2}
&=& 
\langle 0 |  a^\dagger_{\bm p}\, a_{\bm p}
| 0 \rangle \ ,
\end{eqnarray}
where $a^\dagger_{\bm p}$ ($a_{\bm p}$)
denotes the creation (annihilation) operator
of a nucleon with the momentum $\bm p$.
The nucleon momentum distribution (labelled as LDA) shown
in Fig.~\ref{fig:sp_16O} (right) is obtained from the LDA-based spectral function
shown in Fig.~\ref{fig:sp_16O} (left, middle) by using Eq.~(\ref{eq:np}).
This is compared to the one
calculated with a
variational Monte Carlo calculation of Ref.~\cite{pieper}.
Clearly, the LDA-based $n(\bm p)$ is in good agreement with that based on 
the Monte Carlo calculation.
We also show in Fig.~\ref{fig:sp_16O} (right) $n(\bm p)$ from the FG model
corresponding to Fermi momentum $p_F$ = 221~MeV for a comparison.
It is clear that the FG model gives a very different distribution.

After the NuInt12 Workshop~\cite{nuint12}, all the major 
neutrino experimental groups use the spectral function in calculating both QE
and pion production cross sections, 
rather than a simple Fermi-Gas Model~\cite{Moniz}.

\subsection{Final state interaction (FSI)}

The cross section formula based on the IA scheme,
Eq.~(\ref{eq:dsig_qe}), needs to be modified by taking account of the FSI.
For this purpose, we employ the convolution scheme~\cite{conv} where the
IA cross section (Eq.~(\ref{eq:dsig_qe})) is integrated with a folding
function as 
\begin{eqnarray}
\label{eq:xsec_FSI}
\frac{d\sigma^{\rm FSI}_{lA}}{d\omega d\Omega_{l'} }= 
\int d\omega'f_{\bm{q}}(\omega-\omega')\frac{d\sigma^{\rm IA}_{lA}}{d\omega'
d\Omega_{l'}}\ ,
\end{eqnarray}
where $\omega$ is the energy transfer from the lepton, and 
$f_{\bm{q}}(\omega)$ is the folding function through which the FSI
effect is introduced. 
The folding function can be given by 
\begin{eqnarray}
\label{eq:FF}
f_{\bm{q}}(\omega)=\delta(\omega)\sqrt{T_A}+\big(1-\sqrt{T_A}\big)F_{\bm{q}}(\omega),
\end{eqnarray}
where $T_A$ denotes the nuclear transparency, and 
$F_{\bm{q}}(\omega)$ is a finite-width function.
In the limit of full nuclear transparency, $T_A\to 1$, the IA cross
section is recovered in Eq.~(\ref{eq:xsec_FSI}).

Now let us connect the above convolution scheme for describing the FSI
with an optical potential scattering.
The FSI for the QE process can be described with an optical potential 
$U = U_V\ +\ iU_W$, originally proposed by Horikawa et al.~\cite{ref:Horikawa} in the
context of $(e,e')$ processes. 
The real part of the potential, $U_V$, 
modifies the energy spectrum of the final-state nucleon, thereby
shifting the cross section by $\omega\to \omega - U_V$.
In the convolution scheme,
this effect of $U_V$ can be accounted for by modifying the folding
function as
\begin{eqnarray}
\label{eq:rOP}
f_{\bm{q}}(\omega-\omega')\to f_{\bm{q}}(\omega-\omega'-U_V).
\end{eqnarray}
Meanwhile, the imaginary part, $U_W$, 
re-distributes the transition strength of some of one-particle
one-hole final states to more complex final states, leading to the 
quenching of the QE peak and the associated enhancement of its tails. 
In the convolution scheme,
this effect of $U_W$ can be taken care of by using
\begin{eqnarray}
\label{eq:fq}
F_{\bm{q}}(\omega)=\frac{1}{\pi}\frac{U_W}{U_W^2+\omega^2} \ ,
\end{eqnarray}
in Eq.~(\ref{eq:FF}).
The remaining piece is the nuclear transparency $T_A$ in
Eq.~(\ref{eq:FF}).
In the calculation shown below, we use experimentally determined
nuclear transparency of $^{12}$C reported in Ref.~\cite{ref:Rohe}.
We also neglect the $|\bm q|$-dependence of 
$F_{\bm{q}}(\omega)$ in Eq.~(\ref{eq:FF}) and evaluate it at 
$|\bm q|$ = 1~GeV.

\begin{figure}
\begin{center}
\includegraphics[width=0.4\textwidth]{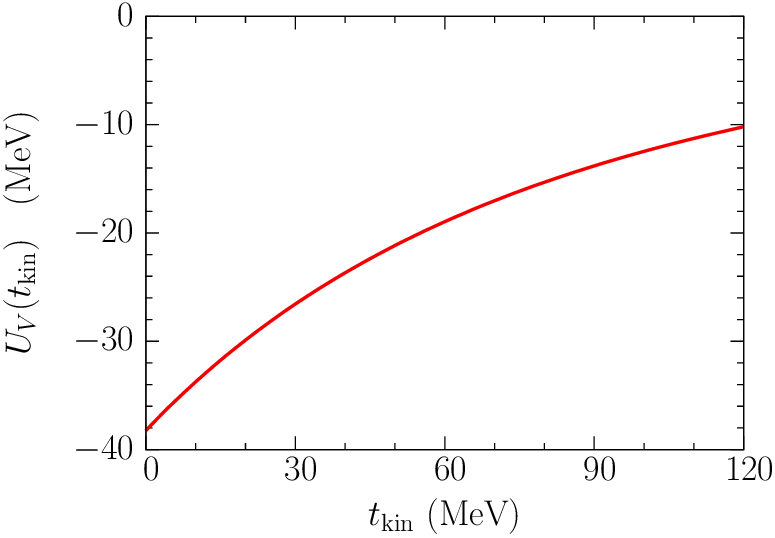}
\end{center}
\caption{\label{fig:opt-ptl}(color online).
Real part of the proton-$^{12}$C optical potential
as a function of kinetic energy of the proton.
The potential is related to 
the Dirac phenomenological fit of Cooper
 \etal{}~\cite{Cooper} through Eqs.~(\ref{eq:sv}) and (\ref{eq:opt-def}).
Figures taken from Ref.~\cite{QE3}. 
Copyright (2015) APS.}
\end{figure}

Having seen the relation between 
the optical potential $U$ and the folding function
$f_{\bm{q}}(\omega)$,
we now employ an optical potential that has been tested by data, and
then plugged it into the folding function following the relation.
With the folding function, the differential cross sections including the
FSI effects are obtained from Eq.~(\ref{eq:xsec_FSI}).
We choose the proton-$^{12}$C optical potential 
due to Cooper et al.~\cite{Cooper} based on the Dirac phenomenology. 
In this framework, the optical potential consists of the scalar ($S$)
and vector ($V$) parts appearing in the Dirac equation.
Their dependence on the nucleon kinetic energy ($t_{\rm kin}$) and the
radial coordinate $r$ has determined by fitting the data in the range of 
$29\leq t_{\rm kin}\leq1040$~MeV.
The scalar and vector potentials are related to the total energy of the
proton 
$E'_\textrm{tot}=E'_\textrm{tot}(t_{\rm kin},r)$ by 
\begin{equation}
\label{eq:sv}
E'_\textrm{tot}(t_{\rm kin},r)=\sqrt{(m_N+S(t_{\rm kin},r))^2+
p^{\prime 2}}+V(t_{\rm kin},r),
\end{equation}
with $p^{\prime}$ being the nucleon's momentum and related to 
$t_{\rm kin}$ by
$t_{\rm kin}=p^{\prime\, 2}/2m_N$.
Here, as seen in Eqs.~(\ref{eq:rOP}) and (\ref{eq:fq}),
the optical potential $U$ is treated as a $r$-independent 
quantity.
Therefore, $U$ is defined as an 
average shift of the nucleon energy from the on-shell energy
 $E_{p'}=\sqrt{m_N^2+p^{\prime 2}}$,
\begin{equation}
\label{eq:opt-def}
U = \int  d^3r\rho(r) E'_\textrm{tot} - E_{p'}\ ,
\end{equation}
which is calculated from the optical potential,
$S$ and $V$, of the Dirac phenomenology.
Regarding the density distribution of $^{12}$C,
we extract it from the measured
charge density~\cite{ref:densityDistributions}
following the procedure in Ref.~\cite{ref:densityUnfolding}.
The real part of the optical potential ($U_V$) is presented in 
Fig.~\ref{fig:opt-ptl}.
In the low $t_{kin}$ region, $U_V$ is large and negative 
as seen in the figure while 
the imaginary part ($U_W$) is small.
Reference~\cite{QE3} 
and numerical results in the next subsection are
mostly concerned with this energy region, and
the above-described procedure was used in taking account of the FSI.
On the other hand, in higher energy region ($>$ 200~MeV),
$U_V$ becomes negligible and $U_W$ dominates. 
In this energy region, a different prescription 
based on a generalization of Glauber theory for including the FSI
effects was taken in Ref.~\cite{QE1}.

Since the nucleon kinematics is integrated over in Eq.~\eqref{eq:dsig_qe},
$t_{\rm kin}$ for $E'_{\rm tot}$ and $T_A$ in using Eq.~\eqref{eq:xsec_FSI}
is determined from the lepton kinematics
\begin{eqnarray}
\label{eq:t}
t_{\rm kin}=\frac{E_l^2(1 - \cos\theta_{l'})}{ m_N + E_l(1 - \cos\theta_{l'})},
\end{eqnarray}
where $E_l$ and $\theta_{l'}$ denote the incident lepton energy and the
lepton scattering angle, respectively.
This relation corresponds to scattering of a massless particle on a nucleon at rest.

\subsection{Numerical results, comparison with electron scattering data}

\begin{figure}
\begin{center}
\includegraphics[width=0.9\textwidth]{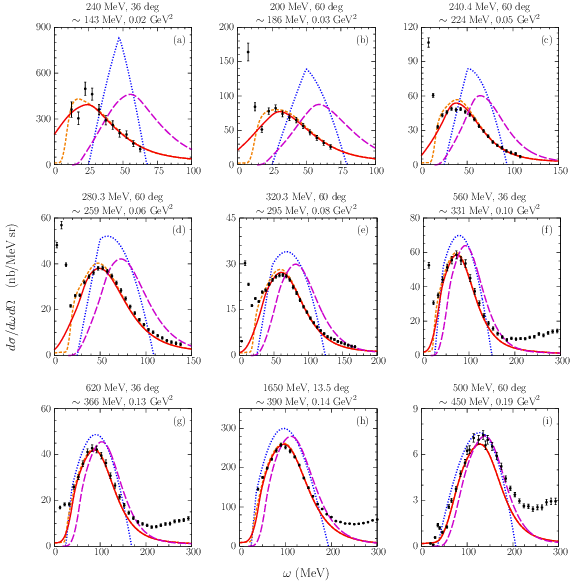}
\end{center}
\caption{\label{fig:e-12C}(color online).
Differential cross sections, $d\sigma/d\omega d\Omega_{e'}$, for
electron-$^{12}$C scattering; $\omega$ is the energy transfer to the nucleus.
The results obtained with Pauli blocking accounted for in the local-density (solid lines) 
and step-function (short-dashed lines) approximations 
are compared to the experimental data from
(a)-(g) Barreau et al.~\cite{Barreau}, 
(h) Baran et al.~\cite{Baran}, 
and (i) Whitney et al.~\cite{Whitney}.
The IA (long-dashed lines) and 
RFG calculations (dotted lines) are also shown for reference. 
The panels are labeled according to beam energy, scattering angle, 
and values of $|\bm q|$ and $Q^2$ at the quasi-elastic peak.
Figures taken from Ref.~\cite{QE3}. 
Copyright (2015) APS.
}
\end{figure}
Comparison of the calculations with all existing inclusive $^{12}{\rm C}(e,e')$
data in the energy region from $E_e$ = 0.2 to 2.0 GeV is shown in
Fig.~\ref{fig:e-12C}~\cite{QE3}.
These figures display
double differential cross sections
as a function of the energy transfer to the nucleus $\omega$
in the increasing order of the momentum transfer $|\bm{q}|$. 
The low $|\bm{q}|$ data show the
contributions from discrete nuclear excitations including a giant 
dipole resonance and the elastic electron scattering
in the small $\omega$ region, which
are absent in the model. As $|\bm{q}|$ increases,
the data start to cover the dip region 
between the QE peak and $\Delta(1232)$ peak
in the large $\omega$ region.

The results obtained with the RFG model 
are shown by the dotted curves in Fig.~\ref{fig:e-12C}.
Parameters of the RFG model,
Fermi momentum 221~MeV and separation energy $\epsilon=25$ MeV,
were determined in Ref.~\cite{Whitney} to fit the data shown
in Fig.~\ref{fig:e-12C}(i).
Finding of the analysis is that even though the RFG model fits the data 
at this kinematics, the model predictions start to deviate from the data 
around the QE peak as $|\bm q|$ decreases.
Within the RFG model, it is not possible to find a parameter set with which the
model can fairly reproduce the $(e,e')$ data in the broad kinematical
region of interest for neutrino oscillation studies.

The results of the IA scheme without FSI are shown
by the long-dashed curves while
those including FSI and with 
the particle SF of Eq.~(\ref{eq:oldPB}) [Eq.~(\ref{eq:LDAPB})]
are shown by the short-dashed [solid] curves in Fig.~\ref{fig:e-12C}.
Large contributions of FSI bring the IA results to fairly agreement with the data,
especially around the QE peak from low to high $|\bm{q}|$ region.
The effects of FSI are
the redistribution of the strength from the peak to the
tails due to the imaginary part of the optical potential
and the shift of the cross section toward lower energy
transfers due to the real part of the optical potential.
The shift of the cross section is 
getting smaller for higher $|\bm q|$
(larger $t_{\rm kin}$), as expected from Fig.~\ref{fig:opt-ptl}.
The two prescriptions of the Pauli blocking,
Eqs.~(\ref{eq:oldPB}) and (\ref{eq:LDAPB}),
give very similar results near the QE peak. In the low $|\bm{q}|$ region,
where Pauli blocking plays an important role,
the LDA prescription of Eq. (\ref{eq:LDAPB}) may be slightly favored.

It is noticed that, at higher $|\bm{q}|$ and $\omega$ region,
the results tend to underestimate the data beyond the QE peak, as in
Figs.~\ref{fig:e-12C}(d)-(i).
In this dip region, 
two-particle--two-hole (2p2h) final states induced by
two-nucleon correlations and two-nucleon currents
as well as the other inelastic processes (e.g., pion production) would contribute to fill
the gap~\cite{alberico1984,Gil1997,Nieves2016}.
Although initial (final) nucleon correlations have been taken into account in
Ref.~\cite{QE3} through the nuclear spectral function (FSI), more
elaborate treatment of FSI and inclusion of two-nucleon currents are
needed.
Analysis in this direction done with the spectral function formalism
is reported in~\cite{Rocco2016}.
Also, recent works on 2p2h mechanisms and effects are reviewed in Ref.~\cite{Luis2014,KM_review}.

\clearpage
\section{Neutrino-nucleus reactions in the RES region}

Recently, 
neutrino-nucleus scattering in the RES region, single pion
productions in particular~\cite{AguilarArevalo:2010bm,Eberly:2014mra,miniboone_pion,minerva_pion}, 
has been actively studied experimentally for
better understanding the process and for developing a better model to
be implemented in a generator. 
A microscopic description of this process is as follows.
In the initial state, 
nucleons are bound in a nucleus under a certain energy and momentum
distribution. 
The distribution is given by a Fermi gas model in the simplest treatment, and
a more realistic distribution is given by a spectral function based on a
shell model or an {\it ab initio} calculation~\cite{LDA}.
Then an electroweak current interacts with one of the nucleons to
produce a meson and the recoiled baryon.
This process is described by elementary amplitudes such as those
discussed in Sec.~\ref{sec:dcc}.
The meson and baryon then propagate in the nucleus.
In the course of the propagation, they can change their momenta and
charges by interacting with the surrounding nucleons. 
The meson can even be absorbed by the nucleus, and a few
nucleons are kicked out from the nucleus at the same time. 
These final state interactions (FSI) can be described, in principle, with a
multiple scattering theory as those formulated, e.g., in Ref.~\cite{KMT}.
More practically manageable formulation is the
so-called $\Delta$-hole
model~\cite{annal99,annal108,annal120,taniguchi},
more details of which will be
discussed later in Sec.~\ref{sec:coh}.
However, these quantum mechanical calculations become a formidable task
when dealing with a typical situation of the neutrino experiments where 
a few nucleons (and one or a few pions)
are emitted from the nucleus.
Because of the complexity of the problem, 
the most common treatment of 
the FSI
has been to use a (semi)classical hadron transport model 
so far~\cite{gibuu,gibuu_miniboone,spain-cascade}.

In the rest of this section, 
we discuss our own works on two particular cases that are relatively
straightforward to deal with.
One of them is neutrino-induced single pion productions off the
deuteron.
The old bubble chamber data are available for these
processes~\cite{Kitagaki:1986ct,anl}, and
it has been well recognized that the data are valuable information from
which the axial form factors associated
with the $N$-$\Delta(1232)$ transition can be extracted.
However, in the previous analyses, the nuclear effects, the FSI
in particular, have not been taken into account in
extracting the axial form factors. 
The deuteron is the simplest nucleus and thus the nuclear effects
can be
taken into account by explicitly dealing with interactions among
all of mesons and baryons that appear in the process; no need to
introduce a mean field nor phenomenological many-body effects. 
We will discuss our formalism and show significant nuclear effects in the
following subsection.
Another subject to be discussed in the subsequent subsection is coherent
pion productions in the neutrino-nucleus scattering. 
In the coherent processes where the nucleus in the final state stays in its ground
state, the FSI can be described with multiple
iterations of an optical potential between the pion and the nucleus in
the ground state. This is the situation where 
the $\Delta$-hole model is the most suitably applied.
It is also timely to study the coherent process
because this process has been studied experimentally 
recently~\cite{Hasegawa:2005td,Hiraide:2008eu,Higuera:2014azj,miniboone_coh,kurimoto,t2k_coh}, and thus
we can study how the $\Delta$-propagation is modified in nuclei
by confronting calculations with the data.

\subsection{Neutrino induced pion production reaction on deuteron}
\label{sec:deuteron}

The neutrino-induced pion production reaction through $\Delta(1232)$
resonance plays a dominant
role in the RES region as discussed in Sec.~\ref{sec:dcc}. 
This process is interesting from hadron physics point of view because we can study
the axial vector $AN\Delta$ transition form factor,
$C_5^A$ [Eq.~(\ref{eq:del_ffs})].
One can relate the $\pi N\Delta$ coupling constant with  the $AN\Delta$ 
form factor assuming the PCAC, which however needs to be examined against cross section data.
Information of the axial vector form factor can be obtained from a combined analysis of
pion electroproductions and neutrino induced pion production reactions.
Currently  available data of the pion production reactions on the nucleon in the
RES region are bubble chamber data from ANL~\cite{Radecky:1981fn,anl},
BNL~\cite{Kitagaki:1986ct} and BEBC~\cite{Allasia:1990uy,Allen1986}, which are
mainly data for neutrino-deuteron reactions.
Consider the charged-current (CC) single pion production reactions on the deuteron:
$\nu_\mu + d \rightarrow \mu^- + \pi^+ + p + n$ and $\nu_\mu + d \rightarrow \mu^- + \pi^0 + p + p$.
The pion production cross sections for three channels
\begin{eqnarray}
 \nu_\mu + p & \rightarrow &\mu^- + \pi^+ + p \label{eq:ccpip-1}\ ,\\
 \nu_\mu + n & \rightarrow &\mu^- + \pi^+ + n \label{eq:ccpip-2}\ ,\\
 \nu_\mu + n & \rightarrow &\mu^- + \pi^0 + p  \label{eq:ccpi0}\ ,
\end{eqnarray}
have been extracted by assuming 'quasi-free' pion production mechanism with the spectator nucleon. 
The first channel Eq.~(\ref{eq:ccpip-1}) is purely isospin $3/2$ reaction and
is dominated by the $\Delta$ excitation. Especially the data for restricted
$\pi N$ invariant mass region, $W<1.4$GeV, are useful to study the $\Delta$ resonance excitation.
It has been recognized that there are some tension between the data of ANL and 
BNL~\cite{Paschos2004,Wascko2006}. Recently, it has been demonstrated
the ratios between the pion production process and the non-pion production quasi-elastic process
of ANL and BNL are consistent with each other, which raises the questions on the
normalization of the neutrino flux~\cite{reanalysis,Graczyk2009,Rodrigues2016}.

\begin{figure}[htbp] \vspace{-0.cm}
\begin{center}
\includegraphics[width=0.8\columnwidth]{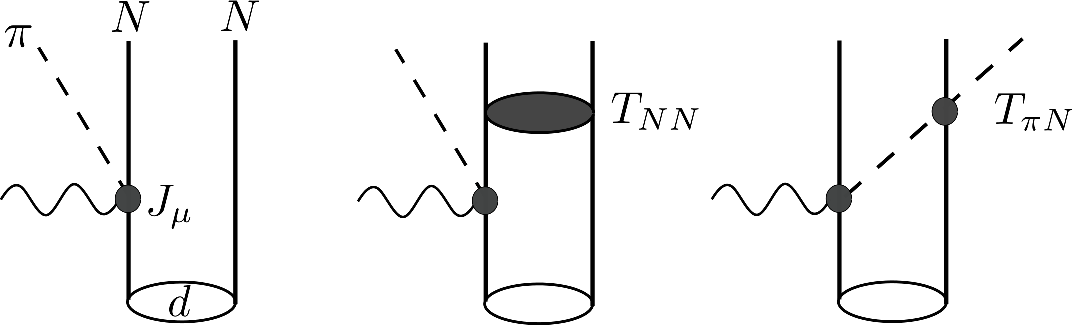}
\caption{
The impulse (left), $NN$ rescattering (middle) and $\pi N$ (right) rescattering mechanisms
of Eq.~(\ref{eq:resc}).
}
\label{fig:resc}
\end{center}
\end{figure}

The cross section data for the three channels in Eqs.~(\ref{eq:ccpip-1})-(\ref{eq:ccpi0})
will be modified by 
additional reaction mechanisms such as FSI of $\pi NN$ system, 
where a simple interpretation with a 'quasi-free' reaction mechanism does not work. 
In fact, it is known the FSI among pion and nucleons
are important in pion photoproductions on the
deuteron~\cite{Darwish2003,Fix2005,Levchuk2006,Schwamb2010}.
The FSI for the $\nu d$ reactions
has been studied only recently by Wu et al.~\cite{wsl}.
Here we briefly show the role of FSI following~\cite{wsl}.
Within the first order correction of the multiple scattering
theory~\cite{KMT}, the transition amplitude of the neutrino-induced pion production
can be written as a sum of impulse,
nucleon-nucleon ($NN$) rescattering and pion-nucleon ($\pi N$) rescattering mechanisms,
\begin{eqnarray}
T^\mu(d \rightarrow \pi N N) & = & <\pi N N |(1 + (T_{NN} + T_{\pi N})G_0 ) J^\mu|d>.
\label{eq:resc}
\end{eqnarray}
Here $J^\mu$ is a pion production weak current $N + J^\mu \rightarrow \pi + N$.
The rescattering correction is taken into account by the second and the third terms
of Eq.~(\ref{eq:resc}) using
 the nucleon-nucleon ($T_{NN}$), the pion-nucleon ($T_{\pi N}$)
scattering t-matrix and the $\pi NN$ Green's function ($G_0$) 
as shown in Fig.~\ref{fig:resc}.

Here we adopt a dynamical model developed in Refs.~\cite{sul,msl,sl,sl-1} (called the SL model). 
to describe electroweak single pion production on the nucleon in the $\Delta$(1232) resonance region.
The SL model has been well tested~\cite{sl,sl-1} against the data of $\pi N$ scattering and
electromagnetic pion production reactions on the nucleon in the $\Delta$(1232) resonance region. 
It also describes~\cite{sul} well the cross sections of neutrino-induced single 
pion productions on the proton and neutron.
We used  the  high precision Bonn nucleon-nucleon potential~\cite{bonn}
to describe the deuteron and NN scattering state.

\begin{figure}[htbp] \vspace{-0.cm}
\begin{center}
\includegraphics[width=0.45\columnwidth]{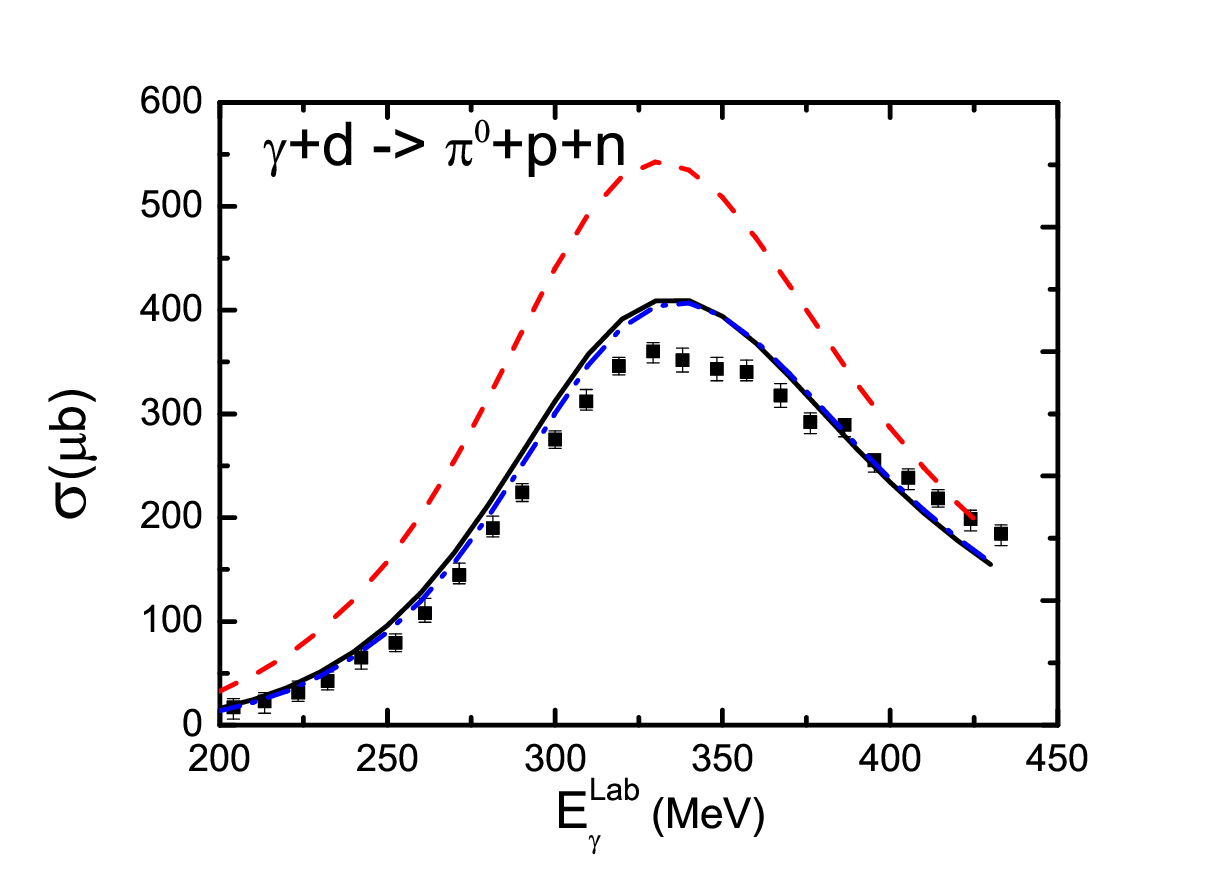}
\includegraphics[width=0.40\columnwidth]{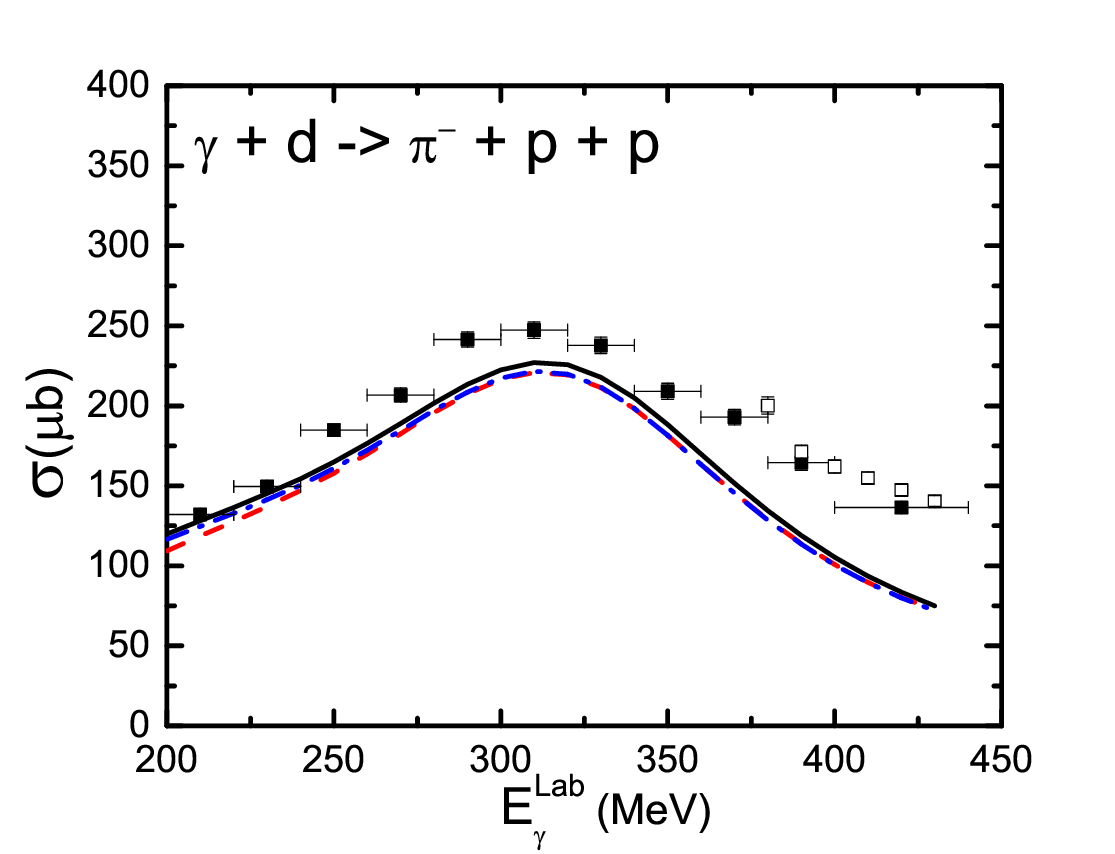}
\caption{(Color online) The total cross sections of
 $\gamma + d \to \pi^0 + n + p$ (Left) and $\gamma + d \to \pi^- + p + p$ (Right).
The red dashed, blue dash-dotted, and black solid
curves represent the impulse term , the impulse + 
($NN$ FSI),
and the impulse + ($NN$ + $\pi N$ FSI), respectively.
This figure is taken from Ref.~\cite{wsl}.}
\end{center}
\label{fig:pitot}
\end{figure}

At first, we show that the current model describes well 
the available data  of incoherent pion photo-production reactions on the deuteron.
The results for the total cross sections of
$\gamma+ d \rightarrow \pi^0+n+ p$ and $\gamma + d \to \pi^- + p + p$ 
as a function of the photon energy are shown in 
Fig.~\ref{fig:pitot}.
For the $\gamma+ d \rightarrow \pi^0+n+ p$ reaction,
the cross section is greatly reduced when the
$np$ FSI ($T_{NN}$ term) is added (dot-dashed curve)
to the impulse term (dashed curve).
The $\pi N$ FSI ($T_{\pi N}$ term)
is also included in our full calculation (solid curve), which is a small effect.
A similar comparison for the total cross sections
of $\gamma+ d \rightarrow \pi^- +p+p$ is shown.
Here both the $pp$ and the $\pi N$ FSI are weak.
Clearly, the $np$ re-scattering effects are very large  for the $\pi^0$ production reaction,
while effects of FSI is small for $\pi^-$ production reaction.
Our full calculations are in reasonable
agreement with the data in both $\pi^0$ and $\pi^+$ production reactions,
while some improvements are still needed
in the future.  The result on the pion photoproductions shows that our calculational
procedure is valid for predicting 
the $\nu+ d \rightarrow \mu +\pi+ N+ N$ cross sections.

\begin{figure}[htbp] \vspace{-0.cm}
\begin{center}
\includegraphics[width=0.45\columnwidth]{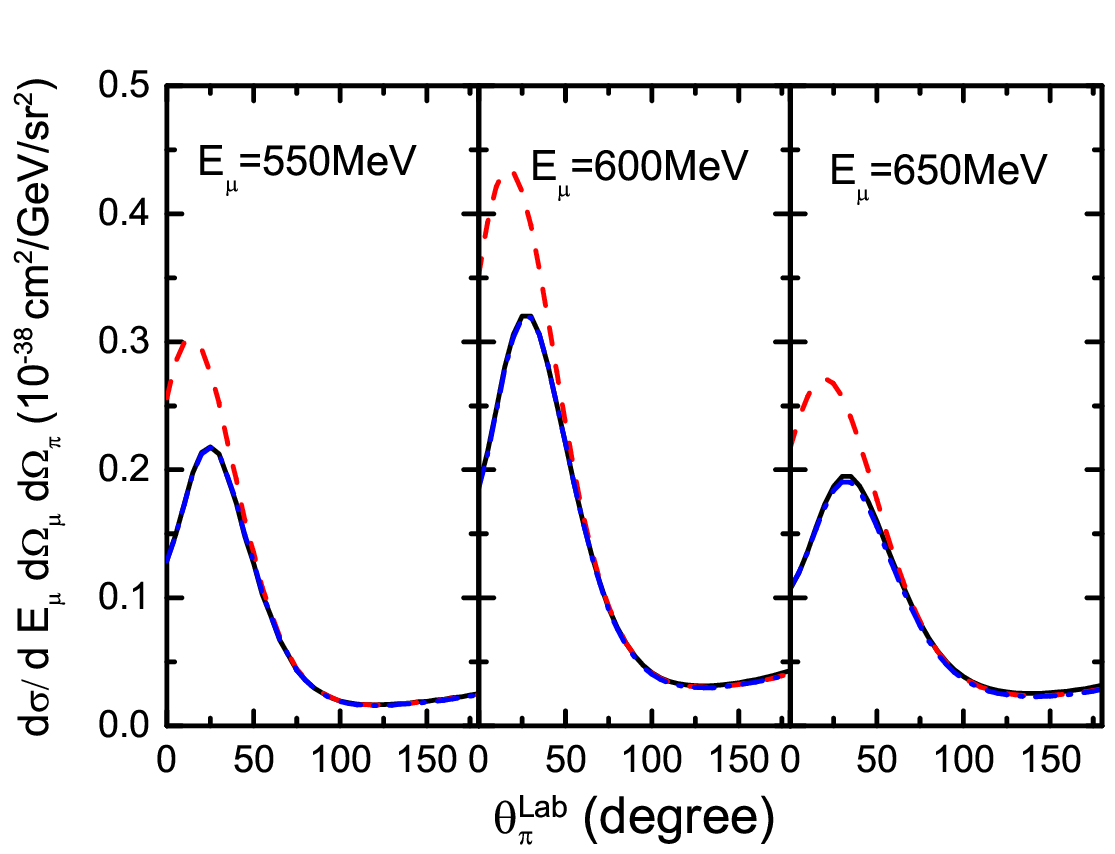}
\includegraphics[width=0.45\columnwidth]{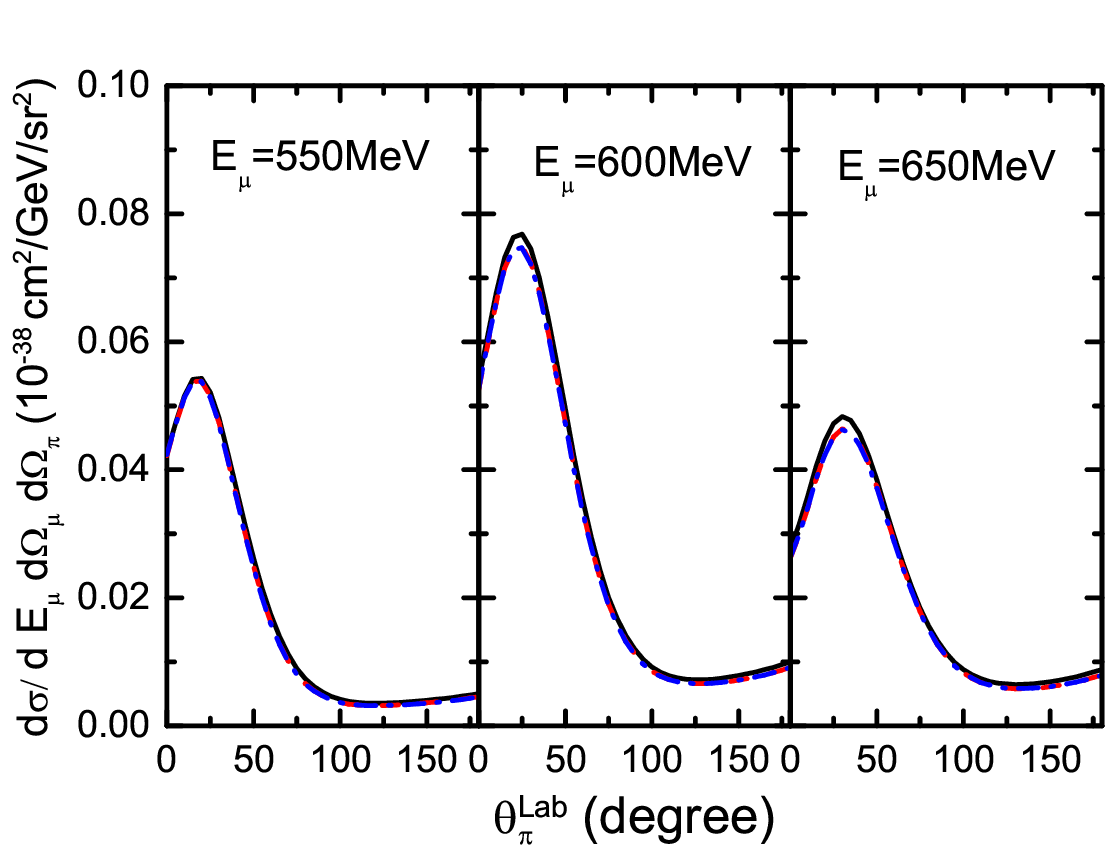}
\caption{The differential cross sections 
$d\sigma / dE_{\mu^-} d\Omega_{\mu^-} d\Omega_{\pi}$ of $\nu_\mu + d \to \mu^- + \pi^+ + p + n$ (left)
and $\nu_\mu + d \to \mu^- + \pi^0 + p + p$ (right) as function of
 $\theta_\pi^{LAB}$ in the laboratory system.
See caption of Fig.~\ref{fig:pitot}. The figure is taken from Ref.~\cite{wsl}.}
\label{fig:nu}
\end{center}
\end{figure}
Within the same calculational procedure as the pion photoproduction, we have studied the
neutrino-induced pion productions on the deuteron:
 $\nu_\mu + d \rightarrow \mu^- +\pi^+ + n +p$ and
 $\nu_\mu + d \rightarrow \mu^- +\pi^0 + p +p$.
The neutrino energy is chosen as $E_{\nu_\mu}= 1$GeV
with the angle between $\nu_\mu$ and $\mu^-$ set at $\theta_{\mu^-}= 25^\circ$
and $E_\mu=550,600,650$ MeV. This kinematics is chosen to simulate the 'quasi-free'
$\Delta$ production process, where maximum values of the predicted cross sections are expected.
The predicted pion angular distributions
 $d\sigma / dE_{\mu^-} d\Omega_{\mu^-} d\Omega_{\pi}$ are shown in  Fig.~\ref{fig:nu}.
For $\nu_\mu + d \rightarrow \mu^- +\pi^+ + n +p$ reaction,
we have included contribution of pion production amplitudes from both proton and neutron.
The $NN$ FSI is large at forward pion
for $\nu_\mu + d \rightarrow \mu^- +\pi^+ + n +p$ reaction, while 
it is small for the  $\nu_\mu + d \rightarrow \mu^- +\pi^0 + p +p$ reaction.
The $\pi N$ FSI is small for both channels.
The situation here is similar to what we have observed in the pion photoproductions that 
the FSI effects from the $np$ scattering are much larger than that 
from the $pp$ scattering. It may be understood by the orthogonality relation
between bound state and scattering state of the $^3S_1$-$^3D_1$ partial wave, 
which affects overlap integrals for low momentum transfer reactions.

In conclusion, the results strongly suggest that
the spectator approximation used to extract the pion production cross sections 
on the nucleon from the data on the deuteron is not valid 
for the $\nu + d \rightarrow \mu^-+ \pi^+  +n+p$, but is a
good approximation for $\nu + d \rightarrow \mu^-+ \pi^0  +p+p$.
It will be important to extend this analysis to cover the whole kinematical region and
examine the FSI effects on the total cross sections for the neutrino-induced pion
production reactions on the deuteron. 
It will be also important to apply our approach to investigating the neutrino-deuteron reactions
in the higher $W$ region where the higher mass nucleon resonances play important roles.
Such  an investigation can be performed with the coupled-channels model
discussed in Sec.~\ref{sec:dcc}.


\subsection{Coherent pion productions}
\label{sec:coh}

Two different theoretical approaches have been taken to study
the neutrino-induced coherent pion production.
One of them is to use a model based on the PCAC 
relation~\cite{Rein:1982pf,RS2_coh,hernandez,paschos2,bs_pcac,paschos3}.
Because of the nuclear form factor, the coherent process is 
strongly suppressed when $Q^2$ departs from zero.
In this situation, the amplitude is dominated by the divergence of the
axial current that is related to the pion-nucleus elastic scattering
amplitude through the PCAC relation. 
Another approach is to use a dynamical microscopic model~\cite{kelkar,singh,valencia1_coh,valencia2_coh,amaro,martini,sxn,hernandez2}.
The model consists of 
ingredients such as elementary amplitudes, a nuclear form
factor, and an optical potential for pion-nucleus elastic scattering. 
The nuclear effect on the $\Delta$ propagation also needs to be under control.
Our model discussed below is classified into the latter approach.
We do not go into details of each of the models, and refer the readers to
a compact summary of the theoretical status given in Ref.~\cite{coh_summary}.

Now we discuss our dynamical model for the coherent pion productions.
It is based on the Sato-Lee (SL) model~\cite{sul,msl} combined with a
$\Delta$-hole model. 
The SL model is a prototype of the DCC model discussed in
Sec.~\ref{sec:dcc}; it has only the $\pi N$ channel coupled to the
$\Delta(1232)$ and is designed to work well in the $\Delta(1232)$ region.
The $\Delta$-hole model accounts for the nuclear effects.
Here we give a brief explanation of the $\Delta$-hole model for the
elastic pion-nucleus scattering;
for a full account, consult Refs.~\cite{annal99,annal108,annal120,taniguchi}.

In this formulation, the nuclear Fock space is divided
into the nuclear ground state and a
pion ($P_0$), one-particle one-hole and a pion ($P_1$),
one-$\Delta$ one-hole ($D$), and all the other ($Q$);
the symbols in the parentheses are the projection operators onto the
corresponding nuclear Fock space, and thus $P_0+P_1+D+Q=1$.
With the projection operator, we write a projected Hamiltonian as, for example,
$H_{P_0D}=P_0HD$.
Applying the projection operator method to 
Schr\"odinger equation in the full space, 
we can derive a Schr\"odinger equation
defined in the subspace $P_0$ that describes the pion-nucleus
elastic scattering. 
In the $\Delta$-hole model, we assume that the $D$-space is the doorway
from the $P=P_0+P_1$ space to the $Q$ space, i.e., 
$H_{PQ}=H_{QP}=0$.
Then we can write
the pion-nucleus scattering amplitude that includes the
$\Delta$ excitation as
\begin{eqnarray}
\label{eq:elastic_amp}
 T_{P_0P_0}(E) = H_{P_0D} G_{\Delta h}(E) H_{DP_0} \ ,
\end{eqnarray} 
where the total energy in the center-of-mass frame is denoted by $E$.
The quantity $G_{\Delta h}$ in Eq.~(\ref{eq:elastic_amp}) is
 $\Delta$-hole propagator that is more explicitly written by
\begin{eqnarray}
\label{eq:nuclear-delta-prop}
G_{\Delta h}^{-1} = 
D(E-H_\Delta)-W_{\rm el} - \Sigma_{\rm Pauli} - \Sigma_{\rm spr} \ .
\end{eqnarray}
We denoted the propagator of $\Delta$ in the vacuum 
with the invariant mass $W$ by $D(W)$,
and write more explicitly as
\begin{eqnarray}
\label{eq:delta-prop}
D(W) = W - m_\Delta^0 - \Sigma_\Delta(W) \ ,
\end{eqnarray}
where 
$m_\Delta^0$ and $\Sigma_\Delta$ are the bare mass and
self energy of the $\Delta$-resonance, respectively.
In Eq.~(\ref{eq:nuclear-delta-prop}),
$H_\Delta$ is the Hamiltonian for the $\Delta$-particle in the nucleus.
The effects associated with
complicated configurations belonging to the $Q$-space such as multi-particle
multi-hole states are very difficult to deal with microscopically, and
thus the common practice is to 
squash them into the phenomenological $\Delta$
spreading potential [$\Sigma_{\rm spr}$ in Eq.~(\ref{eq:nuclear-delta-prop})]
that includes adjustable parameters to fit
pion-nucleus scattering data.
The model takes account of couplings to the $P_1$-space through
the $\Delta$ self energy $\Sigma_\Delta(W)$ included in $D(E-H_\Delta)$ 
[see Eq.~(\ref{eq:delta-prop})]
with a correction $\Sigma_{\rm Pauli}$ associated with the Pauli blocking.
Couplings to the $P_0$-space denoted by $W_{\rm el}$ in
$G_{\Delta h}$ describe the elastic rescattering.
A calculation of the pion-nucleus elastic amplitude in
Eq.~(\ref{eq:elastic_amp}) needs a diagonalization of $G_{\Delta h}$
which is numerically rather involved.
Here, we employ a simplified treatment of $G_{\Delta h}$ 
proposed by Karaoglu and Moniz~\cite{karaoglu},
based on the local density approximation.
A resonant amplitude for coherent pion production induced by an
electroweak current can be obtained from Eq.~(\ref{eq:elastic_amp})
by replacing $H_{DP_0}$ with $H_{DP_0'}$
where $P_0'$ is the space spanned by
the external electroweak current and the nucleus in the ground state.
We emphasize here that 
our framework treats the medium effect on the $\Delta$
propagation in the pion-nucleus scattering 
and also in electroweak pion productions on the same footing. 
Because of this consistency,
we can make a parameter-free prediction for electroweak coherent pion
production cross sections, once we determine
parameters associated with the medium effects, i.e., those included in
$\Sigma_{\rm spr}$, by
fitting the pion-nucleus scattering data.
Our numerical analysis will also be done in this ordering. 
We refer the readers to Ref.~\cite{sxn}
to find more explicit and detailed formulas for the quantities in 
Eqs.~(\ref{eq:elastic_amp})-(\ref{eq:delta-prop}) and
for coherent pion production amplitudes that are actually implemented in
numerical calculations.

\begin{figure}
\begin{center}
 \includegraphics[width=80mm]{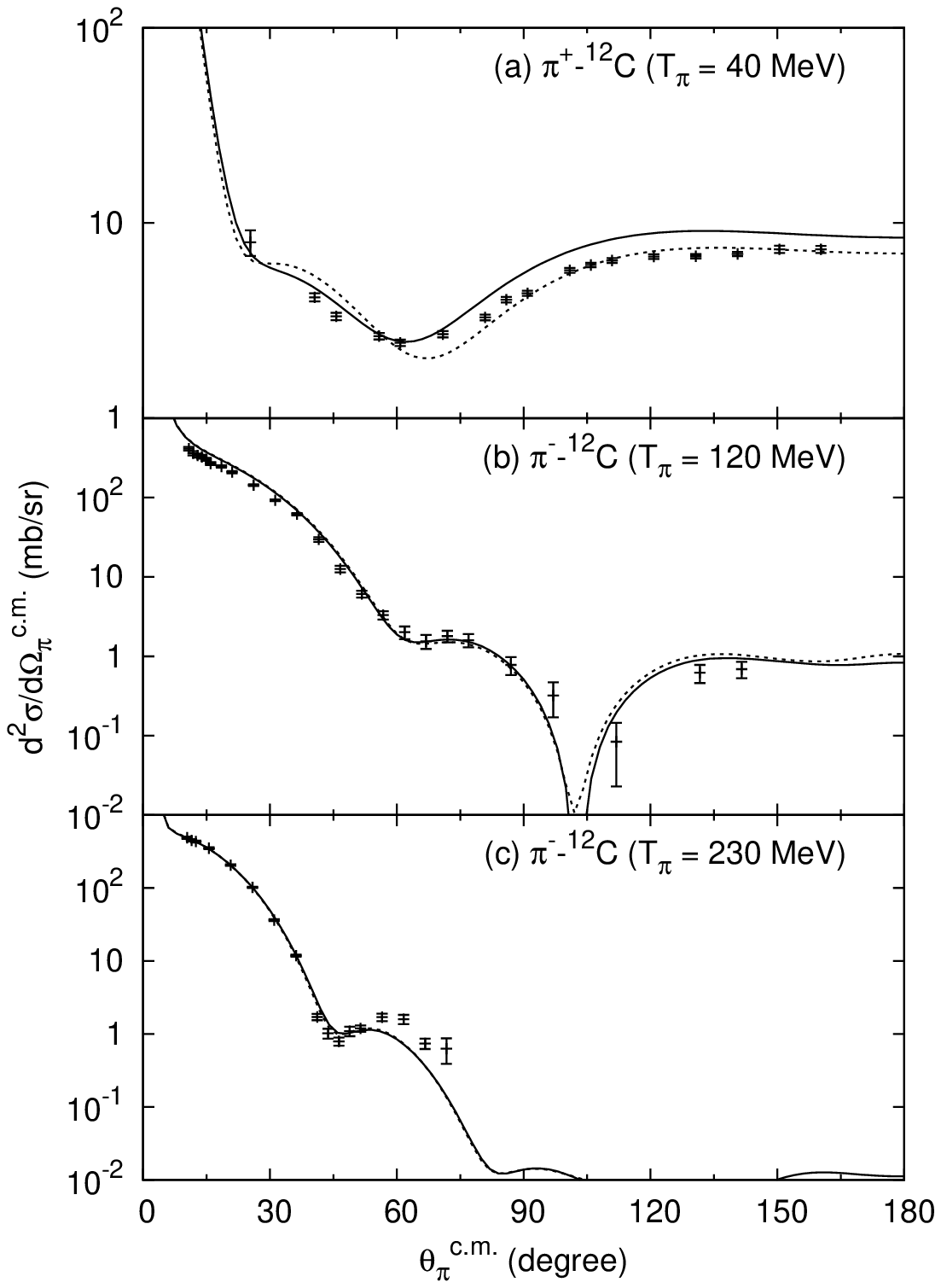}
\end{center}
\caption{\label{fig_elastic} 
$\pi$-$^{12}$C elastic differential cross sections.
The full calculations are shown by 
the solid curves while the results obtained without
the phenomenological $\rho^2$-term
are shown by the dashed curves.
The data are from Ref.~\cite{pi-nucleus2} for (a) and
Ref.~\cite{pi-nucleus1} for (b) and (c).
Figures taken from Ref.~\cite{sxn}. 
Copyright (2010) APS.
}
\end{figure}
From now we present our numerical results. 
The first thing to do is to analyze pion-nucleus scattering data. 
The pion-nucleus optical potential derived from the SL combined with the
$\Delta$-hole model is the main driving force for this process.
We have two complex adjustable parameters in $\Sigma_{\rm spr}$.
In addition, we also consider a phenomenological term that are
proportional to the square of the nuclear density; this term simulates
the pion absorption by two nucleons through non-$\Delta$ mechanism.
Then we have additional two complex adjustable parameters.
After all,
we fit the data for both elastic and total cross sections in and
around the $\Delta$ region by adjusting totally eight free parameters.
The quality of the fit can be seen in Fig.~\ref{fig_elastic} where our
calculations for the $\pi$-${}^{12}$C elastic differential cross
sections are compared with the data. 
The agreement between our calculations and the data are reasonably
good. 
We note that a good agreement is also obtained between our calculation
and pion-nucleus total cross section data. 

\begin{figure}[t]
\begin{center}
 \includegraphics[width=77mm]{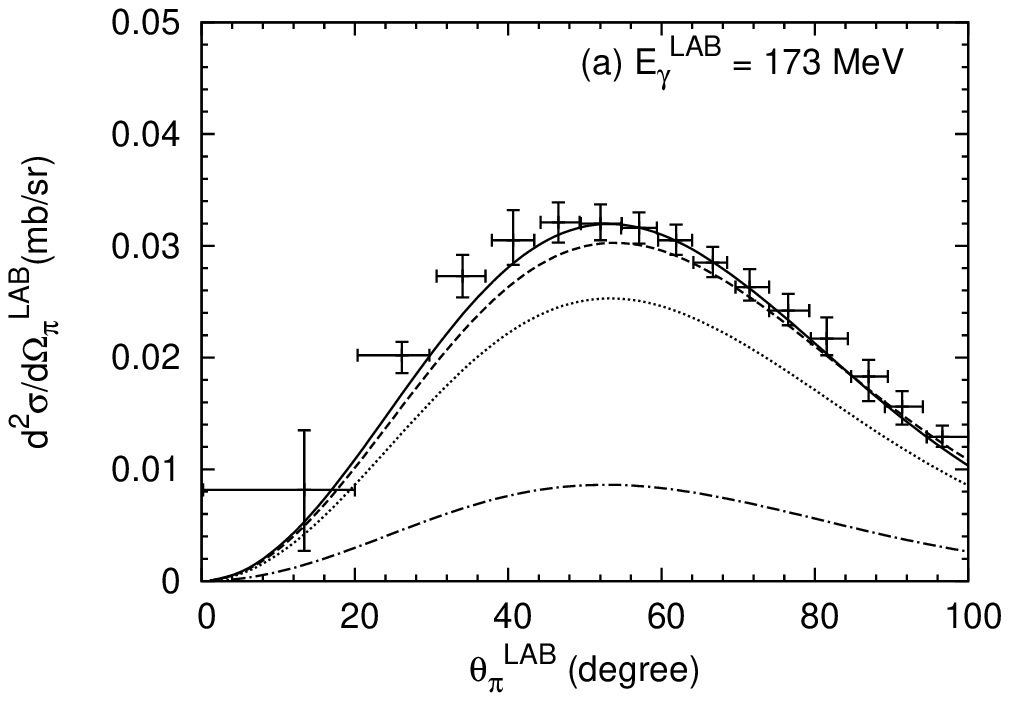}
 \includegraphics[width=77mm]{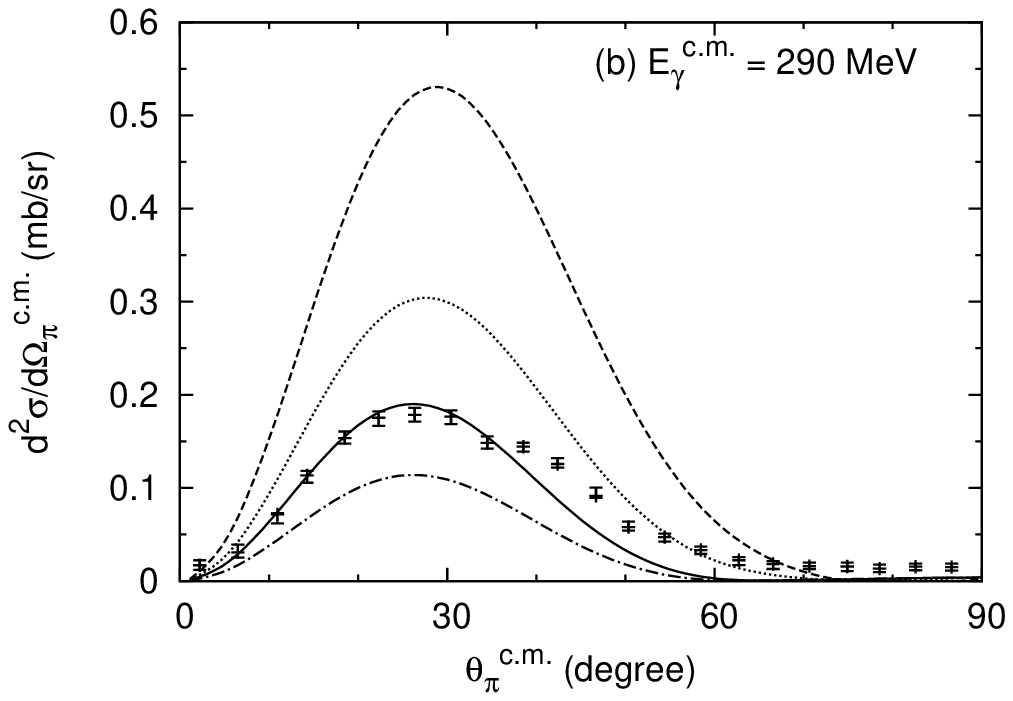}
\caption{\label{fig_krusche}
Differential cross sections for 
$\gamma + {}^{12}{\rm C}_{g.s.} \to \pi^0 + {}^{12}{\rm C}_{g.s.}$.
The incident photon energy is $E_{\gamma}^{\rm LAB}$=173 MeV for (a) and
$E_{\gamma}^{\rm CM}$=290 MeV for (b). 
The solid curves are from the full calculations while
the dashed curves are obtained without the FSI and without
the medium effects on the $\Delta$-propagation.
The dotted curves are obtained with the medium effects 
on the $\Delta$ included.
By considering only the $\Delta$ mechanism in 
the pion production operator, we obtained
the dash-dotted curves.
The data are from Ref.~\cite{gothe} for (a) and 
from Ref.~\cite{krusche} for (b).
Figures taken from Ref.~\cite{sxn}. 
Copyright (2010) APS.
}
\end{center}
\end{figure}
With the parameters fixed in the above analysis,
we can then make a parameter-free prediction for coherent pion productions.
First, we calculate photon-induced coherent pion production for which
data with a good quality are available. 
The photoproduction is a good testing ground to examine the reliability
of our model.
In Fig.~\ref{fig_krusche},
we show a comparison of 
our numerical results for the differential cross sections for
$\gamma + {}^{12}{\rm C}_{g.s.} \to \pi^0 + {}^{12}{\rm C}_{g.s.}$
with data~\cite{gothe,krusche}. 
Different curves in each panel include different dynamical contents. 
The dashed curves include neither FSI nor
the medium effects on $\Delta$-propagation.
By including the medium effects on the
$\Delta$, the dotted curves are obtained.
Finally our full calculation gives the solid curves.
Clearly, the medium effects are large 
and important in achieving a good agreement with the data. 
Particularly in the $\Delta$ region [Fig.~\ref{fig_krusche} (b)],
the medium effects work to absorb pions, leading to
the drastic reduction of the cross sections.
The obtained good agreement seen in Fig.~\ref{fig_krusche}
indicates the soundness of our approach, and encourages us to apply the
same approach to the neutrino-induced coherent pion productions. 
An interesting feature seen in Fig.~\ref{fig_krusche} is that the
non-resonant contribution (the difference between the solid and dash-dotted curves)
is significant even in the energy near the
$\Delta(1232)$ peak [Fig.~\ref{fig_krusche}(b)]. 
This is partly due to the fact that the resonant contribution is significantly
reduced by the pion absorptions.

\begin{figure}[t]
\begin{center}
 \includegraphics[width=85mm]{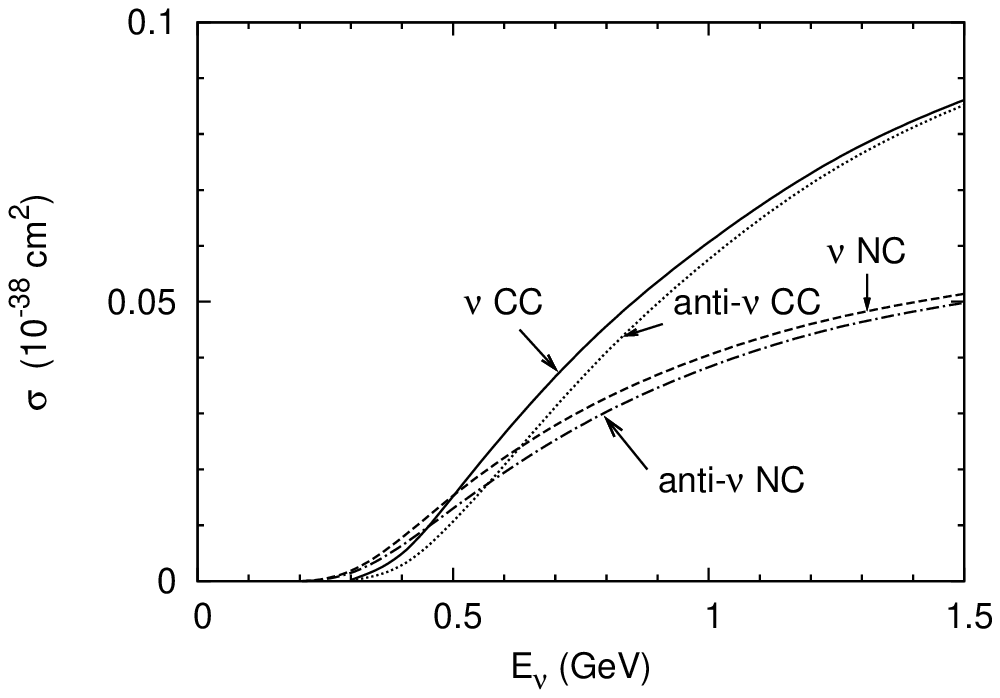}
\caption{\label{fig_tot}
The total cross sections 
as a function of $E_\nu$ for
$\nu_\mu + {}^{12}{\rm C}_{g.s.} \to \mu^- + \pi^+ + {}^{12}{\rm
 C}_{g.s.}$ (solid curve),
$\nu + {}^{12}{\rm C}_{g.s.} \to \nu + \pi^0 + {}^{12}{\rm C}_{g.s.}$
 (dashed curve),
$\bar{\nu}_\mu + {}^{12}{\rm C}_{g.s.} \to \mu^+ + \pi^- + {}^{12}{\rm C}_{g.s.}$ 
(dotted curve) and
$\bar{\nu} + {}^{12}{\rm C}_{g.s.} \to \bar{{\nu}} + \pi^0 + {}^{12}{\rm
 C}_{g.s.}$ (dash-dotted curve).
Figure taken from Ref.~\cite{sxn}. 
Copyright (2010) APS.
}
\end{center}
\end{figure}
We now move on to 
neutrino-induced coherent pion productions
on $^{12}$C target.
We consider the CC processes,
$\nu_\mu (\bar\nu_\mu) + {}^{12}{\rm C}_{g.s.} \to
 \mu^-(\mu^+) + \pi^+(\pi^-) + {}^{12}{\rm C}_{g.s.}$, 
and the NC processes,
$\nu(\bar\nu) + {}^{12}{\rm C}_{g.s.} \to
 \nu(\bar\nu) + \pi^0 + {}^{12}{\rm C}_{g.s.}$.
The total cross sections for these processes are shown in 
Fig.~\ref{fig_tot}
as functions of the incident neutrino (anti-neutrino) energy 
in the laboratory system.
As expected from the isospin factor, we find 
$\sigma_{\rm CC}/\sigma_{\rm NC}\sim 2$ for higher $E_\nu$.
On the other hand, $\sigma_{\rm NC}$ becomes larger than 
$\sigma_{\rm CC}$ at low $E_\nu$ because the massless lepton in the
final state gives the NC processes a larger phase space.
The difference between the neutrino and antineutrino processes is from
the interference between the vector and axial currents. 
Because the axial current dominates in the coherent processes, this
interference gives a rather small contribution.

We compare our result with available data. 
In order to do so, we average the cross sections by convoluting
with the neutrino flux that was used in the experiment. 
We use the neutrino fluxes of $E_\nu \le$ 2 GeV, 
and neglect the fluxes beyond that limit.
To compare with a K2K result for the CC process~\cite{Hasegawa:2005td}, 
we use the flux reported in Ref.~\cite{K2K} 
and obtain
$\sigma_{\rm ave}^{\rm CC} = 6.3 \times 10^{-40} {\rm cm}^2$.
Our result is consistent with the upper limit from 
the K2K experiment; $\sigma_{\rm K2K}  <  7.7 \times 10^{-40} {\rm cm}^2$.
This upper limit was obtained with some kinematical cuts which we also
applied to our calculation.
A similar upper limit was also reported from a 
SciBooNE experiment~\cite{Hiraide:2008eu}.
For other theoretical calculations compared with the K2K result,
see Table~\ref{tab:coh}.
Nonzero CC coherent pion productions at low energies ($E_\nu\ltap$ 1~GeV)
have been recently reported by the T2K Collaboration~\cite{t2k_coh}.
They performed two analyses in each of which a different coherent pion
production model was used:
the models of Rein-Sehgal~\cite{Rein:1982pf} and
Alvarez-Ruso et al.~\cite{valencia1_coh,valencia2_coh}.
The data were collected within a restricted phase space, defined by  
$p_\mu >$ 0.18~GeV, $0.18 < p_\pi < 1.6$~GeV, and
$\theta_{\mu (\pi)}< 70^\circ$, where the detector has a good acceptance;
$p_{\mu (\pi)}$ is the momentum of $\mu$ ($\pi$)
and 
$\theta_{\mu (\pi)}$ is the angle between the $\mu$ ($\pi$) momentum and
the incident neutrino direction.
The flux-averaged cross section to the restricted phase space was
found to be 
$3.2\pm 0.8 ({\rm stat})^{+1.3}_{-1.2}({\rm sys})$
($2.9\pm 0.7^{+1.1}_{-1.1}$) $\times 10^{-40}\, {\rm cm}^2$
when the model of 
Rein-Sehgal~\cite{Rein:1982pf} 
(Alvarez-Ruso et al.~\cite{valencia1_coh,valencia2_coh})
was used.
Then, the flux-averaged total cross section was estimated using 
the model predictions for the unmeasured phase space.
$3.9\pm 1.0^{+1.5}_{-1.4}$
($3.3\pm 0.8^{+1.3}_{-1.2}$) $\times 10^{-40}\, {\rm cm}^2$
was obtained for the model of 
Rein-Sehgal~\cite{Rein:1982pf} 
(Alvarez-Ruso et al.~\cite{valencia1_coh,valencia2_coh}).
We update our calculation of Ref.~\cite{sxn} to compare with this new
data covering the full phase space.
For this purpose, cross sections are convoluted with the neutrino
flux for the T2K experiment~\cite{T2K-flux} that peaks at $E_\nu\sim 0.6$~GeV.
Our result, $3.1\times 10^{-40}\, {\rm cm}^2$, is in a good agreement
with the T2K result.

\begin{table}[t]
 \begin{tabular}[t]{llll}
Channel (EXP.)
  &  CC$\pi^+$(K2K) &  CC$\pi^+$(T2K) &  NC$\pi^0$(MiniBooNE) \\ \hline
Data & $<7.7$~\cite{Hasegawa:2005td} &$3.3\pm 0.8^{+1.3}_{-1.2}$~\cite{t2k_coh}&$7.7\pm 1.6\pm 3.6$~\cite{raaf}\\
 &  & $3.9\pm 1.0^{+1.5}_{-1.4}$~\cite{t2k_coh}&\\
Alvarez-Ruso et al.~\cite{valencia1_coh,valencia2_coh}& 8.3 (4.4)& 5.3~\cite{t2k_coh}&3.9 (2.0)\\
Berger et al.~\cite{bs_pcac} &0.62$\times$12 & -- & -- \\
Nakamura et al.~\cite{sxn} & 6.3 & 3.1 {\small [This work]}&2.8 \\
Hern\'andez et al.~\cite{hernandez2}&$6.1\pm 1.3$ & -- &$2.6\pm 0.5$ \\
 \end{tabular}
\caption{
CC and NC coherent pion production cross sections from several
 theoretical calculations in comparison with data.
The unit is $10^{-40}{\rm cm}^2$.
In the second, third, and fourth columns, 
theoretical cross sections have been convoluted
 with the neutrino fluxes used in K2K~\cite{K2K}, T2K~\cite{T2K-flux}, and 
MiniBooNE~\cite{miniboone-flux} experiments, respectively.
The two results of the
 T2K~\cite{t2k_coh} are from different analyses
in which different coherent pion production models were used.
See the text for more explanation on the T2K data.
The numbers in the parentheses
from Alvarez-Ruso et al.
are obtained with different form factors.
}
\label{tab:coh}
\end{table}
Regarding the NC process, the MiniBooNE Collaboration reported the
flux-averaged cross section, 
$\sigma_{\rm  MiniBooNE}=7.7\pm 1.6\pm 3.6\times 10^{-40} {\rm cm}^2$~\cite{raaf}.
Using the same flux as given in Ref.~\cite{miniboone-flux}, we obtain 
$\sigma_{\rm NC}=2.8\times 10^{-40} {\rm cm}^2$ which is barely
consistent with the MiniBooNE result.
A similar tendency is also found in comparisons of other theoretical
calculations tabulated in Table~\ref{tab:coh} with the data.
Furthermore, the SciBooNE Collaboration published the ratio
between CC and NC coherent pion production cross sections~\cite{kurimoto}:
$\sigma_{\rm CC}/\sigma_{\rm NC} = 0.14^{+0.30}_{-0.28}$.
On the other hand, our as well as all the other 
theoretical calculations after 2005 gave rather different results,
$\sigma_{\rm CC}/\sigma_{\rm NC} = 1.5\sim2$, which is expected from the isospin
factor.
Several authors~\cite{amaro,sxn,hernandez2} have 
suspected that this puzzling situation could arise from the fact that 
the RS model~\cite{Rein:1982pf,RS2_coh} was used in the analyses of the NC data.
They showed that $\eta$-distribution
$(\eta\equiv E_\pi (1-\cos\theta_\pi))$
of the RS model is rather different 
from those of their microscopic models.
In the NC data analyses, the $\eta$-distribution has been proven useful
in separating $\pi^0$ events into each production mechanism.
Thus one may suspect this analyses 
could have overestimated the NC cross sections.
It would be interesting to re-analyze the data with a more realistic
coherent pion production model.

\begin{figure}[t]
 \begin{center}
 \includegraphics[width=77mm]{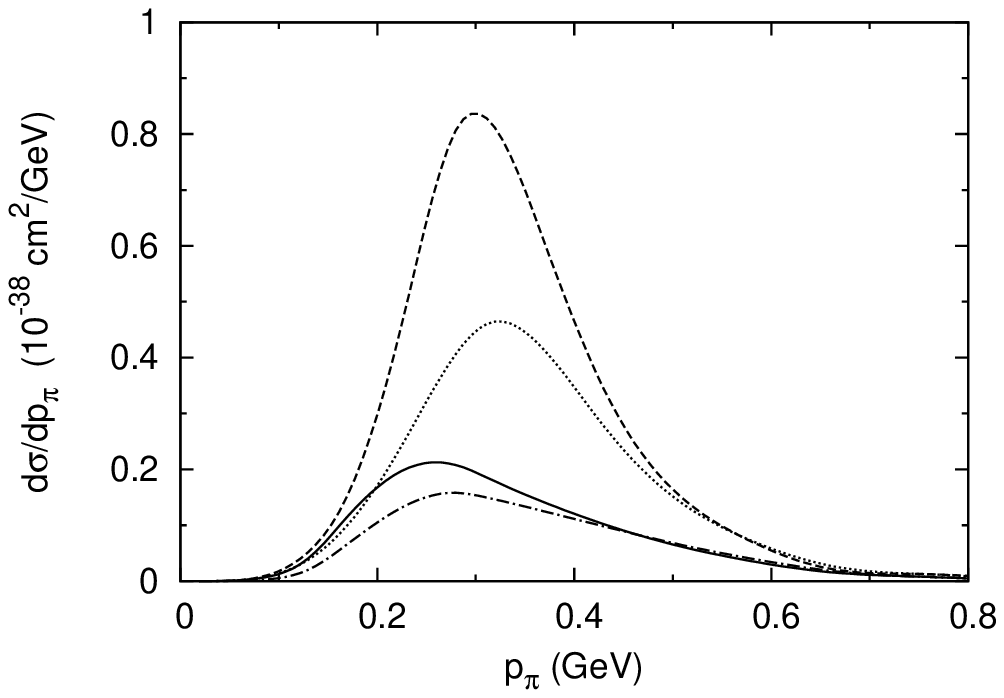}
 \caption{\label{fig_tpi.1gev}
The pion momentum distribution in
 $\nu_\mu +\! {}^{12}{\rm C}_{g.s.} \to \mu^- \!+ \!\pi^+\!+ \!{}^{12}{\rm
  C}_{g.s.}$ at $E_\nu$ = 1 GeV;
$p_\pi$ is the pion momentum in the laboratory frame.
The solid, dashed, dotted and dash-dotted curves include the same
  dynamical features as those in Fig.~\ref{fig_krusche}.
Figure taken from Ref.~\cite{sxn}. 
Copyright (2010) APS.
}
 \end{center}
\end{figure}
Finally, we discuss the pion momentum distribution in the
neutrino-induced coherent pion production to see an impact of the
nuclear effects on the observables.
In Fig.~\ref{fig_tpi.1gev}, we show our result for
 $\nu_\mu +  {}^{12}{\rm C}_{g.s.} \to \mu^-  +  \pi^+ +  {}^{12}{\rm
  C}_{g.s.}$ at $E_\nu$=1~GeV.
As we have seen in the photon-induced coherent pion production
(Fig.~\ref{fig_krusche}), we again find large medium effects here. 
The pion is significantly absorbed near the $\Delta$ peak, 
and the peak position is shifted by the FSI.
It is also shown in the figure that
the non-resonant amplitudes contributes to enhance the cross section by
18\% in this case.
We note that the rescattering effect significantly enhances
the non-resonant contribution.
This is in contrast with the other microscopic calculations
for the neutrino-induced coherent pion productions that found little role of
(tree-level) non-resonant mechanisms.

\clearpage

\section{Neutrino-nucleus reactions in the DIS region}
\label{nu-A-dis}

\subsection{Nuclear modifications of structure functions}

The structure function $F_2$ for the nucleon has been measured 
in a wide kinematical region from small $x$ ($\sim 10^{-4}$)
to large $x$ ($\sim 0.8$) and from small $Q^2$ ($\sim 1$ GeV$^2$)
to large $Q^2$ ($\sim 10^4$ GeV$^2$). In 1970's when the DIS 
experiments started, people expected that the nuclear structure
functions were simple additions of proton and neutron contributions. 
Therefore, it was rather surprising that a nuclear modification of $F_2$ 
was found by the European Muon Collaboration (EMC) in 1983~\cite{emc1983},
although the Fermi-motion part in the large $x$ region was
theoretically discussed before this experiment. 

\begin{wrapfigure}[11]{r}{0.38\textwidth}
\vspace{-1.05cm}
   \begin{center}
       \epsfig{file=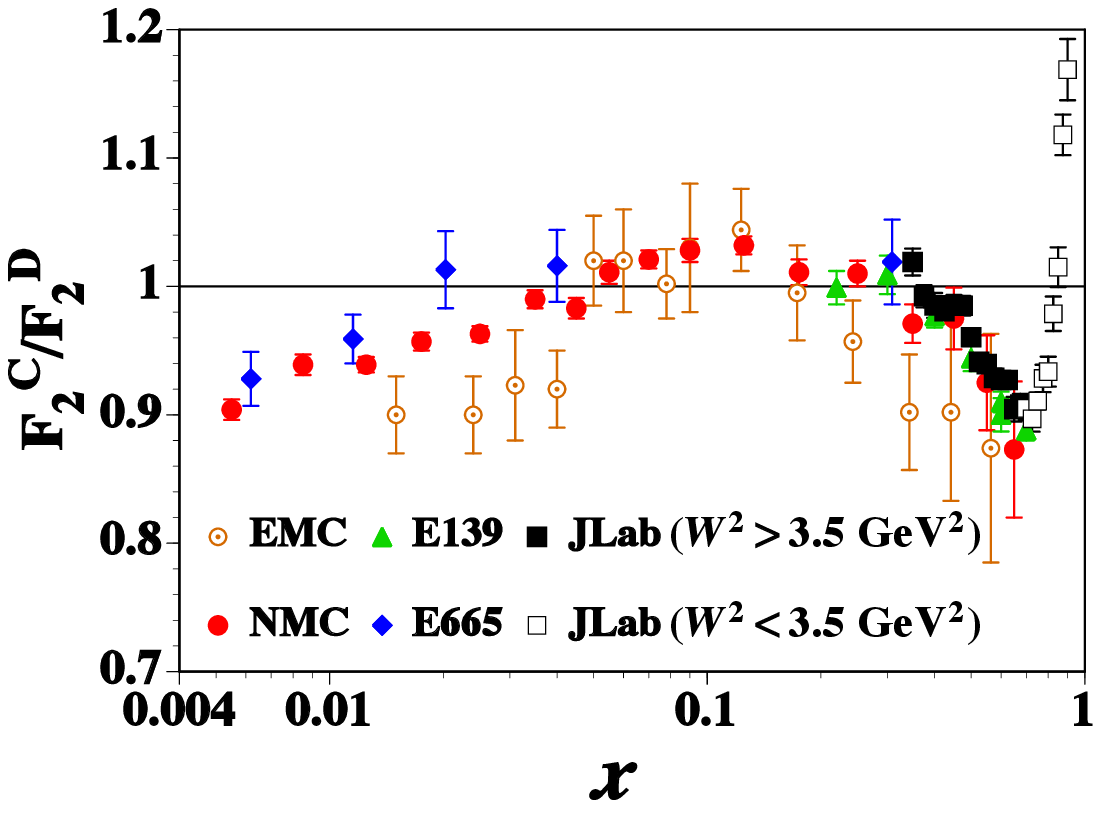,width=0.36\textwidth} 
   \end{center}
\vspace{-0.50cm}
\hspace{-2.40cm}
   \begin{minipage}{0.52\textwidth}
\caption{
Measurements for the nuclear modification ratio $F_2^C /F_2^D$,
where $C$ and $D$ indicate carbon and deuteron, respectively.}
\label{fig:f2c-f2d-ratio}
   \end{minipage}
\end{wrapfigure}

Since nuclear binding energies are negligible
in comparison with DIS energies of more than multi GeV, noticeable 
nuclear effects were not expected except for the large-$x$ region
before the EMC's discovery. The current situation of
nuclear modifications are shown in Fig.\,\ref{fig:f2c-f2d-ratio}
for the carbon nucleus by the ratio $F_2^C /F_2^D$.
At small $x$ ($<0.05$), the ratio is smaller than one and it is caused by
nuclear shadowing, whereas the enhancement at $x \sim 0.1$ is called
anti-shadowing. The modifications are also negative in the medium $x$
($0.3<x<0.7$) due to nuclear binding and possibly nucleon's internal
modification, and the ratio increases at large $x$ ($>0.8$) due to 
the Fermi motion of the nucleon. The nuclear modification mechanisms 
are explained in Ref.~\cite{nuclear-summary}.

The range of the scaling variable $x$ is $0<x<A$
for a nucleus of the mass number $A$ while
$0<x<1$ for the nucleon. If the variable is defined by 
$x_A = Q^2 / (2 p_A \cdot q) = Q^2 / (2 \bar M_A \nu)$,
the range is given by $0<x_A<1$ in the same way as the nucleon. 
Here, $\bar M_A$ and $p_A$ are nuclear mass and momentum, respectively.
The mass difference from the nucleon gives rise to the relation,
$x = (\bar M_A / m_N) x_A \simeq A x_A$, 
so that we have $0<x<A$.
Since there is no DIS data in the region $x>1$, the extremely
large-$x$ region is neglected in our analysis of the NPDFs.

\begin{wrapfigure}[9]{r}{0.38\textwidth}
\vspace{-1.2cm}
   \begin{center}
       \epsfig{file=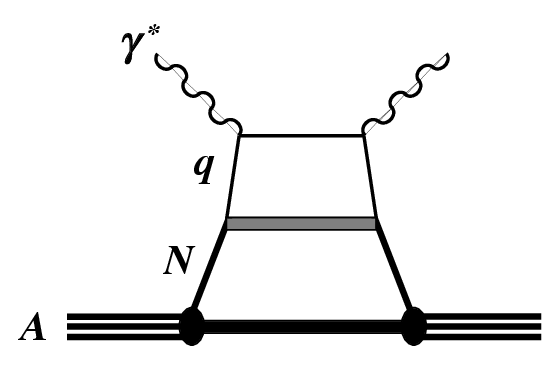,width=0.33\textwidth} 
   \end{center}
\vspace{-0.30cm}
\hspace{-2.40cm}
   \begin{minipage}{0.52\textwidth}
\caption{Convolution description for nuclear structure functions.}
\label{fig:convolution-2}
   \end{minipage}
\end{wrapfigure}

For describing nuclear structure functions at medium and large $x$,
we may use a simple convolution description as shown in 
Fig.\,\ref{fig:convolution-2}. 
The hadron tensor for a nucleus 
$W_{\mu\nu}^A$ is expressed by the one for the nucleon
$W_{\mu\nu}^N$ convoluted with
the spectral function $S(p_N)$ for a nucleon in the nucleus~\cite{nuclear-summary,convolution}:
\vspace{-0.10cm}
\begin{equation}
W_{\mu\nu}^A (p_A, q) = \int d^4 p_N \, S(p_N) \, W_{\mu\nu}^N (p_N, q) .
\label{eqn:w-convolution-w}
\vspace{-0.10cm}
\end{equation}
Namely, the quark distributions in the nucleon are modified 
by the effect of the nucleon energy-momentum distribution in the nucleus
$S(p_N)$, and they are estimated by the convolution integral.
Then, the projection operator
$
\hat P_2^{\, \mu\nu}  = 
   - \bar M_A
   \, p_A \cdot q / (2\, \tilde p_A^{\, 2})
  \left ( g^{\mu\nu} - 3 \, \tilde p_A^{\, \mu} \, 
                 \tilde p_A^{\, \nu} / \tilde p_A^{\, 2}\right ) 
$,
$\tilde p_A^{\, \mu} = p_A^{\, \mu} - (p_A \cdot q) q^{\, \mu} /q^2$
is applied to extract the $F_2$ structure function from $W_{\mu\nu}^A$:
\begin{align}
F_{2}^A (x, Q^2) & = \int_x^A dy \, f(y) \, F_{2}^N (x/y, Q^2) ,
\nonumber \\
f(y) & =  \int d^4 p_N \, S (p_N) 
     \, y \, \delta \left( y - \frac{p_N \cdot q}{m_N \nu} \right) ,
\ \ 
y   =    \frac{\bar M_A \, p_N \cdot q}{m_N \, p_A \cdot q} 
  \simeq \frac{A \, p_N^+}{p_A^+} ,
\label{eqn:conv-F2}
\end{align}
where the lightcone momentum is defined by $p^+ = (p^0 + p^3 )/\sqrt{2}$.
The spectral functions $S(p_N)$ have been investigated in electron 
scattering studies. Using them in the convolution integral, 
we can calculate the effects of binding, Fermi motion, 
and short-range correlations in $F_2^A/F_2^D$. However, 
they may not be enough to explain the whole 
nuclear modifications observed experimentally, and 
modifications of the internal structure of the nucleon 
in $F_{2}^N$ could be needed~\cite{ST1994}.

On the other hand, the small-$x$ part is modified by a different mechanism
of nuclear shadowing. In charged-lepton scattering, virtual photon ($\gamma^*$)
transforms into a vector meson or corresponding $q\bar q$ state because
they have the same quantum numbers. Using the photon momentum
$ q = (\nu ,{\rm{ }}0,{\rm{ }}0,{\rm{ }} - \sqrt {\nu ^{\rm{2}}  + Q^2 } ) $,
we have the propagation length of the vector meson as
$ \lambda  = 1 / | E_V  - E_\gamma  | \simeq 2\nu /(M_V^2  + Q^2 )
= 0.2 \ {\rm{fm}} / x > 2 \ {\rm{ fm \ at \ }}x < 0.1 $.
It is larger than the average nucleon separation in a nucleus
(2.2 fm).
The $F_2$ structure function is expressed by the $\gamma^* A$ 
cross section as 
$ F_2^A(x,Q^2) \simeq Q^2 \sigma_{\gamma^* A} / (4\pi^2 \alpha) $
at small $x$, where $\alpha$ is the fine structure constant.
We consider the picture that the virtual photon
interacts as the vector meson or $q\bar q$ state ($h$) with a nucleus,
and then structure function $F_2$ is expressed by the 
hadron-nucleus cross section $\sigma_{hA}$ and spectral
function of the hadronic state $\Pi(s)$ as
\begin{align}
\! \! \! 
F_2^A(x,Q^2) & =
 \frac{Q^2}{\pi}
 \int_{4m_{\pi}^2}^\infty  dM^2
 \frac  {M^2 \,\Pi(M^2)}
        {\left(M^2+Q^2\right)^2}\;
 \sigma_{hA}(M^2) ,
\ \ 
\Pi(s) & = \frac{1}{12 \pi^2}
         \frac{\sigma_{e^+e^-\rightarrow {\rm hadrons}}(s)}
          {\sigma_{e^+e^-\rightarrow \mu^+\mu^-}(s)} ,
\label{eqn:f2a-vmd}
\end{align}
where $\Pi(s)$ contains vector-meson and $q\bar q$-continuum contributions.
Because the hadron propagation length is longer than
the average nucleon separation, the hadron $h$ interacts 
with a nucleon not only once but also multiple times.
This process is described by multiple scattering theory,
and it gives rise to the nuclear shadowing.
In the weak interactions, the axial vector current contributes
to the cross section in addition to the vector one, so that
the axial-vector mesons (or corresponding $q\bar q$ states)
contribute to the shadowing.
Here, we do not explain more about the nuclear modifications 
by the convolution picture and the shadowing phenomena 
so one may read Ref.~\cite{nuclear-summary} for the details.

\subsection{Parton distribution functions in nuclei}

As discussed in the previous section, we know the $x$ dependence
of nuclear modifications from experimental measurements for various
nuclei, and major theoretical ideas have been proposed to explain the
nuclear modifications observed experimentally. 
If the order of nuclear effects is merely needed,
or if only the gross properties of nuclear structure functions are studied,
such theoretical models are good enough to predict 
the nuclear structure functions
for estimating neutrino-nucleus cross sections.
However, it is not the case in the current neutrino experiments.
Accurate theoretical cross sections are needed for 
neutrino oscillation experiments where 
uncertainties associated with the current understanding of
the neutrino-nucleus interactions 
are the dominant sources of systematic errors.
The theoretical models are valuable for us to understand 
the physics behind the nuclear modifications; however,
they may not provide predictions accurate enough
for high-energy neutrino reactions. 
Therefore, it is usual to use a global analysis result
for the NPDFs in the same way that the 
global-analysis PDFs are 
used for Large Hadron Collider (LHC) analysis for finding
new physics beyond the standard model.

One could determine the NPDFs by parametrizing them in
the same way as the nucleonic PDF analysis. However,
it is practically more useful to express the NPDFs 
in terms of the nucleonic PDFs by assigning nuclear modification
factors $ w_{i} (x,A,Z)$ at the initial $Q^2$ scale
($\equiv Q_0^2$) as~\cite{hkn-npdfs}
\begin{align}
u_v^A (x,Q_0^2) & = w_{u_v} (x,A,Z) \, 
\frac{1}{A} \left[ Z\, u_v (x,Q_0^2) + N d_v (x,Q_0^2) \right] ,
\nonumber \\
d_v^A (x,Q_0^2) & = w_{d_v} (x,A,Z) \, 
\frac{1}{A} \left[ Z\, d_v (x,Q_0^2) + N u_v (x,Q_0^2) \right] ,
\nonumber \\
\bar u^{\,A} (x,Q_0^2) & = w_{\bar u} (x,A,Z) \, 
\frac{1}{A} \left[ Z\,\bar u (x,Q_0^2) + N \bar d (x,Q_0^2) \right] ,
\nonumber \\
\bar d^{\,A} (x,Q_0^2) & = w_{\bar d} (x,A,Z) \, 
\frac{1}{A} \left[ Z\,\bar d (x,Q_0^2) + N \bar u (x,Q_0^2) \right] ,
\nonumber \\
\bar s^{\,A} (x,Q_0^2) & = w_{\bar s} (x,A,Z) \, \bar s (x,Q_0^2) ,
\nonumber \\
g^{\,A} (x,Q_0^2) & = w_{g} (x,A,Z) \, g (x,Q_0^2) ,
\label{eqn:fia}
\end{align}
because the nucleonic PDFs have been determined accurately
in a wide kinematical region of $x$ and $Q^2$ and 
nuclear modifications are of the order of 10-20\%.
Furthermore, nuclear data are often shown in ratio
forms such as $F_2^A/F_2^D$, so that the modifications
could be obtained more easily rather than the absolute 
distributions. Here, 
$w_i$ is the nuclear modification factor to be determined
by a global analysis, $i$ is a parton species,
and $A$, $Z$, and $N$ are mass, atomic, and neutron numbers.
The functional form of Eq.\,(\ref{eqn:fia}) cannot
describe the NPDFs at $x>1$; however, it is not an issue
at this stage because there is no DIS data in such an extremely
large-$x$ region. The scale $Q_0^2$ is taken as $Q_0^2 =1$ GeV$^2$
in our analysis, so that the charm-quark modification factor 
$w_c (x,A,Z)$ does not exist. Although we found an effect 
on the NuTeV anomaly 
about $\sin^2 \theta_W$ from the nuclear modification
difference between $u_v$ and $d_v$ in Ref.~\cite{nutev-anomaly},
it is impossible to determine it from experimental measurements
at this stage~\cite{flavor-depedent-emc}. Therefore, 
the parameters in $w_{u_v}$ and $w_{d_v}$ are assumed to be the same
except for one-type parameters (such as $a_{u_v}$ and $a_{d_v}$
in Eq.\,(\ref{eqn:wi})), which are fixed independently
by the baryon-number and charge conservations,
in our current parametrization. 
Furthermore, the antiquark modifications are assumed equal 
$w_{\bar u}=w_{\bar d}=w_{\bar s} \equiv w_{\bar q}$
although they could be different~\cite{sk-ubar-dbar-nucleus,ubar-dbar}.

We try to determine the nuclear modification parts $w_i (x,A,Z)$
by a global analysis of world data on high-energy nuclear reactions.
The functions are parametrized in the form
\begin{equation}
w_i (x,A,Z) = 1 + \left( {1 - \frac{1}{{A^{\alpha} }}} \right)
\frac{{a_i  + b_i x + c_i x^2  + d_i x^3 }}{{(1 - x)^{\beta} }} ,
\label{eqn:wi}
\end{equation} 
with the parameters $\alpha$, $\beta$, $a_i$, $b_i$, $c_i$, and $d_i$.
At this stage, there is no solid data
to separate flavor-dependent nuclear effects, 
and thus we take $i=u_v$, $d_v$, $\bar q$, and $g$ as the parton species.
The cubic functional form is employed in the numerator of
the second term of Eq.\,(\ref{eqn:wi}) in order to have similar
$x$-dependent variations as in Fig.~\ref{fig:f2c-f2d-ratio}.
The initial scale $Q_0^2$ is arbitrary as long as it is taken
in the region where perturbative QCD can be applied.
In order to avoid higher-twist effects, it is desirable to 
take it more than a few GeV$^2$. However, the electron-ion collider
like HERA does not exist for nuclei at this stage, which limits 
the kinematical region of measured structure functions. This fact 
inevitably leads to small $Q^2$ values at small $x$ in 
the structure-function measurements. 
In order to include valuable shadowing information
at small $x$, we may take $Q_0^2=1 \ {\rm GeV}^2$. However,
the small $Q^2$ region could contain the higher-twists effects,
which are neglected so far in our analysis.
There are three obvious constraints for the NPDFs from
the conservations of baryon number, charge, and momentum:
\begin{equation}
\begin{array}{ll}
{\text{Baryon number:}} & 
A \int dx \left[ \frac{1}{3} u_v^A (x) 
               + \frac{1}{3} d_v^A (x) \right] = A, \\[6pt] 
{\text{Charge:}} & 
A \int dx \left[ \frac{2}{3} u_v^A (x) 
               - \frac{1}{3} d_v^A (x) \right] = Z, \\ [6pt] 
{\text{Momentum:}} & 
A \sum\limits_{i = q,\bar q,g} \int dx \, xf_i^A (x)  = 
A,
\end{array}
\vspace{-0.10cm}
\label{eqn:conservations}
\end{equation}
where $f_i^A (x)$ indicates the PDFs 
$f_u^A (x)= u^A(x)$, $f_d^A (x)= d^A(x)$, and so on.
Therefore, the parameters in Eq.~(\ref{eqn:wi}) are determined in
a global $\chi^2$ analysis under the three constraints of Eq.~(\ref{eqn:conservations}).

\begin{wrapfigure}[10]{r}{0.38\textwidth}
\vspace{-0.5cm}
   \begin{center}
       \epsfig{file=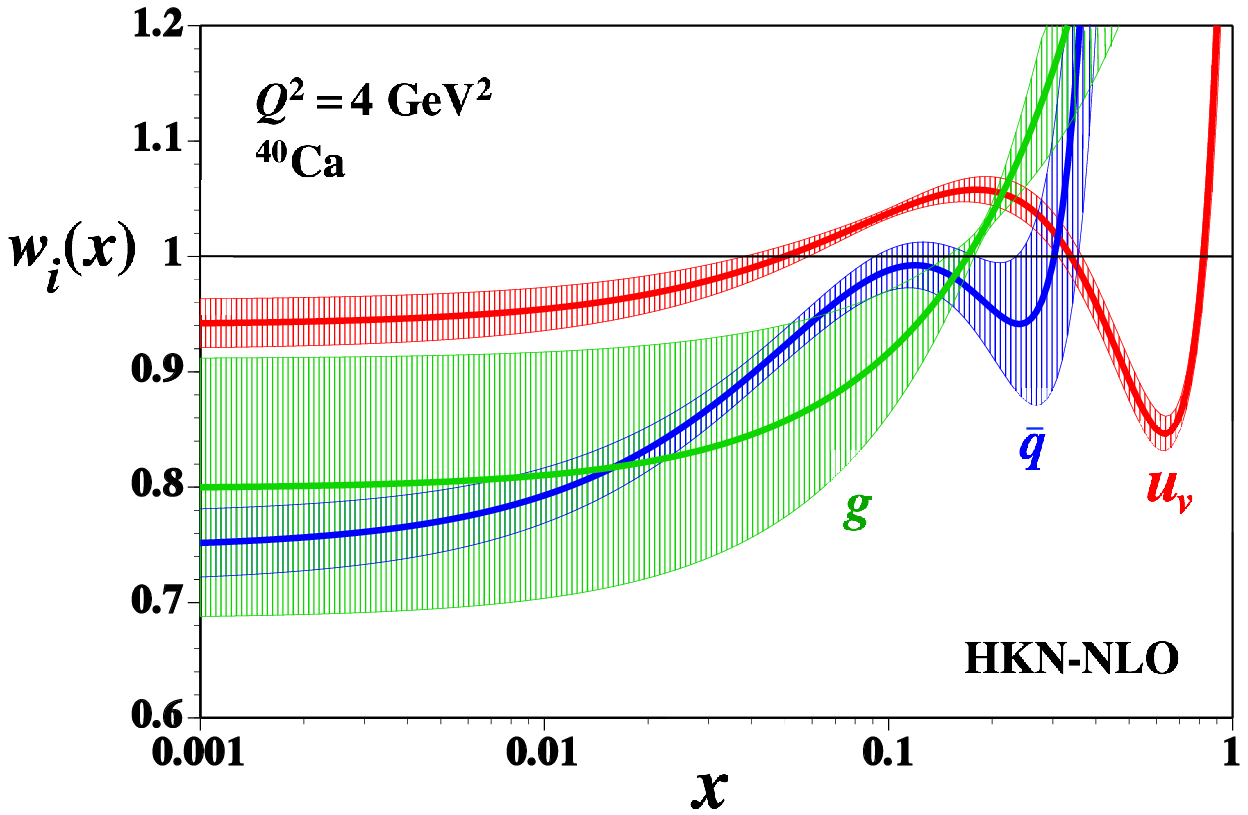,width=0.36\textwidth} 
   \end{center}
\vspace{-0.40cm}
\hspace{-2.40cm}
   \begin{minipage}{0.52\textwidth}
\caption{HKN nuclear modifications at $Q^2 = 4$ GeV$^2$.
}
\label{fig:hkn-npdfs}
   \end{minipage}
\end{wrapfigure}

The nuclear modification functions $w_i (x,A,Z)$, more specifically 
the parameters introduced in Eq.~(\ref{eqn:wi}),
are determined by a global analysis of charged-lepton DIS measurements 
on the ratios $F_2^A /F_2^D$ and $F_2^A /F_2^{A'}$
and Drell-Yan ratios $\sigma_{DY}^{pA} / \sigma_{DY}^{pD}$~\cite{hkn-npdfs}.
The NPDFs of Eq.~(\ref{eqn:fia}) are evolved to the experimental
$Q^2$ points by the standard DGLAP 
(Dokshitzer-Gribov-Lipatov-Altarelli-Parisi)
evolution equations~\cite{bf1} for calculating the $\chi^2$ values.
The functions $w_i$ are shown in Fig.\,\ref{fig:hkn-npdfs}
with the uncertainty bands estimated
by using the Hessian matrix obtained in the $\chi^2$ fit.
The valence-quark distributions are well determined for $x>0.3$ because
the valence-quark distributions dominate the structure function
$F_2^A$ at medium and large $x$ ($>0.3$) and there are many
experimental data, as typically seen in Fig.~\ref{fig:f2c-f2d-ratio},
to constrain the valence-quark modification $w_{u_v}$ and 
$w_{d_v}$ ($=w_{u_v}$ for an isoscalar nucleus $^{40}$Ca) at $x>0.3$.
The antiquark distributions dominates $F_2$ at small $x$, so that
the antiquark modification $w_{\bar q}$ is also determined
well in the shadowing region at $x<0.05$ by the $F_2$-ratio data.
The function $w_{\bar q}$ is also fixed by the Drell-Yan data, 
which indicated almost no nuclear modification in the region $0.05<x<0.2$.
If there is no nuclear modification for the antiquark distributions
at $0.1 < x <0.2$, the anti-shadowing data, typically seen
as the positive modification at $x \sim 0.1$ in Fig.~\ref{fig:f2c-f2d-ratio},
should be interpreted as the positive nuclear effects in 
the valence-quark distributions. This is the reason why 
the function $w_{u_v}$ is also determined well at $0.1 < x <0.2$.
Because of the baryon-number and charge conservations in 
Eq.~(\ref{eqn:conservations}), the valence-quark modifications 
at small $x$ are constrained by the same functions determined for $x>0.1$.
Therefore, the valence-quark functions have small uncertainties
in the whole-$x$ region.
The antiquark distributions at $x>0.2$ are not determined by
the current measurements and have large errors as shown in 
Fig.~\ref{fig:hkn-npdfs}. The gluon distribution in the nucleon
is determined mainly by the scaling violation of $F_2$. However,
the scaling violation is not obvious experimentally in the ratios
$F_2^A / F_2^{A'}$, which makes it difficult to pin down
the gluon nuclear modification $w_g$ as shown by the large
uncertainty band in Fig.~\ref{fig:hkn-npdfs}. Including LHC data,
we hope to find more accurate gluon distributions in nuclei.

In our studies, the distributions defined by Eq.\,(\ref{eqn:fia}) 
are called NPDFs by considering a nucleus as a whole system
in the same way as the nucleonic PDF.
However, the distributions $f_i^{p/A}$ are sometimes 
called NPDFs for the parton type $i$ by other groups~\cite{other-npdfs}
with the definition
$ f_i^A (x,Q_0^2) = [ \, Z\,f_i^{p/A} (x,Q_0^2) 
                + N f_i^{n/A} (x,Q_0^2) \, ] \, /A $
by considering proton and neutron contributions in
a nuclear medium, and they should not be confused with 
our NPDFs $f_i^A (x,Q_0^2)$.
If both definitions are not confused,
the choice does not matter. 
In comparing the nuclear modifications of various groups, 
we should be careful about this difference in the definition.

\begin{figure}[b]
\vspace{-0.0cm}
   \begin{center}
       \epsfig{file=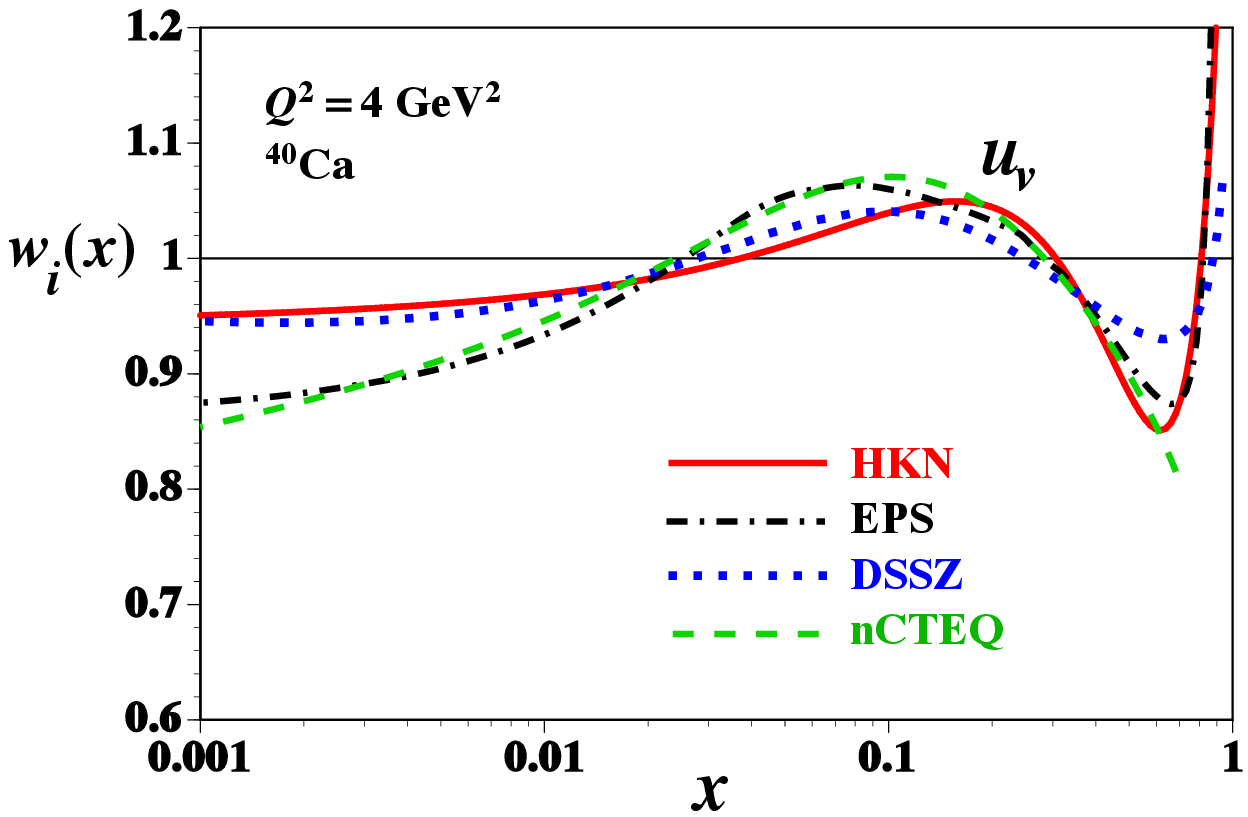,width=0.32\textwidth} 
       \epsfig{file=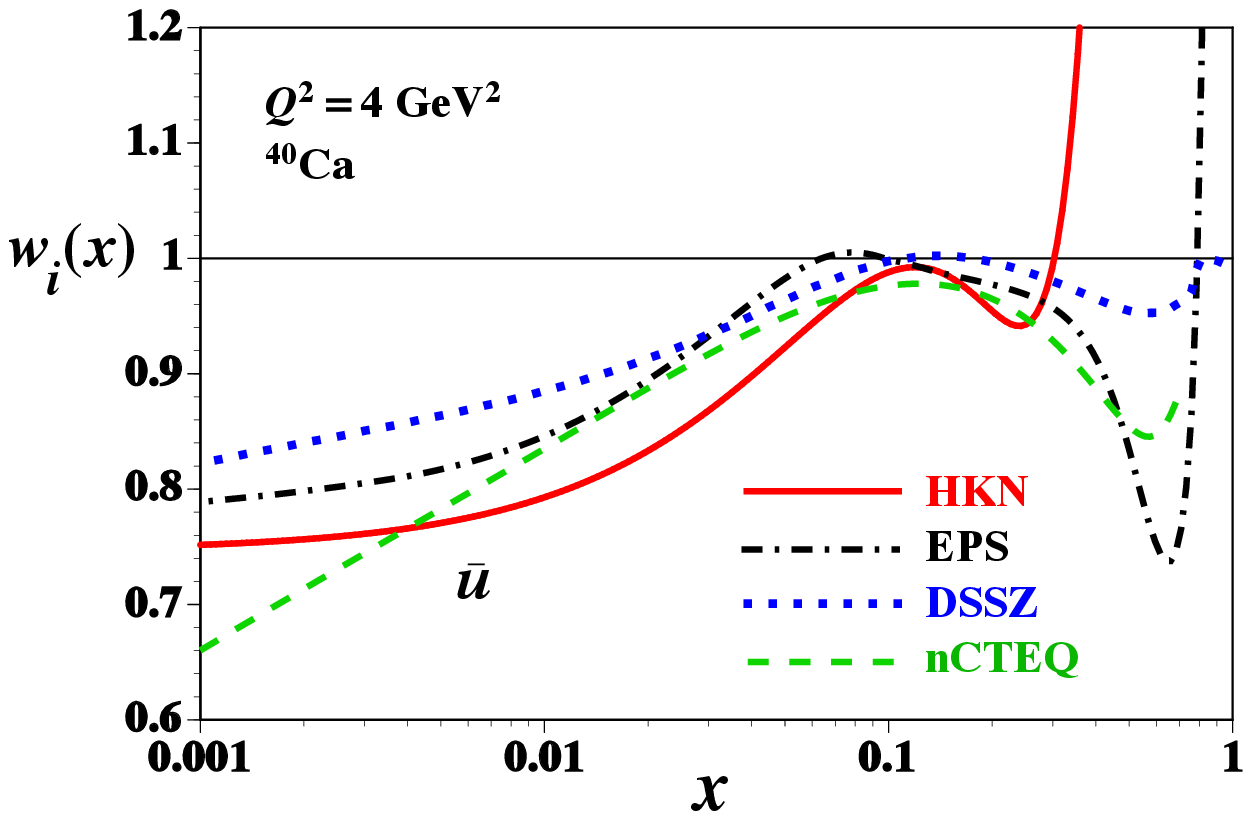,width=0.32\textwidth} 
       \epsfig{file=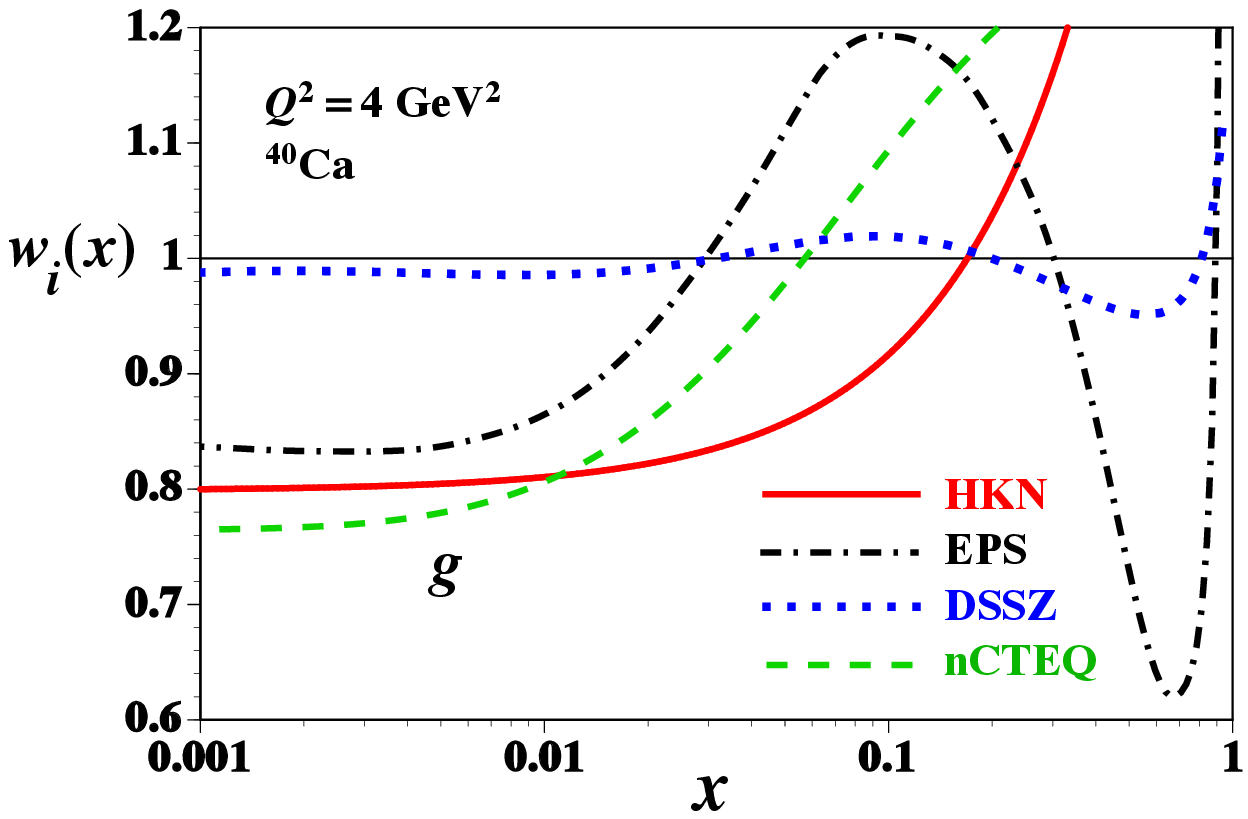,width=0.32\textwidth} 
   \end{center}
\vspace{-1.20cm}
\hspace{0.00cm}
\begin{center}
\hspace{-1.70cm}
   \begin{minipage}{0.90\textwidth}
\caption{Comparison of various nuclear PDFs at $Q^2 =4$ GeV$^2$
for $^{40}$Ca.}
\label{fig:npdfs-all}
   \end{minipage}
\end{center}
\end{figure}

Determined HKN NPDFs are compared with other analysis
results~\cite{other-npdfs} in Fig.\,\ref{fig:npdfs-all}.
All the distributions are similar except for the gluon modifications
at small $x$ and antiquark/gluon modifications at medium $x$ because 
they are not well constrained by experimental measurements.
However, the LHC experiments are in progress and they are generally
sensitive to small-$x$ NPDFs, so that the gluon shadowing should
be determined more reliably by including LHC measurements.
In fact, 
a large gluon shadowing is suggested by studying
productions of vector mesons at LHC~\cite{guzey}.
On the other hand, in comparison with the LHC data on 
charged-hadron, dijets, and direct-photon productions in p+Pb, 
the current NPDFs seem to be consistent~\cite{lhc-npdfs}.
There is a possibility that nuclear modifications are 
flavor dependent as studied in Refs.~\cite{flavor-depedent-emc,sk-ubar-dbar-nucleus},
and such effects may be found in neutrino DIS data.
In fact, flavor-dependent nuclear modifications were recently
investigated by the nCTEQ collaboration~\cite{other-npdfs}.
However, there is no clear data at this stage to indicate the 
flavor dependence for valence-quark and antiquark modifications. 
There is also Drell-Yan experiment in progress at Fermilab to probe
the flavor dependence and nuclear modifications in antiquark distributions 
by the E906-SeaQuest collaboration~\cite{Fermilab-E906}.
The flavor dependence could be investigated by a future JLab
experiment~\cite{jlab-flavor-dependence}, and the full analysis
of nuclear EMC ratio $F_2^A/F_2^D$ is also in progress at JLab.
In future, JLab and Fermilab Drell-Yan measurements will shed
light on the flavor separation of the nuclear modifications.

\subsection{Nuclear modifications in neutrino-nucleus DIS}

\begin{wraptable}{r}[0.0cm]{0.50\textwidth}
\vspace{-0.20cm}
\hspace{-1.90cm}
\begin{minipage}{0.80\textwidth}
\caption{\normalsize Neutrino DIS experiments.}
\label{tab:nu-experiments}
\end{minipage}
       \begin{minipage}[c]{0.3cm}
       \ \ 
       \end{minipage}
\footnotesize
\begin{center}
\vspace{-0.5cm}
\begin{tabular}{lcccc}
\hline
\           & \      & \   $\nu$ energy   & 
\multicolumn{2}{c}{\# of data} \\
Experiment	& \hspace{-0.2cm} Target &  (GeV) & $\nu$ & $\bar\nu$ \\
\hline
CDHSW~\cite{nu-cdhsw}   & Fe     & 20$-$212 &  465 & 464  \\
CHORUS~\cite{nu-chorus} & Pb     & 10$-$200 &  412 & 412  \\
NuTeV~\cite{Berge:1991hr}  & Fe     & 30$-$500 & 1168 & 966  \\
\hline
\end{tabular}
\end{center}
\normalsize
\end{wraptable}

Next, we discuss impacts of neutrino data on the determination of the NPDFs. 
The CDHSW, CHORUS, and NuTeV collaboration data~\cite{nu-cdhsw,nu-chorus,Berge:1991hr}
in Table~\ref{tab:nu-experiments} could be used in a global analysis.
In addition, there are measurements by WA25, WA59, Serpukhov, CDHS, and CCFR. 
However, since their errors are larger than those of CDHSW, CHORUS, and NuTeV,
they may not play a major role for determining NPDFs in a global analysis. 
There is no small $x$ ($<0.01$) data, and the kinematical range 
of the CDHSW, CHORUS, and NuTeV data is roughly comparable to 
the current charged-lepton DIS data. 
The advantages of the neutrino reactions are to probe 
flavor-dependent distributions and valence-quark distributions
by the structure functions $F_2$ and $F_3$
as shown in 
Eqs.~(\ref{eqn:f2-nu}) and (\ref{eqn:f3-nu}).
However, neutrino interactions are weak, so that a huge heavy-nuclear
target should be used. Nuclear corrections need to be taken into account
appropriately to discuss nucleonic PDFs from the neutrino DIS measurements.
For the charged-lepton DIS, the $F_2^A/F_2^D$-ratio data  are available,
whereas there is no deuteron measurement for the neutrino DIS. 
Therefore, instead of the neutrino DIS structure functions, the cross sections 
could be directly taken into account in the global $\chi^2$ analysis.
For example, the NuTeV collaboration supplied their data by both
cross sections and structure functions measured with the iron target.
There, the ``isoscalar corrections'' are applied to the structure function 
data, whereas raw data without the corrections are given for the cross
sections. 

\begin{figure}[t]
 \includegraphics[width=\textwidth]{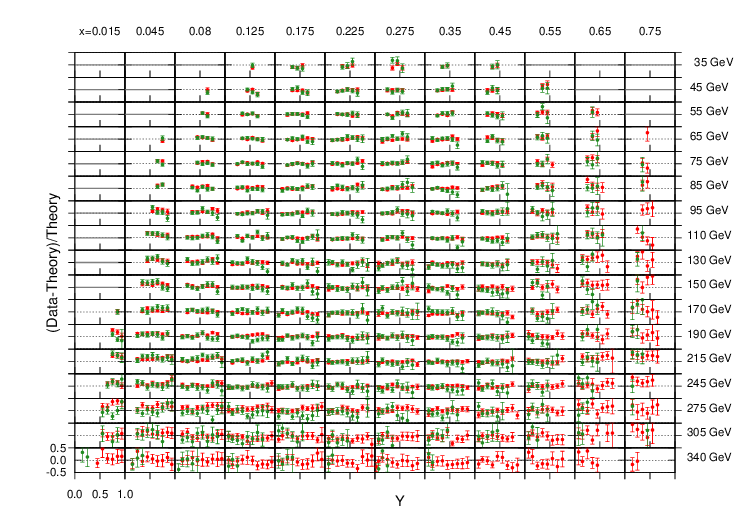}
\caption{Comparison of HKN calculations with NuTeV data
on $\nu$-Fe and $\bar\nu$-Fe cross sections~\cite{HKS-in-progress}.}
\label{fig:nutev-comparison}
\end{figure}

According to the nCTEQ analysis~\cite{nCTEQ-nu}, there are significant
nuclear modification differences between charged-lepton DIS and neutrino ones. 
Our preliminary analysis also tends to obtain a result similar
to the nCTEQ modifications which are different from the charged-lepton 
ones, especially in the region $x<0.4$.
If it is true, we should be careful in calculating neutrino-nucleus
cross sections in terms of the NPDFs determined mainly by 
the charged-lepton DIS measurements.
So far, we take the same nuclear modifications for granted 
in discussing neutrino and charged-lepton DIS processes for nuclei.
In the NPDF analyses of other groups (EPS, DSSZ)~\cite{other-npdfs}, there is no 
conspicuous difference, so that it is desirable to examine this 
issue by an independent analysis group. However, according to nCTEQ,
the different results are caused by using different neutrino data 
sets. The nCTEQ used the raw cross section data while
others used the corrected structure-function data.

We compare the cross sections data with the HKN calculations
at the same kinematical points. For example, the comparison with
the NuTeV data is shown in Fig.~\ref{fig:nutev-comparison}.
Almost all the data are compatible with the HKN except that
there are discrepancies at $x = 0.015$ and 0.045 columns. 
These differences indicate that shallower shadowing effect is 
needed for explaining the NuTeV data. Actually, the shallow shadowing 
and suppression of the antishadowing effect are reported by 
the nCTEQ analysis~\cite{nCTEQ-nu}. On the other hand,
the HKN parametrization seems to be consistent in the region 
$0.125 \le x \le 0.55$, $E_\nu \le 245$ GeV, so that nuclear 
modifications would be similar to the HKN ones in this medium-$x$ region.
Therefore, if the neutrino data are included in the global analysis,
the nuclear modifications could be rather different
from the ones obtained mainly from the charged-lepton DIS
at small $x$. Our analysis is in progress and more complete results 
will be reported elsewhere~\cite{HKS-in-progress}.
There is also an effort to compare the neutrino-$F_2$ data directly
with the charged-lepton $F_2$ for the iron nucleus~\cite{Kalantarians-2015},
and some differences seem to exist in the small-$x$ region ($x<0.05$).

Aside from this issue, the neutrino DIS experiment is in progress
by the MINER$\nu$A collaboration~\cite{Mousseau:2016snl} for various
nuclei such as carbon, iron, and lead. They reported 
negative nuclear modifications at $0.05<x<0.15$ 
significantly larger than current expectations from typical simulations. 
It means that nuclear effects in neutrino DIS are much different from our 
understanding by the current NPDFs at $x \sim 0.1$. It is an
interesting result to be investigated theoretically. 
At small $x$, the $W$ boson from the neutrino could propagate 
as axial-vector meson states in addition to vector meson ones,
which causes a shadowing phenomena by multiple scattering of
the mesons in a nucleus. Due to the additional axial-vector meson 
contributions, the shadowing could be different in the neutrino DIS 
from the one in the charged-lepton DIS. Although the MINER$\nu$A measurement
at $x \sim 0.1$ is still not in the shadowing region, 
such an effect may influence phenomena in this region. 

\subsection{Neutrino-nucleus reactions at $Q^2 \to 0$}

Since the neutrino energies are of the order of 1-10 GeV
for the current neutrino oscillation experiments, we need to have accurate
neutrino cross sections in a wide kinematical range shown in 
Fig.~\ref{fig:kinem}. There is a region, 
$0 < Q^2 < 1$ GeV$^2$ and $W^2 \ge 4$ GeV$^2$, 
where both the DIS and resonance descriptions do not apply. 
Due to the condition $Q^2 < 1$ GeV$^2$, the cross section 
cannot be described by partons. It also cannot be described 
by nucleon resonances due to the condition $W^2 \ge 4$ GeV$^2$. 
Before stepping into neutrino-nucleus reactions, we need to understand
neutrino-nucleon reactions at $Q^2 \to 0$.

In the next two paragraphs, we show that the neutrino cross sections at $Q^2\to 0$ are
related to the pion-nucleon cross sections thanks to the 
partial conservation of the axial-vector current (PCAC).
This observation will be a guidance to develop a model for the
kinematical region in question.
In general, the transverse ($T$) and longitudinal ($L$) structure functions 
are defined by corresponding total neutrino cross sections 
$\sigma_{T,\,L}$ as~\cite{Q2->0-theory}
\begin{align}
F_{T,\,L} (x,Q^2) & = \frac{1}{\pi} \, \sqrt{1+Q^2/\nu^2}
                    \, Q^2 \, \sigma_{T,\,L} \, ,
\nonumber \\
\sigma_{T,\,L}  & = \frac{(2 \pi)^4}{4\sqrt{(p\cdot q)^2-p^2 q^2}}
     \sum_f  \delta^{\,(4)} (p+q-p_f) 
     \left | \, \langle \, f \, | \, \varepsilon_{T,\,L} \cdot J (0) \, 
           | \, p \, \rangle \,  \right |^2 ,
\label{eqn:f-tl}
\end{align}
where $\varepsilon_{T,\,L}$ is the polarization vector of $W$ or $Z$.
In weak interactions, there are vector ($V$) and axial-vector ($A$) currents, 
which have transverse and longitudinal components. 
We know that the transverse cross section is finite at $Q^2 =0$. 
From Eq.~(\ref{eqn:f-tl}), this fact suggests that 
the transverse structure function should vanish ($F_T \to 0$) 
at $Q^2 \to 0$.

The vector-current part of the hadron tensor should satisfy 
$q^{\,\mu} W_{\mu\nu} =0$ because the vector current is conserved.
Therefore, the relation between the transverse and longitudinal ones 
($F_L^{\,V} \sim Q^2 F_T^{\,V}$) for the vector part is the same as
the charged-lepton DIS. The axial-vector current is not conserved,
so that a special attention needs to be paid at $Q^2 \to 0$.
It should be taken so as to satisfy the PCAC, 
$\partial_\mu \, A_a ^{\,\mu} (x) = f_\pi \, m_\pi^2 \, \pi_a (x)$,
where $f_\pi$ is the pion-decay constant, $m_\pi$ is the pion mass,
and $\pi_a (x)$ is the pion field with the isospin index $a$.
The PCAC leads to the relation that the axial-vector part of the
structure function is finite, and it is given by the pion-scattering 
cross section $\sigma_\pi$ as $F_L^{\,A} \to f_\pi^{\,2} \, \sigma_\pi /\pi$
at $Q^2 \to 0$.

In order to describe the neutrino cross section in this region
of $0 < Q^2 < 1$ GeV$^2$ and $W^2 \ge 4$ GeV$^2$, we need to
express the structure functions by separating them into 
vector and axial-vector components. Then, we use the PCAC 
for the axial-vector part. However, the extrapolation
of the structure functions into region $Q^2 <1$ GeV$^2$ is not
obvious, although they are known at $Q^2 = 0$.
Considering experimental data, we may determine the $Q^2$ dependence
with some theoretical guidance. We have not completed such an analysis
at this stage. There were some 
studies, and Bodek and Yang~\cite{by-2011}, for example, supplied 
a widely-used model that is obtained by extending 
the DIS structure functions so as to be consistent with the data
and the PCAC. However, a much simpler model without considering
the PCAC is sometimes used, for example, 
in the FLUKA simulation~\cite{fluka-2009} by taking
$F_{2,\, 3} (x,Q^2) = (2 Q^2)/(Q_0^2 + Q^2) \, F_{2,\, 3} (x,Q_0^2)$.
Further theoretical investigations 
in this kinematical region are highly called for.
Particularly, the current studies in this kinematical region 
are restricted to the neutrino-nucleon interactions. 
Because we have information on nuclear structure functions
$F_2^A$ and $F_3^A$, we should be able to estimate the 
neutrino-nucleus interactions and such theoretical efforts are needed.


\section{Perspective}
\label{sec:perspective}

We have seen the latest development of neutrino-nucleus reaction models
for the low-energy, QE, RES, and DIS regions;
our own developments are particularly highlighted.
Towards a unified description of the neutrino-nucleus reactions
for all the relevant energy region, further developments are needed in
each of the kinematical regions.
To close the paper, we remark below possible future developments.

\noindent
{\bf Few-body}:
The presented framework, correlated Gaussian combined with complex
scaling method, would be applied to electron- and
neutrino-induced reactions on light nuclei in the QE
region. Implementation of a modern nuclear force including genuine
three-body interaction will be important for more accurate descriptions
of nuclear responses with various momentum transfer as suggested in
Ref.~\cite{bacca-2009}. It would also be
interesting to study two-body current contributions. 
An {\it ab initio} calculation done with 
the Green's function Monte Carlo method
showed a rather large two-body current contributions~\cite{gfmc}.
An independent {\it ab initio} calculation is desired to evaluate
possible uncertainties associated with theoretical models and
calculational methods.

\noindent
{\bf QE}: 
It was shown that the inclusive electron scattering data in the QE region are well described
with the impulse approximation scheme. In the scheme,
the hole and particle propagations in nuclei are described by 
the nuclear spectral function, and FSI are accounted for by
the the convolution method.
On the other hand, the limitation of the approach is found
in describing the dip region for which more mechanisms must be taken into account~\cite{Luis2014,KM_review,martini,martini2,2p2h-spain}.
There are a large body of 
electron scattering data available over a broad range of energy and
momentum transfer that can be used 
to test various mechanisms such as 2p2h, meson-exchange current, and transparency.
The analysis of the electron scattering data can serve not only to determine
 the vector current part of the neutrino cross sections, but also 
to constrain the nuclear effects more stringently than the neutrino scattering data.
In interpreting data from long-baseline
neutrino oscillation experiments like T2K that mainly utilize
low-energy neutrino beam ($E_\nu \ltap$ 1~GeV),
a solid understanding of the CCQE(-like) processes is essential.
Therefore, only theoretical models that have been validated by the
electron scattering data should be used in analyzing the data.

\noindent
{\bf RES}: 
There are still remaining issues in the elementary neutrino-nucleon
reaction amplitudes.
All models have used cross section data for the elementary
single pion production processes measured in the bubble-chamber
experiments~\cite{Kitagaki:1986ct,anl}, in order to fix the axial
coupling for the $N$-$\Delta(1232)$ transition.
However, the experiments used the deuterium target and, 
as discussed in Ref.~\cite{wsl} and Sec.~\ref{sec:deuteron},
the FSI could be important to extract
the elementary-process cross sections from the data
although it was neglected in the published
analyses~\cite{Kitagaki:1986ct,anl}.
Thus it will be important to extend the analysis of Ref.~\cite{wsl}
to analyze the bubble-chamber data to extract the elementary cross
sections, and this is currently underway.
Another issue is $Q^2$ dependence of axial $N$-$N^*$ form factors for
higher resonances beyond $\Delta(1232)$.
Although the $Q^2$ dependence has been assumed to be a dipole form with
the axial mass of $\sim$ 1~GeV, 
this could be improved for $N^*$ near the DIS region
by matching the structure functions of
the DCC model with those calculated with the nucleonic PDF in the region
where the RES and DIS regions are overlapping; $W\sim$ 2~GeV and 
$Q^2>$ 1~GeV$^2$.
This matching can be done with the DCC model because the model includes
$\pi\pi N$ channel that would give a dominant contribution in this
region.
\begin{figure}[t]
\begin{center}
 \includegraphics[width=0.4\textwidth]{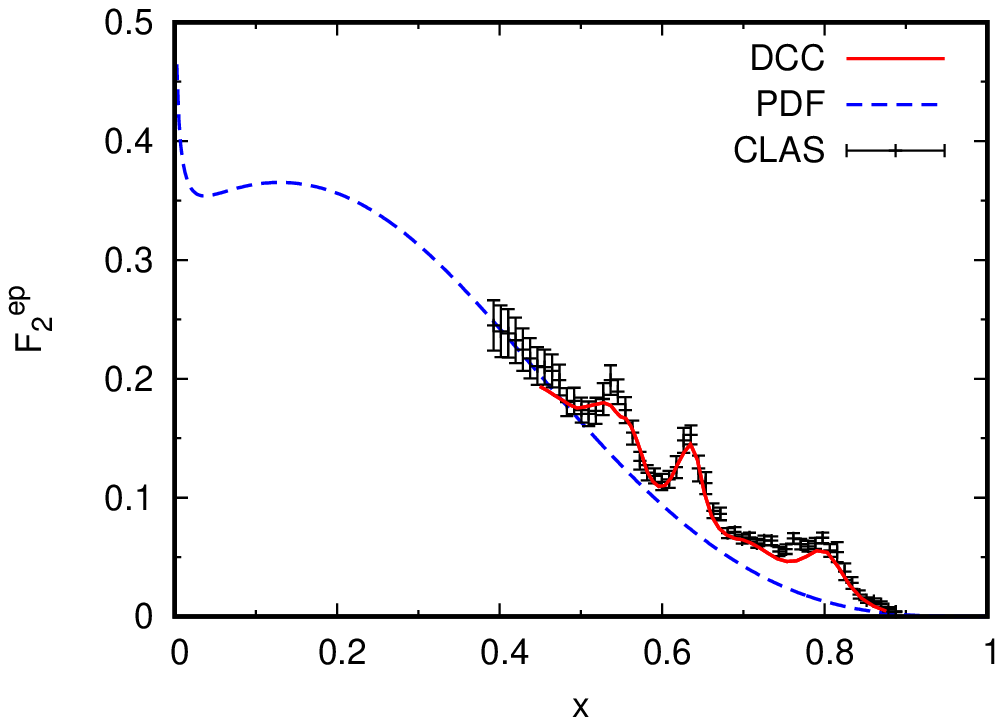}
\end{center}
\caption{\label{fig:matching} (Color online)
$F_2$ for the inclusive electron-proton scattering
at $Q^2=2.425~{\rm GeV}^2$
calculated with the DCC model (red solid curve) and with the nucleonic PDF
(blue dashed curve).
The horizontal axis is Bjorken scaling variable $x$.
The data are from Ref.~\cite{clas_f2}.
}
\end{figure}
Actually, as shown in Fig.~\ref{fig:matching}, the DCC model and the
nucleonic PDF are already reasonably consistent
for the vector (electromagnetic) current at $x\sim 0.45$ (where $W\sim$ 2~GeV), 
because both of them have been fitted to relevant inclusive electron
scattering data.
We expect a result similar to Fig.~\ref{fig:matching} can also be
achieved for the axial current by adjusting the 
$N$-$N^*$ form factors in the DCC model.
In future, lattice QCD will be able to provide us with information on
the axial vector response of the nucleon in the RES region, and this LQCD-based
input will be implemented in the reaction model~\cite{dina11}.

Next important task is to construct a neutrino-nucleus reaction model
in the RES region using the DCC elementary amplitudes as a building block.
In the $\Delta(1232)$ region, 
a quantum mechanical description is possible with the 
$\Delta$-hole model~\cite{annal99,annal108,annal120,taniguchi}.
As discussed in Sec.~\ref{sec:coh}, we have developed such a model for
the coherent pion productions.
It would be desirable to apply the method also to incoherent pion
productions that are relevant to the T2K experiment.
The DCC amplitudes should also be a reasonable input 
for a neutrino-nucleus reaction model for
the whole resonance region, and this development should be pursued.
It is worth emphasizing again that the DCC model is the only available
model that provides: (i) two-pion production amplitudes
with all the resonance contributions taken into account; 
(ii) the well-controlled interference between the resonance and non-resonance
contributions.

\noindent
{\bf DIS}: 
The NPDFs should be determined by including neutrino DIS data.
In particular, the issue of the difference between the nuclear modifications
of charged-lepton DIS and neutrino DIS should be clarified. Furthermore,
experimental information from LHC needs to be considered and it could
constrain the NPDFs in a small-$x$ region. Experimental efforts are
in progress to measure the nuclear modification in neutrino DIS
by the MINER$\nu$A collaboration. Its data will provide new information
in the anti-shadowing region. Because axial-vector mesons contribute
to the shadowing in the neutrino DIS in addition to vector mesons
according to the vector-meson-dominance model, shadowing could be 
different for neutrino DIS from the one for charged-lepton DIS. 
Such effects could affect the nuclear modifications in the neighboring
anti-shadowing region.
Therefore, the shadowing and anti-shadowing phenomena should be
interesting theoretical topics that can be studied with future neutrino
measurements.
Next, a model needs to be developed in the small $Q^2$($<1$ GeV$^2$)
region with large $W^2$($>4$ GeV$^2$).
Because this region cannot be described solely by DIS models or 
resonance models, simple empirical models have been used.
However, such studies are so far only for the neutrino-nucleon interactions,
and a realistic model for the neutrino-nuclear interactions should be
developed by taking into account proper nuclear modifications.

\section*{Acknowledgements}

This work was supported by JSPS KAKENHI Grant Number JP25105010, JPT16K053540
and JP25800149.
The work of M.S. is partly supported by the Grant-in-Aid for Scientific
Research on Innovative Areas (2014-2019, Project No. 26104006).

\end{document}